\documentclass[structabstract]{aa} 
\pdfoutput=1
\usepackage{natbib,amssymb,graphicx,graphics}
\usepackage{hyperref}
\usepackage{float}
\usepackage{graphicx}
\usepackage{txfonts}

\def\bp{$\mathrm{\beta}$~Pictoris\ }

\def\L'{m$_{L'}$\ }

\def\aap{A\&A\ }

\def\apjl{ApJL\ }

\def\apjs{ApJS\ }

\DeclareGraphicsExtensions{.jpg, .png, .bmp, .eps, .pdf, .tiff}

\begin{document}
\sloppy
\title{{\bf The  near-infrared spectral energy distribution of \object{$\mathrm{\beta}$~Pictoris b}}\thanks{Based on observations made with ESO telescopes at the Paranal
Observatory under programs  073.D-0534, 076.C-0339, 078.C-0472, 084.C-0739, 085.D-0625,  088.C-0358, and 090.C-0653}}
\subtitle{ }

\author{
 M. Bonnefoy \inst{1}
 \and
 A. Boccaletti \inst{2}
\and
 A.-M.~Lagrange \inst{3}
 \and
 F.~Allard \inst{4}
 \and
 C.~Mordasini \inst{1}
 \and
H. ~Beust \inst{3}
 \and 
 G.~Chauvin \inst{3}
  \and
 J. H. V. ~Girard \inst{5}
  \and
 D.~Homeier \inst{4}
  \and
 D. Apai \inst{6, 7}
  \and
S.~Lacour \inst{2}
  \and
 D.~Rouan \inst{2} 
}


\institute{
  Max Planck Institute for Astronomy, K\"{o}nigstuhl 17, D-69117 Heidelberg, Germany\\ 
\email{bonnefoy@mpia-hd.mpg.de}
\and
LESIA, Observatoire de Paris, CNRS, University Pierre et Marie Curie Paris 6 and University Denis Diderot Paris 7, 5 place Jules Janssen, 92195 Meudon, France 
\and
 UJF-Grenoble 1 / CNRS-INSU, Institut de Plan\'{e}tologie et d'Astrophysique de Grenoble (IPAG) UMR 5274, Grenoble, F-38041, France
 \and
CRAL, UMR 5574, CNRS, Universit\'{e} de Lyon, \'{E}cole Normale Sup\'{e}rieure de Lyon, 46 All\'{e}e d'Italie, F-69364 Lyon Cedex 07, France
\and
  European Southern Observatory, Casilla 19001, Santiago 19, Chile 	
  \and
 Department of Astronomy, 933 N. Cherrt Avenue, Tuscon, AZ 85721, USA
  \and
 Department of Planetary Sciences, The University of Arizona, 1929 E. University Blvd., Tuscon, AZ 85721, USA
}

   \date{Received December 3rd, 2012; accepted February 5th, 2013}

 
  \abstract
   {A gas giant planet has previously been directly seen orbiting at 8-10 AU within the debris disk of the $\sim$12 Myr old star $\beta$ Pictoris. The $\beta$ Pictoris system offers the rare opportunity to study the physical and atmospheric properties of an exoplanet placed on a wide orbit and to establish its formation scenario.}
   {We aim to build for the first time the 1-5 $\mu$m spectral energy distribution (SED) of the planet. Our goal is to provide secure and accurate constraints on its physical and chemical properties.}
   {We obtained J (1.265 $\mu$m),  H (1.66 $\mu$m), and M' (4.78 $\mu$m) band angular differential imaging of the system between 2011 and 2012.  We use Markov-Chain Monte-Carlo simulations on the astrometric data to revise constraints on the orbital parameters of the planet. Photometric measurements are compared to those of ultra-cool dwarfs and young companions. They are combined with existing photometry (2.18, 3.80, and 4.05 $\mu$m) and compared to predictions from 7 PHOENIX-based atmospheric models in order to derive the atmospheric parameters ($T_{eff}$, log g) of $\beta$ Pictoris b. Predicted properties from (``hot-start", ``cold-start" and ``warm start") evolutionary models are compared to independent constraints on the mass of $\beta$ Pictoris b. We use planet population synthesis models following the core-accretion paradigm to discuss the planet possible origin.}
  {We detect the planetary companion in our four-epoch observations. We estimate $\mathrm{J=14.0\pm0.3}$, $\mathrm{H=13.5\pm0.2}$, and  $\mathrm{M'=11.0 \pm 0.3}$ mag. Our new astrometry  consolidates previous semi-major axis (8-10 AU) and excentricity (e $\leq$ 0.15) estimates of the planet. The location of $\beta$ Pictoris b in color-magnitude diagrams  suggests it has spectroscopic properties similar to L0-L4 dwarfs. This enables to derive $\mathrm{Log_{10}(L/L_{\sun})=-3.87\pm0.08}$ for the companion.  The analysis with atmospheric models reveals the planet has a dusty atmosphere with $\mathrm{T_{eff}=1700\pm100}\:K$ and $\mathrm{log\:g= 4.0\pm0.5}$.  ``Hot-start" evolutionary models give a new mass of $\mathrm{10^{+3}_{-2}\:M_{Jup}}$ from $\mathrm{T_{eff}}$ and $\mathrm{9^{+3}_{-2}\:M_{Jup}}$ from luminosity. Predictions of ``cold-start" models are still inconsistent with independent constraints on the planet mass. ``Warm-start" models constrain the mass to $\mathrm{M\geq6 M_{Jup}}$ and the initial entropies to values ($\mathrm{S_{init}\geq9.3 K_{b}/baryon}$) intermediate between those considered for cold/hot-start models, but likely closer to those of hot-start models.}
  {}

   \maketitle

\section{Introduction}

\begin{table*}
\begin{minipage}[ht]{\linewidth}
\caption{Log for the new J, H, and M'-band observations.}
\label{table:logobs}
\centering
\renewcommand{\footnoterule}{}  
\begin{tabular}{l l l l l l l l l l l l}
\hline \hline
Date & Band & Density filter & Camera &DIT & NDIT & $\mathrm{N_\mathrm{exp}} $ & $\mathrm{\theta_{start}}$\tablefootmark{a}  & $\mathrm{\theta_{end}}$\tablefootmark{b} & $\mathrm{EC\:mean }$\tablefootmark{c} & $\mathrm{\tau_{0}\: mean}$\tablefootmark{d}  \\
    &      &      &&  (s)    &          &                    & ($\mathrm{\degr}$)     & ($\mathrm{\degr}$)     & (\%)      & (ms)       \\
\hline
16/12/2011		&		J		&		ND\_short 	&  S13	&		0.500		&		100		&		8  	&  -34.110 & -30.882   	&  60.95 &  3.26		\\
16/12/2011		&		J		&		\dots		 	&  S13	&		0.500		&		100		&		48    &  -25.394&-2.596  	& 61.11 & 3.87 			\\
16/12/2011		&		J		&		ND\_short 	&  S13	&		0.500		&		100		&		8  	&  -2.309&1.514   		&  59.60 &  3.73			\\
18/12/2011		&		H		&		ND\_short 	&	S13	&		0.500		&		100		&		8		&	-28.267&-24.834	&	 56.86 & 2.79  			\\
18/12/2011		&		H		&		\dots			&	S13	&		0.500		&		100		&		48	&	-24.559&-1.654		&	54.17	&	2.52			\\
18/12/2011		&		H		&		ND\_short 	&	S13	&		0.500		&		100		&		8		&	-1.350&2.525			&	 56.61 & 2.21  			\\
11/01/2012		&		H		&		ND\_short 	&	S13	&		0.200		&		150		&		8		&	-26.530&-24.305	&	 56.60	&	3.42			\\
11/01/2012		&		H		&		\dots			&	S13	&		0.400		&		150		&		84	&	-22.563&12.082		&	61.00	&	3.18			\\
11/01/2012		&		H		&		ND\_short 	&	S13	&		0.200		&		150		&		8		&	14.879&17.268		&	71.55 & 2.80			\\
26/11/2012	&		M'		&		ND\_long		&	L27	&		0.100		&		250		&		8	&	-28.362&-26.063   &  27.93	&	1.27			\\
26/11/2012	&		M'		&		\dots		&	L27 &		0.065	&		300	&	184	&	-19.482&32.322   &  19.72	&	1.10				\\
26/11/2012		&		M'		&	ND\_long		&L27	&	0.130		&		150		&		8		&	32.721&34.411	   &  13.25	&	0.96			\\

\hline
\end{tabular}
\end{minipage}
\tablefoot{ 
\tablefoottext{a}{Parallactic angles ($\theta$) at the begining of the observations.} 
\tablefoottext{b}{Parallactic angles ($\theta$) at the end of the observations.} 
\tablefoottext{c}{Average coherent energy during the observations.} 
\tablefoottext{d}{Average coherence time during the observations.} 
}
\end{table*}

Understanding planetary systems formation and evolution has become one of the outstanding topics in astronomy, since the imaging of a debris disk around the young (~$\simeq$ 12 Myr) and massive star ($M=1.75M_{\odot}$) $\beta$ Pictoris, in the 80's \citep{1984Sci...226.1421S} and the discovery of the first exoplanet around a solar-type star in the 90's \citep{1995Natur.378..355M}. The few dozen of  wide-orbit ($>5$AU) planetary mass companions directly imaged around young and nearby stars suggest that alternative formation mechanisms to the core-accretion scheme prefered for  closer-in exoplanets should be considered.  In particular, the four giants planets discovered at projected separations of 15-68 AU around the young and  massive star HR8799 \citep{2008Sci...322.1348M, 2010Natur.468.1080M} can not be all explained by \textit{in-situ} formation by core-accretion. Instead, disk instability have been proposed to account for the observed properties of these objects \citep{2009ApJ...707...79D, 2011ApJ...731...74B}.

The atmospheric properties of wide-orbit (non-irradiated) and young ($\lesssim 150$ Myr old) ``exoplanets" discovered by direct imaging can be studied based on multiple-band photometry \citep{2004A&A...425L..29C} and even low-resolution spectra  \citep{2008A&A...489.1345V}  in the near-infrared (1-5 $\mu$m). \cite{2011ApJ...735L..39B} and \cite{2011ApJ...732..107S} have used atmospheric models to conclude that non-equilibrium chemistry of $\mathrm{CO/CH_{4}}$ and/or thick clouds of refractory elements could account for the photometric and spectroscopic properties of the planetary mass companion 2M1207b. \cite{2010A&A...512A..52B} and \cite{2012arXiv1206.5519F} also recently showed that the near-infrared  ($1.1-1.4\mu$m and $1.5-2.5\mu$m) spectra of young planetary mass companions AB Pic b and 2M1207 b match those of young and isolated brown dwarfs \citep{2006ApJ...639.1120K, 2009AJ....137.3345C}. This also provided the empirical evidence that some of the peculiar photometric/spectroscopic features of these companions can be interpreted as a consequence of  their lower surface gravity (compared to field L-T dwarfs). The description of the photometric and spectroscopic properties of the cooler planets found around HR8799 requiere a more complex picture possibly involving thick clouds \citep{2011ApJ...737...34M, 2011ApJ...729..128C}, non equilibrium chemistry \citep{2010ApJ...710L..35J, 2011ApJ...733...65B} or even multiple cloud layers \cite{2012ApJ...753...14S}. It is still unclear whether these properties account for the young age - and reduced surface gravity - of the objects \citep{2012ApJ...754..135M} or could be related to the peculiar nature of this system. 

\bp is one of the best laboratories for the study of early phasis of planetary systems formation and evolution. It has been the subject of hundreds of observations over a broad spectral range. With NaCo at VLT, we detected a giant planet orbiting the star \citep{2009A&A...493L..21L, 2010Sci...329...57L}, roughly along the dust disk position angle (PA), with a semi-major axis $a$ between 8 and 14 AU. Its observed L' luminosity (3.8 $\mu$m) indicated it has $\mathrm{T_{eff}\simeq1700 K}$ and a mass of 7-11 M$_{Jup}$. Follow-up observations at 4.05 $\mu$m \citep[NB\_4.05 filter; ][]{2010ApJ...722L..49Q} and  at 2.18 $\mu$m ($\mathrm{K_{s}}$ band) \citep{2011A&A...528L..15B} confirmed these values. Additional astrometric data obtained from October 2010 to March 2011 showed that  the semi-major axis is in the range 8-10 AU, and the eccentricity is less than 0.2 \citep[Fig. 1 of][]{2012A&A...542A..41C}. \bp b is so far the closest planet ever imaged around a star. 

These predictions of \bp b properties ($\mathrm{T_{eff}}$, mass), like those of imaged planets or companions rely however on models that are currently debated: either the so-called "hot-start" models which assume a spherical collapse  from an arbitrary large initial radius \citep{2000ApJ...542..464C, 2003A&A...402..701B}, or "cold-start" models \citep{2007ApJ...655..541M, 2012ApJ...745..174S, 2012A&A...547A.111M} which are supposed to predict the properties of planets produced by accretion of gas through a super-critical accretion shock onto the planet embryo.  "Cold-start" models of \cite{2007ApJ...655..541M} and \cite{2012ApJ...745..174S}  predict giant planets more than 100 times fainter at 10 Myr than the "hot start" model. The main difference is due to the gas initial properties (temperature, entropy), which depends on the fate of the energy released during the gas accretion and shock process.

Based on current models predictions \citep{2008ApJ...673..502K}, and given its separation, $\beta$ Pictoris b could have formed via core accretion \citep{2010Sci...329...57L}.   Using limits coming from radial velocities measurements \citep{2012A&A...542A..40L}, we could constrain its dynamical mass to be $\leq$ 15.5 $\mathrm{M_{Jup}}$ and $\leq$ 12 $\mathrm{M_{Jup}}$ if the semi-major axis is shorter than 10  and 9 AU respectively. Current "cold-start"  models fail to predict masses from the available photometry that respect these dynamical mass constraints.\\

We present new data obtained at J (1.265 $\mu$m), H (1.66 $\mu$m), and M' (4.78 $\mu$m) bands that we combine to previously published data at 2.18, 3.8, and 4,05 $\mu$m to build the near-infrared (1--5 $\mu$m) spectral energy distribution (SED) of the planet. The goal of the study is to achieve a better characterization of the atmospheric ($\mathrm{T_{eff}}$, log g, composition) and physical (radii, mass) properties of the companion needed to understand its formation process and evolution. The paper is organized as follows: Sections \ref{obs} and \ref{dataproc} respectively describe the data and their processing. Our core results are presented in Section \ref{Results}, divided in four main sub-sections. In Sub-section \ref{subsec:empanal} we compare the resulting colors to those of reference young and old objects, found isolated or as companions to star/brown-dwarfs. We analyse  in Sub-section \ref{subsec:specsynth} the SED of the planet with atmospheric models. We present new orbital derived from new astrometry of the planet in Sub-section \ref{orbitpar}. We  derive new model-dependent mass estimates of the companion in Sub-section \ref{newmasses}. The results and their implication on the formation history of the planet are discussed in Section \ref{Discussion}.

\section{Observations}
\label{obs}
We used the Nasmyth Adaptive Optics System (NAOS) coupled to the Near-Infrared Imager and Spectrograph (CONICA) at VLT/UT4 \citep[NACO; ][]{2003SPIE.4841..944L, 2003SPIE.4839..140R} to obtain four new high-contrast observations of $\beta$ Pictoris (see Table \ref{table:logobs}) between fall-2011 and 2012. 

The star was first observed on December 16, 2011 in the J band ($\mathrm{\lambda_{c}=1.265 \:\mu m ;\:width=0.25\:\mu m}$) with the S13 camera. We recorded consecutively 4800$\times$0.5s  exposures stored in 48 datacubes (100 frames each) in pupil stabilized mode. The field orientation changed by 22.8$^{\circ}$ during this sequence. The data integration time was chosen to saturate the core of the flux distribution of $\beta$ Pictoris over a radius of $\sim$150 mas. This and the choice of the appropriate reading mode (double\_rd\_rstd)  ensured we had the necessary detector dynamics to detect the companion. The star was dithered ($\pm$3" amplitude) through the instrument field-of-view every 200 exposures (1 second) during the observations to perform a first order subtraction of  bad pixels, detector bias, and background.  We took 8 additional 100$\times$0.5s  unsaturated exposures of the star with a neutral density (ND\_short) before and after the sequence of saturated images.  These data were later used for the calibration of the flux and position of $\beta$ Pictoris b.   

We repeated this sequence with the H band filter ($\mathrm{\lambda_{c}=1.66 \:\mu m ;\:width=0.33\:\mu m}$)  on December 18, 2011. The parallactic angle variation is close to the one of the J band sequence (22.9$^{\circ}$ variation of field orientation). 

We obtained second epoch observations of $\beta$ Pictoris in the  H band on January 11, 2012. These observations benefited from excellent atmospheric conditions and increased field rotation (34.6$^{\circ}$). We lowered the  exposure time to 0.4s to maintain the saturated area to separations shorter than 70 mas.  The star was kept at a fixed position in the instrument field of view (no dithering). We recorded a series of sky  frames at the end of the sequence to subtract the bias and remove hot pixels.  The suppression of telescope offsets and  point-spread function (PSF) drifts previously observed when tracking the pupil through meridian  (see the NaCo user manual version 91.0) improve the stability of the PSF and simplify the registration of individual frames (see Section \ref{dataproc}). Two sequences of 8$\times$150$\times$0.2s unstaturated exposures of the star were taken at the begining/end of the sequence for the estimation of the photometry and the astrometry. 

We finally collected a sequence of  M' band ($\mathrm{\lambda_{c}=4.78 \:\mu m ;\:width=0.59\:\mu m}$) saturated exposures (184 datacubes with 300 frames each and 0.065s  individual integration times) of the star on November 26, 2012 in pupil-stabilized mode.  The L27 camera ($\sim$27.12 mas/pixel) was used for these observations. We recorded 8 non-saturated exposures with a neutral density (ND\_long) before/after the sequence with a higher integration time (0.1 and 0.13s respectively) for the same purpose as above. The median observing conditions were degraded compared to the J and H band observations. Nevertheless,  the observations benefited from the improved Strehl ratio and point-spread function stability  at these wavelengths.
 
   \begin{figure}
   \centering
   \includegraphics[width=\columnwidth]{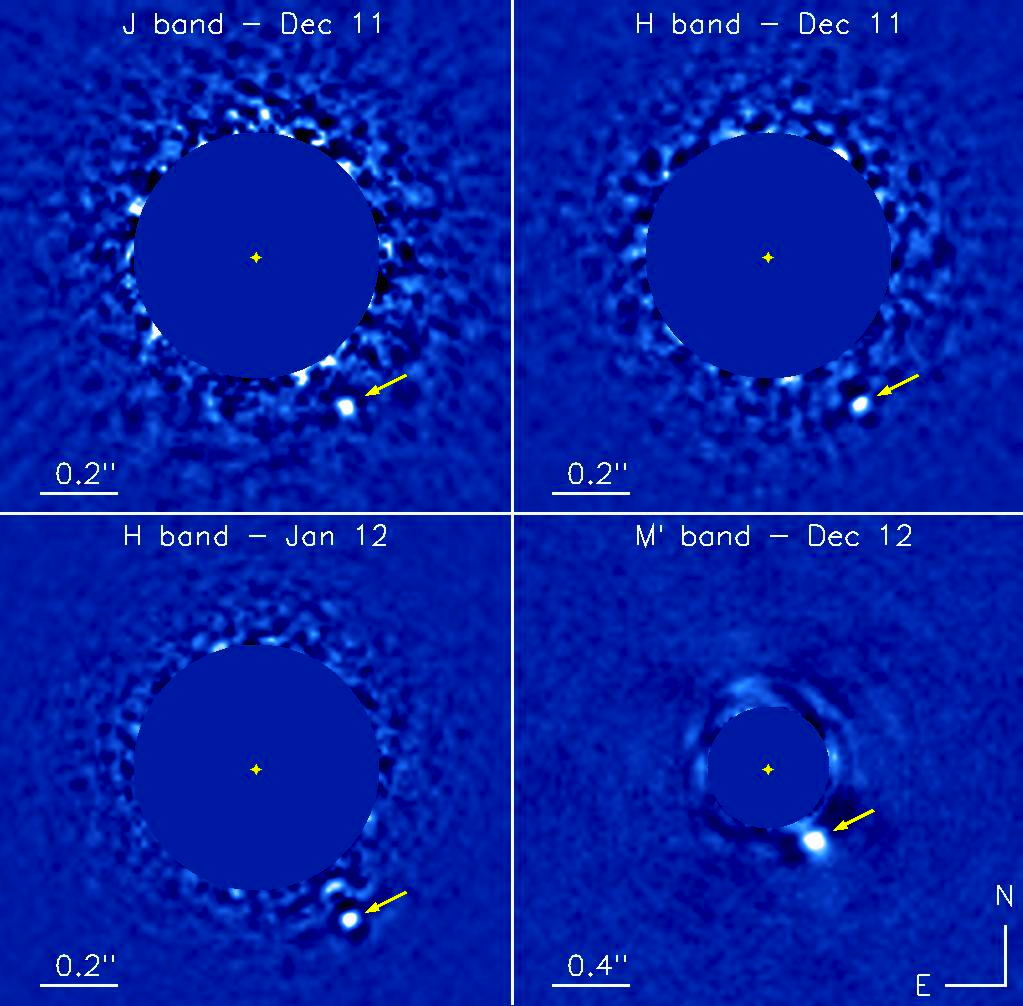}
      \caption{Re-detection of $\beta$ Pictoris b in the J (upper-left panel) band, in the first epoch (upper-right panel) and second epoch (lower-left panel) H band observations, and in the M' band (lower-right panel).  Arrows indicate the planet  position.}
         \label{Fig:Fig1}
   \end{figure}

\section{Data processing}
\label{dataproc}
\subsection{Initial steps}
Each  dataset is composed of a sequence of raw datacubes containing 100 to
 300 frames each (see Table \ref{table:logobs}). We reduced the data using two independent pipelines \citep{2010Sci...329...57L, 2011A&A...528L..15B, 2012A&A...544A..85B, 2012A&A...542A..41C, 2012A&A...542A..40L, 2012A&A...545A.111M}. Pipelines first apply basic cosmetic steps to all frames (sky subtraction, flat fielding, bad pixel interpolation) contained in each raw cubes. They use weighted bi-dimensional Moffat fitting  to find the position of the saturated star for each exposures and center it in the field of view. The parallactic and hour angles associated to each frame is computed using the Universal Time corresponding to the  first and last exposures contained in each raw cube respectively\footnote{We retrieve the sideral time  at Paranal (LST) from the Universal Time. The LST and target coordinate is used to derive the hour angle. The hour angle is then converted to parallactic angle.}. Frames with low encircled energy and small saturated area are flagged and removed from the datacube. The datacube are then binned temporally  and placed in a final master cube. The master cube and the list of parallactic angles are used as input of routines which apply the classical-ADI (CADI; also often called  ``median-ADI"), radial-ADI \citep[RADI; ][]{2006ApJ...641..556M}, and LOCI \citep{2007ApJ...660..770L} algorithms to remove the flux distribution of the star. We also checked our results for three of the most favorable datasets (see Table \ref{table:logobs}) with the innovative algorithm of \cite{2012ApJ...755L..28S} based on Principal Component Analysis (KLIP).

\subsection{Photometry and astrometry}
\label{sec:photastro}
$\beta$ Pictoris b is  detected in the J band with the RADI, LOCI, and KLIP algorithms (see Figure \ref{Fig:Fig1}). We also retrieved the planet in our two epochs H band data and single epoch M' band data. We find nominal SNR of 8, 17, 23, and 11 with  LOCI on the three set of data (for separation criteria of 1.00, 1.00, 0.75, and 0.5 $\times$FWHM respectively). To compute these SNR, we measured the signal in a circular aperture (3-pixels in radius) centered on the planet. The noise per pixel was estimated on a ring at the same radius (but excluding the planet) and translated to the same surface as the circular aperture (quadratically).

\begin{table}
\begin{minipage}{\columnwidth}
\caption{Photometric measurements of $\beta$ Pictoris b.}
\label{measurements}
\centering
\renewcommand{\footnoterule}{}  
\begin{tabular}{llllll}
\hline \hline 
Obs. Date			 		&    Filter	&	  Methods 					&		 Contrast	 				\\   			
								&				&										&		(mag)						\\	
\hline
 16/12/2011		    &   	J		&		RADI		&	$\mathrm{10.4\pm0.5}$		\\
 16/12/2011		    &   	J		&		LOCI		&	$\mathrm{10.5\pm0.3}$		\\
16/12/2011		    &   	J		&		KLIP		&	$\mathrm{10.8\pm0.3}$		\\
\hline
 18/12/2011		    &   	H		&		CADI		& $9.9\pm0.3$	 		\\
 18/12/2011		    &   	H		&		RADI		& $10.2\pm0.4$	 		\\
 18/12/2011		    &   	H		&		LOCI		& $9.9\pm0.3$	 		\\
\hline
 11/01/2012		    &   	H		&		CADI		&		$\mathrm{10.1 \pm 0.3}$ \\
 11/01/2012		    &   	H		&		RADI		&		$\mathrm{10.0 \pm 0.2}$ \\
 11/01/2012		    &   	H		&		LOCI		&	 	$\mathrm{9.9 \pm 0.2}$ \\
11/01/2012		    &   	H	&		KLIP 	&	 	$\mathrm{10.0 \pm 0.3}$ \\
\hline
26/11/2012		    &   	M'		&		CADI		&		$\mathrm{7.5 \pm 0.3}$ \\
26/11/2012		    &   	M'		&		RADI		&		$\mathrm{7.5 \pm 0.3}$ \\
26/11/2012		    &   	M'		&		LOCI		&	 	$\mathrm{7.6 \pm 0.3}$ \\
26/11/2012		    &   	M'		&		KLIP 	&	 	$\mathrm{7.8 \pm 0.3}$ \\
\hline
\end{tabular}
\end{minipage}
\end{table}

The position and flux of the planet  is affected by the inevitable partial point source self-subtraction occuring in ADI  \citep{2007ApJ...660..770L, 2010SPIE.7736E..52M, 2011A&A...528L..15B}. We estimated the level of self-subtraction injecting artificial planets in the raw frames built from renormalized unsaturated exposures of the star taken before and after the sequence of saturated exposures.  We  first injected  artificial planets at seven different position angles (256, 301, 346, 31, 76, 121, and 166$^{\circ}$) and different positive fluxes (flux ratio from $1\times10^{-5}$ to $1\times10^{-4}$ for the J band, and $5\times10^{-5}$ to $1.5\times10^{-4}$ for the H band) at the guessed planet separation in the raw J and H band data and applied the ADI algorithms. We repeated this procedure in the M' band but considered only three position angles (32, 122, 302$^{\circ}$) to avoid a cross-contamination of the angular-differential imaging pattern induced by the artificial planets. We then compared the flux of $\beta$ Pictoris b integrated over a circular aperture (radius of 3 pixels) to those of  artificial planets in order to  calibrate the self-subtraction (hereafter $FPPOS$). We also followed an alternative approach  injecting artificial planets one by one with negative fluxes at the companion position  \citep[$FPNEG$][]{2011A&A...528L..15B, 2012A&A...542A..41C}. We retrieved the flux of $\beta$ Pictoris b minimizating the standard deviation of the residuals at the injection position for each injection flux. We finally derived contrast values using the estimated flux of $\beta$ Pictoris b, the flux of the unstaturated frames corrected from the transmission of the neutral density filters and renormalised to the integration time of the saturated exposures. We give transmissions of NaCo neutral densities for the J, H, $\mathrm{K_{s}}$, L', and M' band filters that we measured from on-sky data for that purpose in Appendix \ref{App:transm}. We repeated the procedure for the CADI, RADI, LOCI algorithms using 6 separation criteria (0.25, 0.5, 0.75, 1, 1.25, and 1.5$\times$FWHM respectively) and   with the KLIP algorithm keeping only the 10 and 20 first Karhunen-Lo\`{e}ve transforms of the references images ($\mathrm{K_{KLIP}}$ parameter). $FPPOS$ and $FPNEG$ give similar constrasts when used with LOCI. Nevertheless, the level of self-subtraction with LOCI is not expected to vary  linearly with the injection flux  and could  biased the contrast derived with $FPNEG$. We then chose to conserve the constrast values derived from $FPPOS$ with LOCI only. 

   \begin{figure}
   \centering
   \includegraphics[width=\columnwidth]{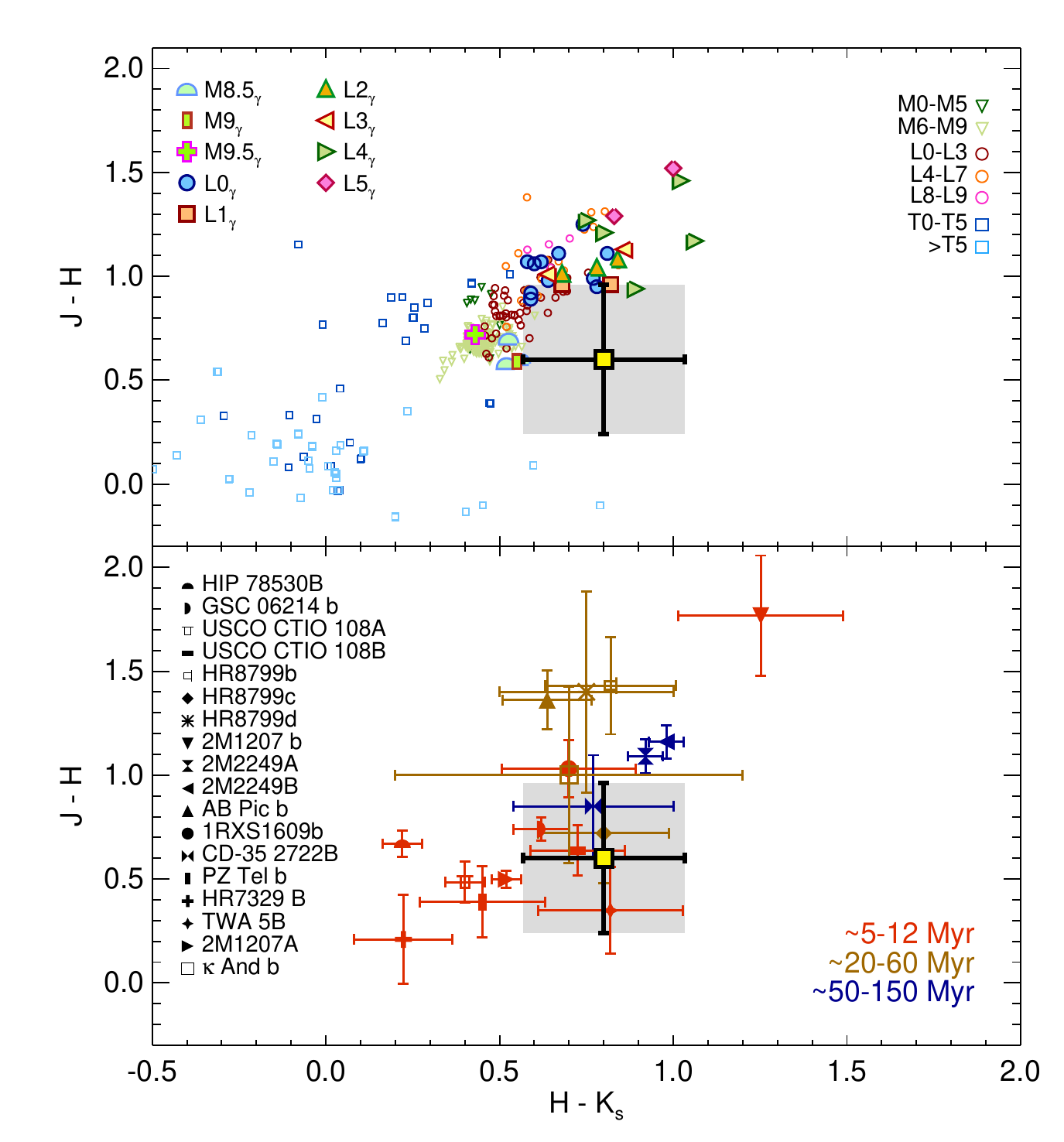}
      \caption{Location of $\beta$ Pictoris b (yellow square) in a J-H versus H-$\mathrm{K_{s}}$ color-color diagrams.}
         \label{Fig:Fig3}
   \end{figure}

The mean contrast values for each ADI algorithm are reported in Table \ref{measurements}. The error is dominated by the variation of the flux of non-saturated exposures (variable Strehl ratio and atmospheric transmission). KLIP confirms previous contrast estimates found with CADI, RADI, and LOCI. The algorithm could provide a more accurate estimate of the planet contrast. Nevertheless, it has not been extensively tested on NaCo data yet. We chose not to use the corresponding contrast values in the following analysis.

We extracted  the position of the planet  with respect to the star in the high signal-to-noise residual frames obtained from January 2012 data using the $FPNEG$ method described in \cite{2012A&A...542A..41C}. Results are reported in Table \ref{astrom}. We used the instrument True North and Platescale values derived from observations of the $\Theta$ Ori C astrometric field \citep{1994AJ....108.1382M}. The uncertainty on the star center dominates the final error budget. We analyze this new astrometry and its implication on the orbital parameters of $\beta$ Pictoris b in Section \ref{orbitpar}.

\begin{table}[t]
\begin{minipage}[ht]{\linewidth}
\caption{Astrometry of $\beta$ Pictoris b for the H-band data obtained on January 11, 2012.}
\label{astrom}
\centering
\renewcommand{\footnoterule}{}  
\begin{tabular}{ll}
\hline \hline
UT Date & 11/01/2012 \\ 
True North\tablefootmark{a} &  $-0.53\pm0.03$ deg \\ 
Platescale\tablefootmark{a} &  $13.22\pm0.02$ mas \\ 
Rotator Offset &  $90.46\pm0.10$ deg\\ 
$\Delta RA$ & $-241\pm13$ mas \\
$\Delta DEC$ & $-387\pm14$ mas \\
Separation & $456\pm11$  mas \\
 PA  & $211.89\pm2.32$ deg \\
\hline
\end{tabular}
\end{minipage}
\tablefoot{ 
\tablefoottext{a}{Derived from observations of the Theta Ori C astrometric field (ESO Large Program 184.C-0567) taken in the H band with the S13 camera on January 2, 2012.} 
}
\end{table}

  \begin{figure*}
   \centering
   \includegraphics[width=\linewidth]{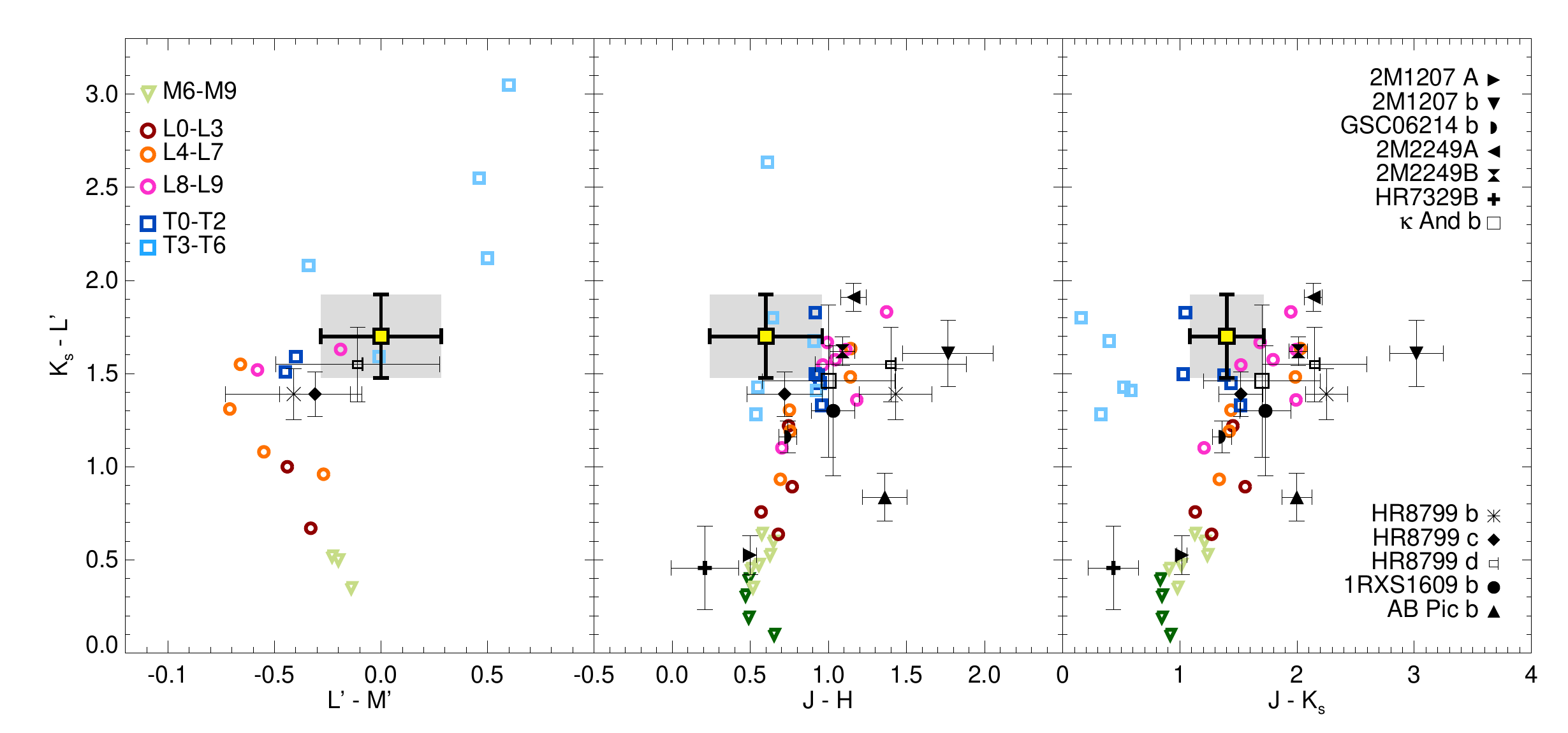}
      \caption{Location of $\beta$ Pictoris b (yellow square) in $\mathrm{K_{s}}$-L' versus L'-M' (left panel), J-H (middle  panel), and J-$\mathrm{K_{s}}$ (right  panel) color-color diagrams with respect to field M (open triangles), L (open circles), and T (open squares) dwarfs.  Colors of known young objects (companions, binaries) with available L'-band and M'-band photometry  are overlaid.}
         \label{Fig:Fig2}
   \end{figure*}

\section{Results}
\label{Results}
We derive contrasts of $\mathrm{\Delta H=10.0\pm0.2}$ mag  and $\mathrm{\Delta M'=7.5 \pm 0.2}$ mag for $\beta$ Pictoris b (see Table \ref{measurements}). The planet is detected at a lower SNR in the J band when using RADI ($\mathrm{SNR\leq6}$). For this reason, we decided to retain the photometric measurement obtained with LOCI only in this band for the following analysis ($\mathrm{\Delta J=10.5\pm0.3}$ mag).

We convert the ESO J and H-band photometry of $\beta$ Pictoris \citep{1996A&AS..119..547V} to the 2MASS photometric system (J=$3.524\pm0.013$ mag, H=$3.491\pm0.009$ mag) using the color transformations  of \cite{2001AJ....121.2851C}. We estimate that the magnitude difference between 2MASS and NaCo systems for a A6 star is less than 0.002 mag using  the mean spectrum of A5V and A7V stars of the \cite{1998PASP..110..863P} library, a flux-calibrated spectrum of Vega \citep{2007ASPC..364..315B}, and the corresponding filter passbands. We then measure $\mathrm{J=14.0\pm0.3}$ and $\mathrm{H=13.5\pm0.2}$ for  $\beta$ Pictoris b. This corresponds to absolute magnitudes $\mathrm{M_{J}=12.6\pm0.3}$ mag and $\mathrm{M_{H}=12.0\pm0.2}$ mag at $19.44\pm0.05$ pc \citep{2007A&A...474..653V}. We combined the M' band contrast of $\beta$ Pictoris b we obtained to the star magnitude \citep[M=$3.458\pm0.009$;][]{1991A&AS...91..409B} and derived $\mathrm{M'=11.0\pm0.2}$ mag and $\mathrm{M_{M'}= 9.5\pm0.3}$ mag. \cite{2011ApJ...736L..33C} previously derived a photometry of the planet at M' band (4.98 $\mu$m) from ADI observations .  We did not used this photometric data point as comparison since the dataset was missing unsaturated exposures.  Without this calibration, it is not possible to estimate properly the level of self-subtraction associated to the data reduction method

We  use this new photometry and the one reported in \cite{Quanz2010} and \cite{2011A&A...528L..15B} for the $\mathrm{K_{s}}$ ($\mathrm{2.18 \mu m}$), $\mathrm{L'}$ ($\mathrm{3.8 \mu m}$), and NB\_4.05 ($\mathrm{4.05 \mu m}$) filters for the following analysis. It is summarized in Table \ref{tab:sum}.

\subsection{Empirical comparison}
\label{subsec:empanal}
\subsubsection{Color-color diagrams}
We  overlaid $\beta$ Pictoris b colors in $\mathrm{J-H}$ vs $\mathrm{H-K_{s}}$ diagrams in Figure \ref{Fig:Fig3} along with additional field dwarfs objects \citep{2006ApJ...637.1067B, 2008AJ....136.1290R, 2012ApJS..201...19D} and low mass companions (see Table \ref{tab:compobj} for a description of the objects). We also overlaid the colors of  M8.5$_{\gamma}$-L5$_{\gamma}$ dwarfs having spectral features indicative of reduced surface gravity and likely or confirmed members of young nearby associations \citep[][and ref. therein]{2007ApJ...669L..97L,  2010ApJ...715L.165R,2012arXiv1206.5519F}.  $\beta$ Pictoris b  has colors compatible with field L dwarfs and redder than T dwarfs. The  comparison also reveals that the L5$_{\gamma}$, and most of the L4$_{\gamma}$  dwarfs are redder  than the planet. $\beta$ Pictoris b colors are close to those of late-M/early-L companions TWA 5B, USCO CTIO 108B, GSC 06214-00210 b, 1RXS J160929.1-210524 b,  and CD-35 2722B in addition to $\kappa$ And b in the diagram. 

We reach different conclusions comparing the  L' and M' band based colors of $\beta$ Pictoris b to those of mature M, L, T field dwarfs \citep{2004AJ....127.3516G}, and young companions or binaries  in Figure \ref{Fig:Fig2}. $\beta$ Pictoris b colors are compatible with those of mid-L to T field dwarfs in the three diagrams. The companion $\mathrm{K_{s}-L'}$  color is $\sim$ 1 magnitude redder than those of late-M field dwarfs. The planet has J-H and $\mathrm{J-K_{s}}$ colors bluer  than most of the young  companions with the exeption of $\kappa$ Andromedae, 1RXS J160929.1-210524 b, and HR8799c. The origin of this apparent inconsistency is discussed in section \ref{subsec:cloudatmp}.
 
\subsubsection{Color-magnitude diagrams}
\label{subsub:cmds}
\begin{figure}
   \centering
   \includegraphics[width=\columnwidth]{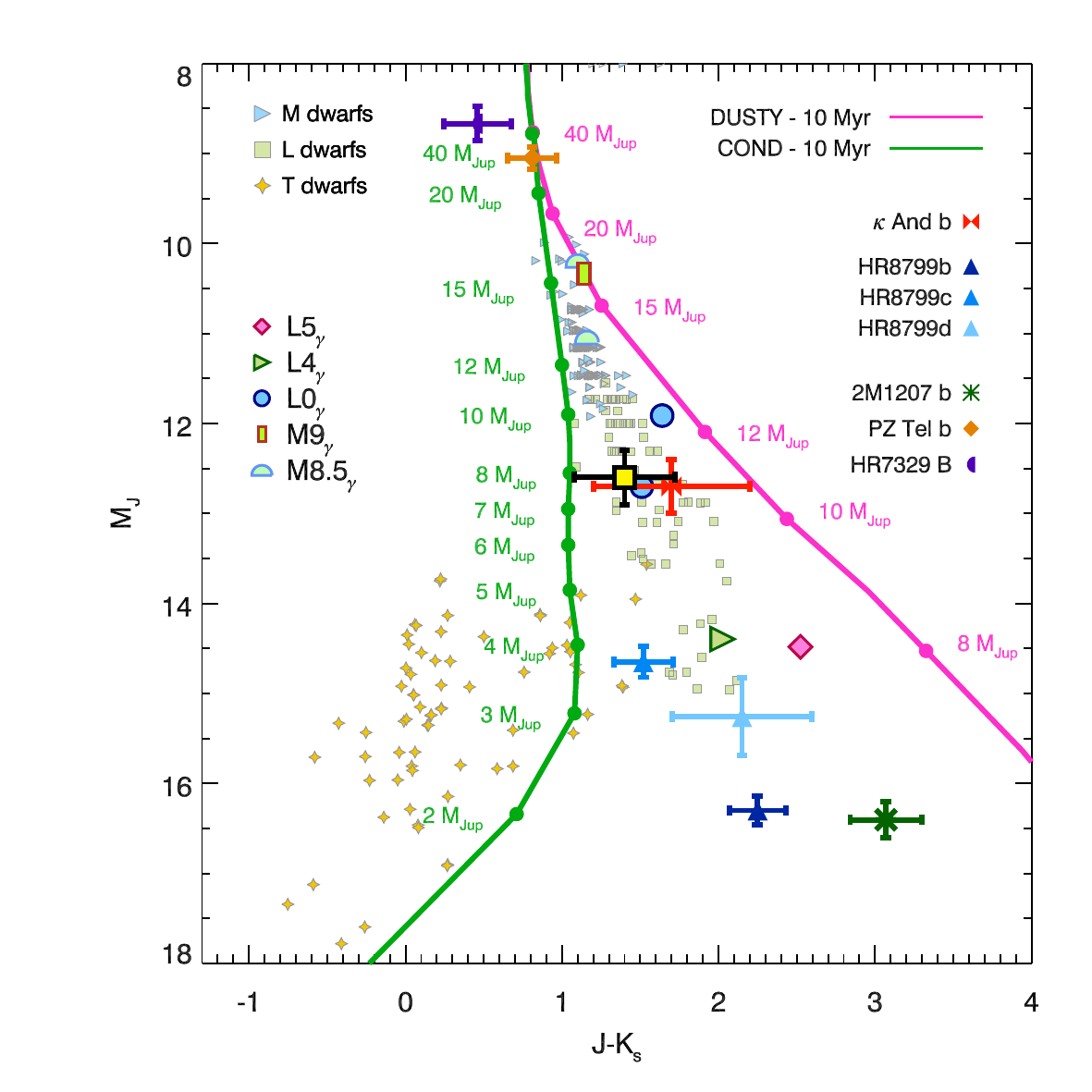}
      \caption{$\beta$ Pictoris b (yellow square) in a $M_{J}$ versus J-$\mathrm{K_{s}}$ color-magnitude diagram.} 
         \label{Fig:Fig4}
   \end{figure}

\begin{figure}
   \centering
   \includegraphics[width=\columnwidth]{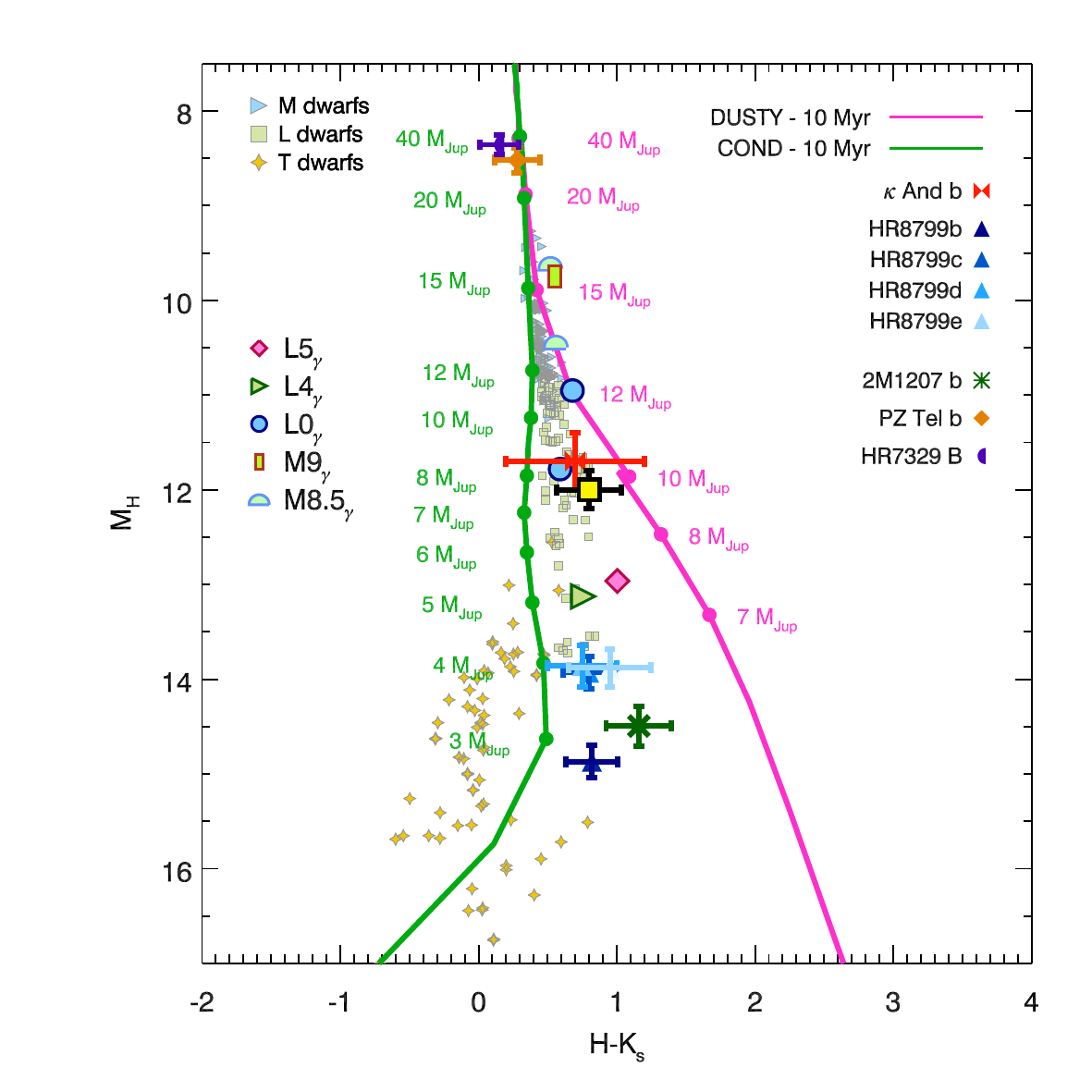}
      \caption{Same as Figure \ref{Fig:Fig4} but for  $M_{H}$ versus H-$\mathrm{K_{s}}$}
         \label{Fig:Fig4bis}
   \end{figure}

The planet is more easily distinguished from comparison objects (mature/young field dwarfs and young companions) in  color-magnitudes diagrams (Figures \ref{Fig:Fig4} and \ref{Fig:Fig4bis}). It lies in the middle of the sequence of field L-dwarfs in  the diagrams.

The two other late-M (M5-M9)  brown dwarf companions orbiting stars of the $\beta$ Pictoris moving group, PZ Tel b \citep[][]{2010ApJ...720L..82B} and HR7329 B \citep[][]{2000ApJ...541..390L}, are  more luminous and bluer than the planet. This is consistent with $\beta$ Pictoris b being less massive and cooler than these objects. We report in the diagrams the photometry of other planet/brown-dwarfs companions to massive-stars. The planet/brown-dwarf companion recently imaged around the $\sim$30 Myr old star $\kappa$ And b \citep{2012arXiv1211.3744C} falls at close positions in color-magnitude diagrams (JHKL') as $\beta$ Pictoris b.   The good match in colors is consistent with ``hot-start"  evolutionary models predictions  which give close $T_{eff}$ ($\sim$1700 K) for the (close) observed luminosity of these companions. Conversely, the  late-L/early-T exoplanets HR8799 bcde \citep{2008Sci...322.1348M, 2010Natur.468.1080M} are all at distinct positions than $\beta$ Pictoris in these diagrams. This agrees with the HR8799 system \citep[30 Myr, Columba association,][]{2011ApJ...732...61Z} being older and the planets cooler than $\beta$ Pictoris b.

 We also show in these Figures the photometry of the young $\mathrm{M_{\gamma}}$ and $\mathrm{L_{\gamma}}$ dwarfs with measured parallaxes \citep{2012ApJ...752...56F, 2012arXiv1206.5519F}. $\beta$ Pictoris is  fainter than $\mathrm{M8.5_{\gamma}-M9_{\gamma}}$ dwarfs, and among them 2MASS J06085283−2753583, the currently lowest-mass isolated member of the $\beta$ Pictoris moving group \citep{2010ApJ...715L.165R}. The planet is more luminous than the $\mathrm{L4_{\gamma}}$ (2MASSJ05012406-0010452)  and  $\mathrm{L5_{\gamma}}$ (2MASSJ035523.51+113337.4) dwarfs. Interestingly, $\beta$ Pictoris b and $\kappa$ And b  are at close locations to the $\mathrm{L0_{\gamma}}$ dwarf 2MASSJ00325584-4405058 in these diagrams. We recently confirm the optical classification   \citep{2009AJ....137.3345C} and the presence of peculiar spectral features in the near-infrared spectrum of this object.  2MASSJ00325584-4405058 might then be a valuable proxy of the spectroscopic properties of $\beta$ Pictoris b and $\kappa$ And b. We can then conclude from this empirical analysis that  most of $\beta$ Pictoris b  photometric  properties are similar to those of  early-L (L0-L4) dwarfs. 

We used this crude constraint on the spectral type of the planet to derive a luminosity estimate. \cite{2012arXiv1206.5519F} shown that the SED of young L-type dwarfs can deviate significantly from those of mature field dwarfs with similar spectral types.  We rather estimated a $\mathrm{BC_{K}=3.22\pm0.16}$ mag for $\beta$ Pictoris b taking the mean of the bollometric corrections measured for the young L0 dwarf 2MASS J01415823-4633574 and L5 dwarf 2MASS J035523.37+113343.7 \citep[3.39 and 3.06 respectively;][]{2010ApJ...714L..84T, 2012arXiv1206.5519F}. We used this correction to derive a luminosity $\mathrm{log_{10}(L / L_{\odot})=-3.87\pm0.08}$ mag for $\beta$ Pictoris b.

\subsection{Spectral synthesis}
\label{subsec:specsynth}
			We compared $\beta$ Pictoris b fluxes to synthetic photometry generated from 7 atmospheric models. All models make use of the multipurpose radiative-transfer code \texttt{PHOENIX} \citep{1999JCoAM.109...41H, 1999STIN...0030685H} assuming 1D radiative transfer, hydrostatic equilibrium, convection based on the mixing length theory, chemical equilibrium, and an opacity sampling treatment of the opacities.   \texttt{PHOENIX} computes the emerging flux at top of the photospheric layers at given wavelengths for a grid of effective temperatures ($\mathrm{T_{eff}}$), surface-gravities (log $g$) and metallicities ([M/H]). Integrated surface fluxes and magnitudes were derived for each combinaison of atmospheric parameter ($\mathrm{T_{eff}}$, log $g$, [M/H]) using the corresponding model spectra, the NaCo filter passbands, and a flux-calibrated model spectrum of Vega \citep{2007ASPC..364..315B}. Model fluxes were multiplied by the dilution factor $\mathrm{R^{2}/d^{2}}$  (with R and d the radius and distance of the object respectively) to match the apparent fluxes of $\beta$ Pictoris b. Our spectral synthesis tool then retrieved simultaneously $\mathrm{T_{eff}}$, log $g$, and R minimising the $\mathrm{\chi^{2}}$ with the planet flux. The analysis was conducted with solar-metallicity models. We also explored the effects of the metallicity on the determination of $\mathrm{T_{eff}}$ and log $g$  with the DRIFT-PHOENIX and BT-Settl 2012 models.
			 
			The empirical study presented in Section \ref{subsec:empanal} and previous temperature estimates \citep{2011A&A...528L..15B, 2011ApJ...736L..33C} suggests that clouds of silicate dust are present in the atmosphere of the planet. The \texttt{PHOENIX} models used in the following sections explore different cloud formation conditions.

	\subsubsection{AMES-Dusty and AMES-Cond}
	\label{subsec:AMES}
AMES-Dusty/Cond  \citep{2001ApJ...556..357A} atmospheric models used \texttt{PHOENIX} to solve the plan-parallel radiative transfer together with the Nasa AMES TiO and H$_2$O molecular line lists \citep{2000ApJ...540.1005A}.  Models relied on the solar abundances by \cite{1993A&A...271..587G}.  These models explored the limiting conditions of dust formation for 30 types of grains in phase equilibrium. The AMES-Dusty models simulate the case of maximum dust formation where the sedimentation (or gravitate settling) is neglected. The AMES-Cond models consider the opposite case where dust forms and is immediately rained-out from the photosphere, producing the depletion of the elements. The grain opacities are obtained by solving the Mie equation for spherical grains assuming the interstellar grain size distribution using complex refraction index of materials as a function of wavelength compiled by the Astrophysikalisches Institut und Universit\"ats-Sternwarte in Jena. 

We showed in \cite{2010A&A...512A..52B} that AMES-Dusty models do not reproduce simultaneously the pseudo-continuum and the water-band absorptions of the near-infrared (1.1-2.5 $\mu$m) spectrum of the L-type planet/brown-dwarf companion AB Pic b. $\beta$ Pictoris b is expected to lie in the same temperature range \citep[][Bonnefoy et al. 2012, A\&A, accepted]{2010A&A...512A..52B}. Our analysis could then be similarly affected by this bias. Nevertheless, AMES-Dusty and AMES-Cond models are used as boundary conditions of  the evolutionary models of \cite{2000ApJ...542..464C} and \cite{2003A&A...402..701B}. We therefore used them to determine $\mathrm{T_{eff}}$, log $g$, and radius estimates in order to preserve the self-consistenty in Section \ref{subsec:Lyon}. 

\begin{table}
\begin{minipage}{\columnwidth}
\caption{Best fitted atmospheric parameters for $\beta$ Pictoris b}
\label{atmoparPHOENIX}
\centering
\renewcommand{\footnoterule}{}  
\begin{tabular}{lllll}
\hline \hline 
Atmospheric model		 		&   $\mathrm{T_{eff}}$		&	 $\mathrm{log\:g}$	& Radius		& $\mathrm{\chi^{2}}$	\\   			
									&				(K)					&		$\mathrm{log(cm.s^{-2}}$)	&	($\mathrm{R_{Jup}}$)	\\	
\hline
AMES-Dusty				&		1700						&	3.5			&	1.22		&	4.96   \\
AMES-Cond		 		&		1400						&	3.5			&	1.89		&	8.04	\\
BT-Settl 2010				&		1600						&	3.5			&	1.61		&	2.99	\\
BT-Cond 2012				&		1800						&	4.0			&	1.16		&	17.80	\\
BT-Dusty 2012\tablefootmark{a}					&		1700:		&	4.5:			&	1.76:		&	3.08	\\
BT-Settl 2012	 [M/H]=0.0		&		1700						&	3.5			&	1.29		&	4.27	\\
BT-Settl 2012	 [M/H]=+0.5		&		1600						&	4.5		&	1.43		&	5.45	\\
DRIFT-P. [M/H]=0.0			&		1700						&	4.0			&	1.38		&	2.02    \\
DRIFT-P. [M/H]=+0.5			&		1700						&	4.0			&	1.38		&	3.06    \\
DRIFT-P. [M/H]=-0.5			&		1700						&	4.0			&	1.38		&	2.14    \\
\hline
\end{tabular}
\end{minipage}
\tablefoot{ 
\tablefoottext{a}{Analysis limited to log g $\geq$ 4.5.} 
}
\end{table}

\begin{figure}[!]
   \centering
   \includegraphics[width=\columnwidth]{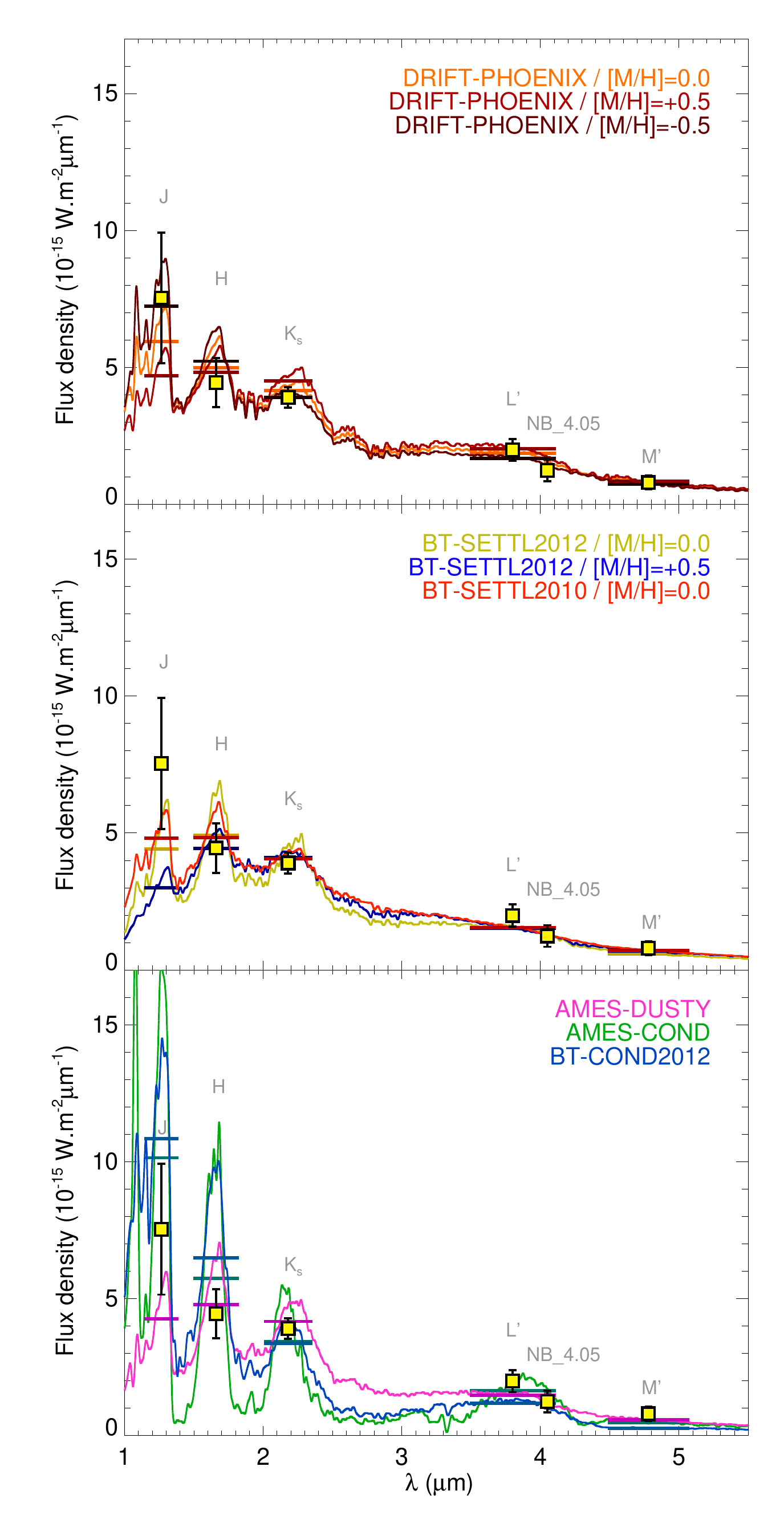}
      \caption{Comparison of the apparent fluxes of $\beta$ Pictoris b   (yellow squares) to  best fitted fluxes (horizontal barres) generated from synthetic spectra.  Bottom: comparison to AMES-Cond (green; $\mathrm{T_{eff}=1400K,\:log\:g=3.5,\: and\:R=}$ 1.89 $\mathrm{R_{Jup}}$), AMES-Dusty (pink; $\mathrm{T_{eff}=1700K,\:log\:g=3.5,\: and\:R=}$ 1.22 $\mathrm{R_{Jup}}$), and BT-Cond2012   (blue; $\mathrm{T_{eff}=1800K,\:log\:g=4.0,\: and\:R=}$ 1.16 $\mathrm{R_{Jup}}$) synthetic fluxes/spectra.  Middle: comparison to BT-Settl 2012 models at solar metallicity (gold; $\mathrm{T_{eff}=1700K,\:log\:g=3.5,\: and\:R=}$ 1.29 $\mathrm{R_{Jup}}$), to BT-Settl 2012 models with [M/H]=+0.5 dex (blue; $\mathrm{T_{eff}=1600K,\:log\:g=4.5,\: and\:R=}$ 1.43 $\mathrm{R_{Jup}}$), and to BT-Settl 2010 (red; $\mathrm{T_{eff}=1600K,\:log\:g=3.5,\: and\:R=}$ 1.61 $\mathrm{R_{Jup}}$) synthetic fluxes at solar metallicity. The corresponding model spectra smoothed at the resolution of the NB\_4.05 filter are also shown. Top: comparison to DRIFT-PHOENIX spectra ($\mathrm{T_{eff}=1700K,\:log\:g=4.0,\: and\:R=}$ 1.38 $\mathrm{R_{Jup}}$) with solar metallicity (M/H=0.0, orange), metal-enriched  (M/H=+0.5, dark red), and metal-depleted (M/H=-0.5, brown) atmospheres.}
         \label{Fig:Fig5}
   \end{figure}

We selected a sub-grid of synthetic spectra with $\mathrm{1000\:K \leq T_{eff} \leq 3000\:K}$, $\mathrm{3.5 \leq  log g \leq 6.0}$. Best fitted atmospheric parameters and radii are reported in Table \ref{atmoparPHOENIX}. We  show in Figure \ref{Fig:Fig5} the best fitted fluxes and the corresponding synthetic spectra. Models reproduce $\beta$ Pictoris b photometry for close $\mathrm{T_{eff}}$  (see Table \ref{atmoparPHOENIX}). However, the quality of the fit as expressed by low $\chi^{2}$ values is significantly better using the AMES-Dusty models.  Visually (Figure \ref{Fig:Fig5}), the J-band flux  of $\beta$ Pictoris b is intermediate between those of best-fit AMES-Cond and AMES-Dusty models. The remaining photometric points of the planet are better reproduced by the AMES-Dusty models with the exception of the L' band (discussed in Section \ref{subsec:cloudatmp}). In general, AMES-Cond models produce a SED slope too blue compared to the planet. Both best-fit AMES-Cond/Dusty spectra fall at the edge of the range of surface gravities covered by the grids of models. The $\chi^{2}$ map of the AMES-Cond models suggest that the fit could still be improved for lower surface gravities for the same temperature.  The corresponding map for AMES-Dusty models shows the fit is mostly sensitive to the effective temperature (Figure \ref{Fig:Figrefchi2}). 
 
\onlfig{7}{
   \begin{figure*}
   \centering
   \begin{tabular}{cccc}
   \includegraphics[width=5.8cm]{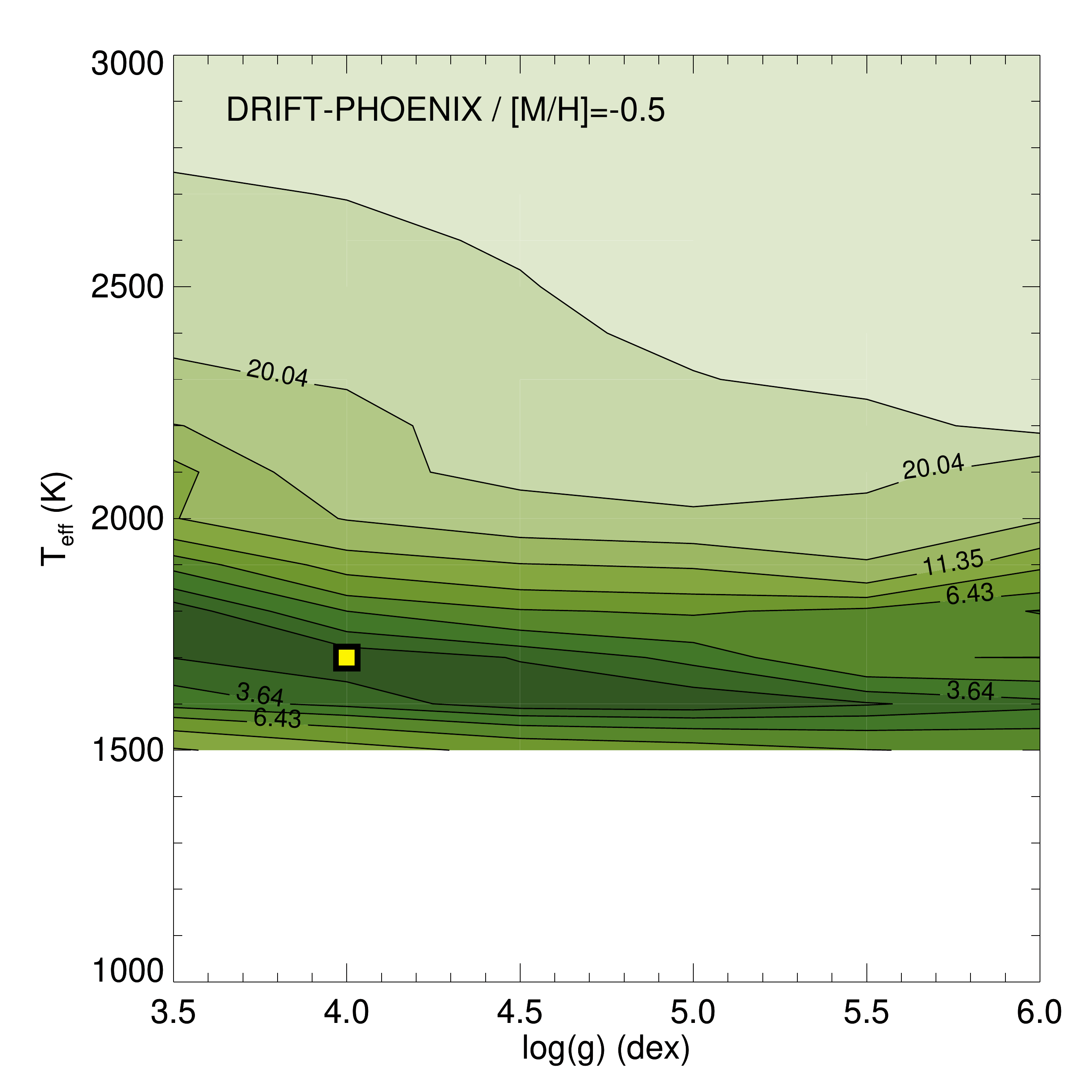} &
   \includegraphics[width=5.8cm]{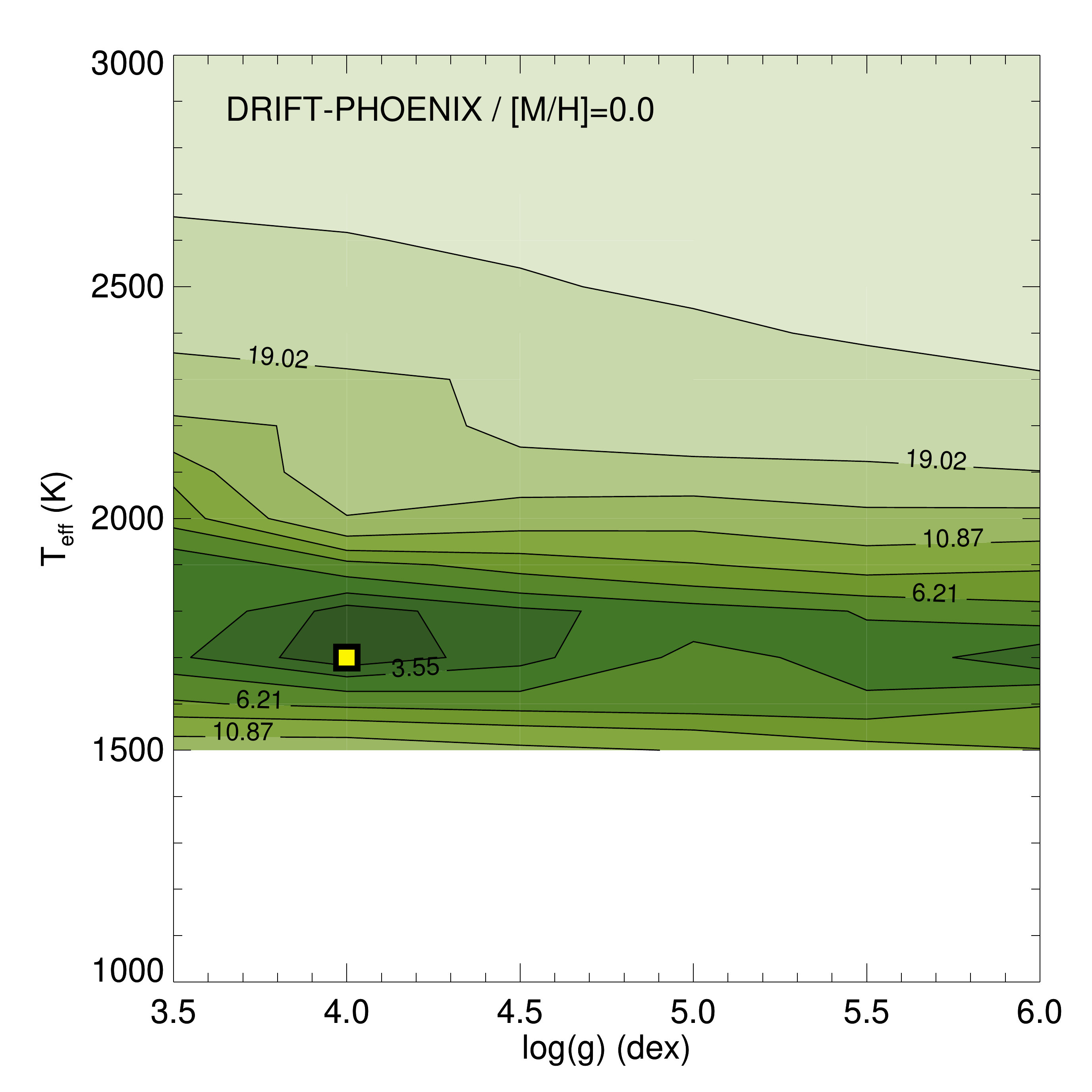} &
   \includegraphics[width=5.8cm]{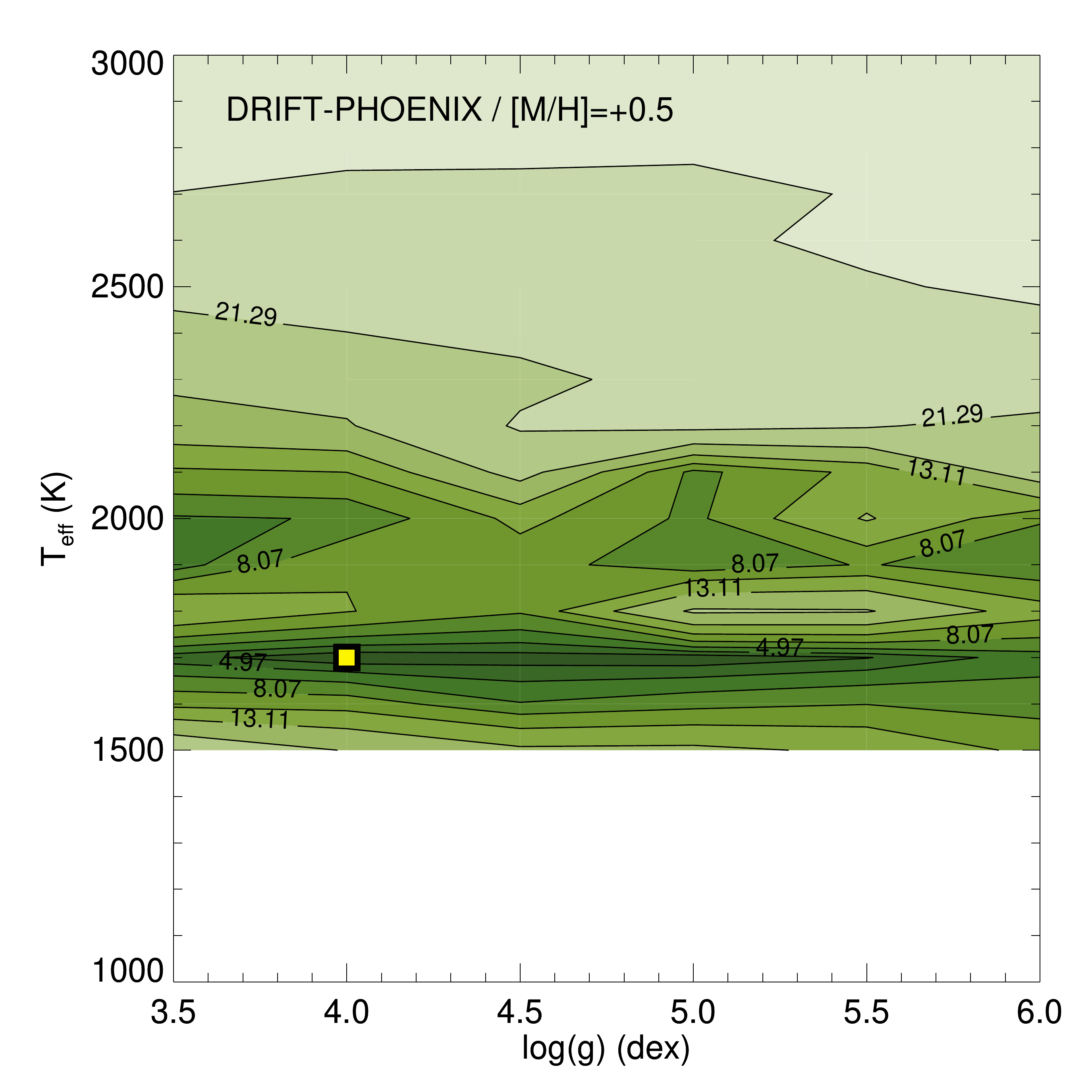} \\
   \includegraphics[width=5.8cm]{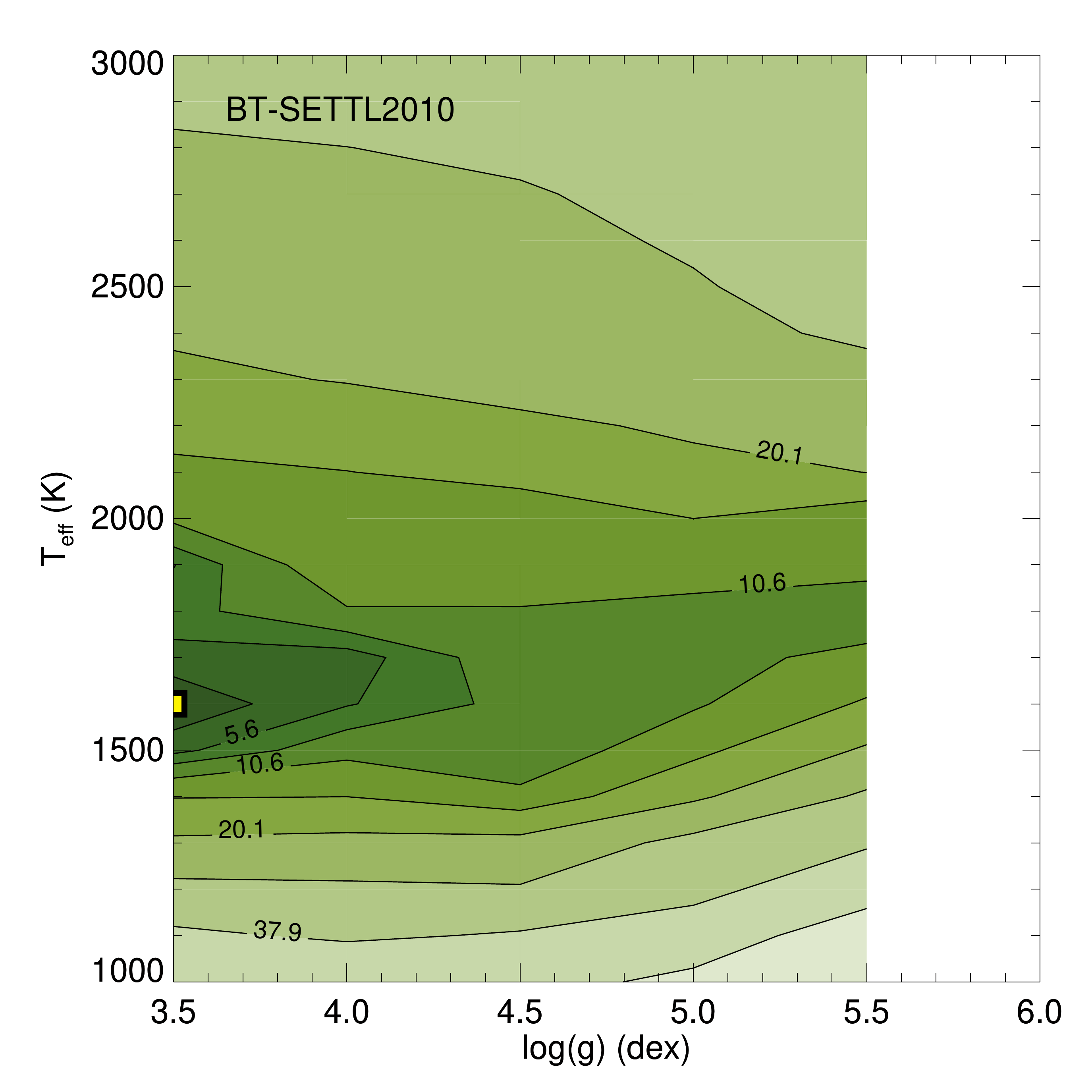} &
   \includegraphics[width=5.8cm]{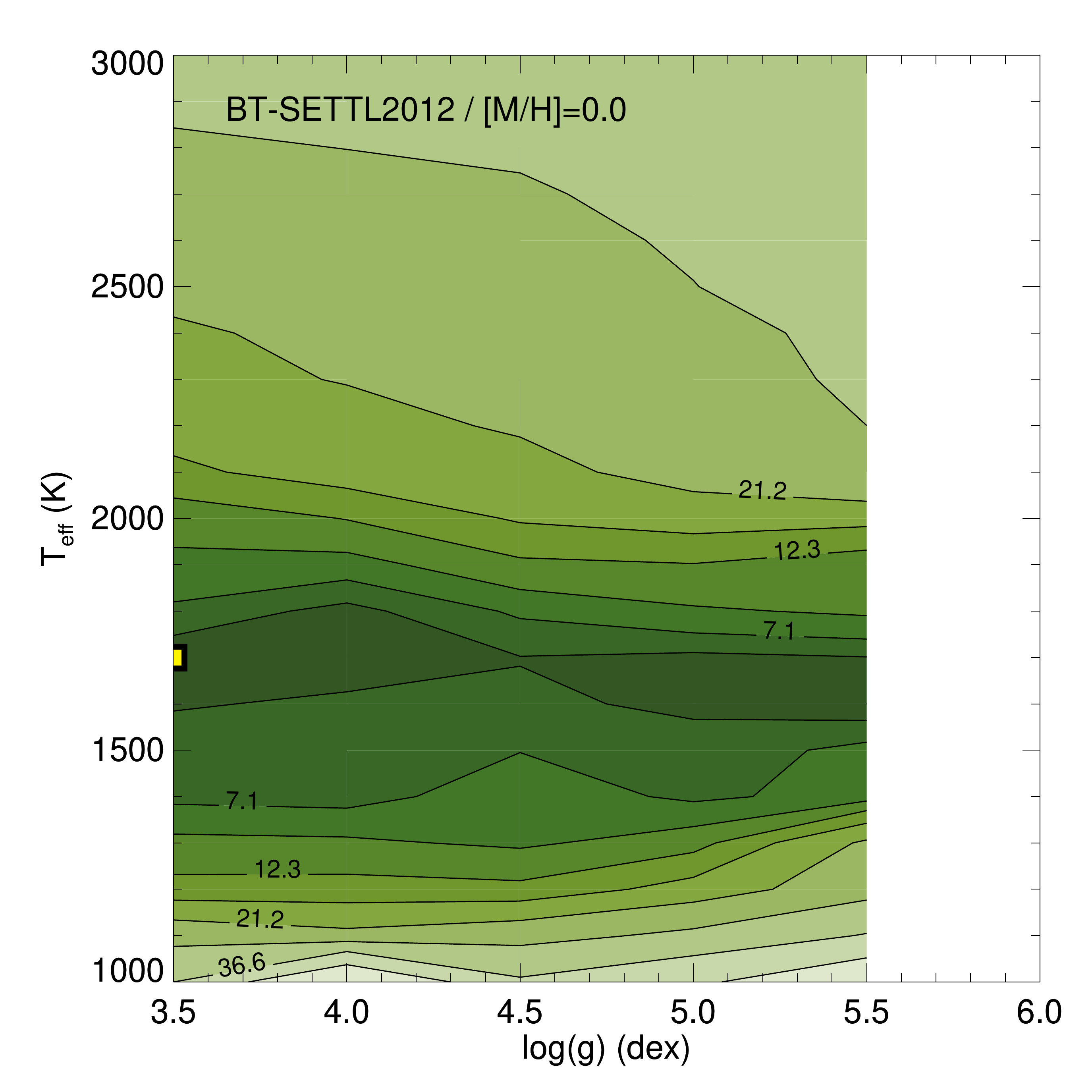} &
   \includegraphics[width=5.8cm]{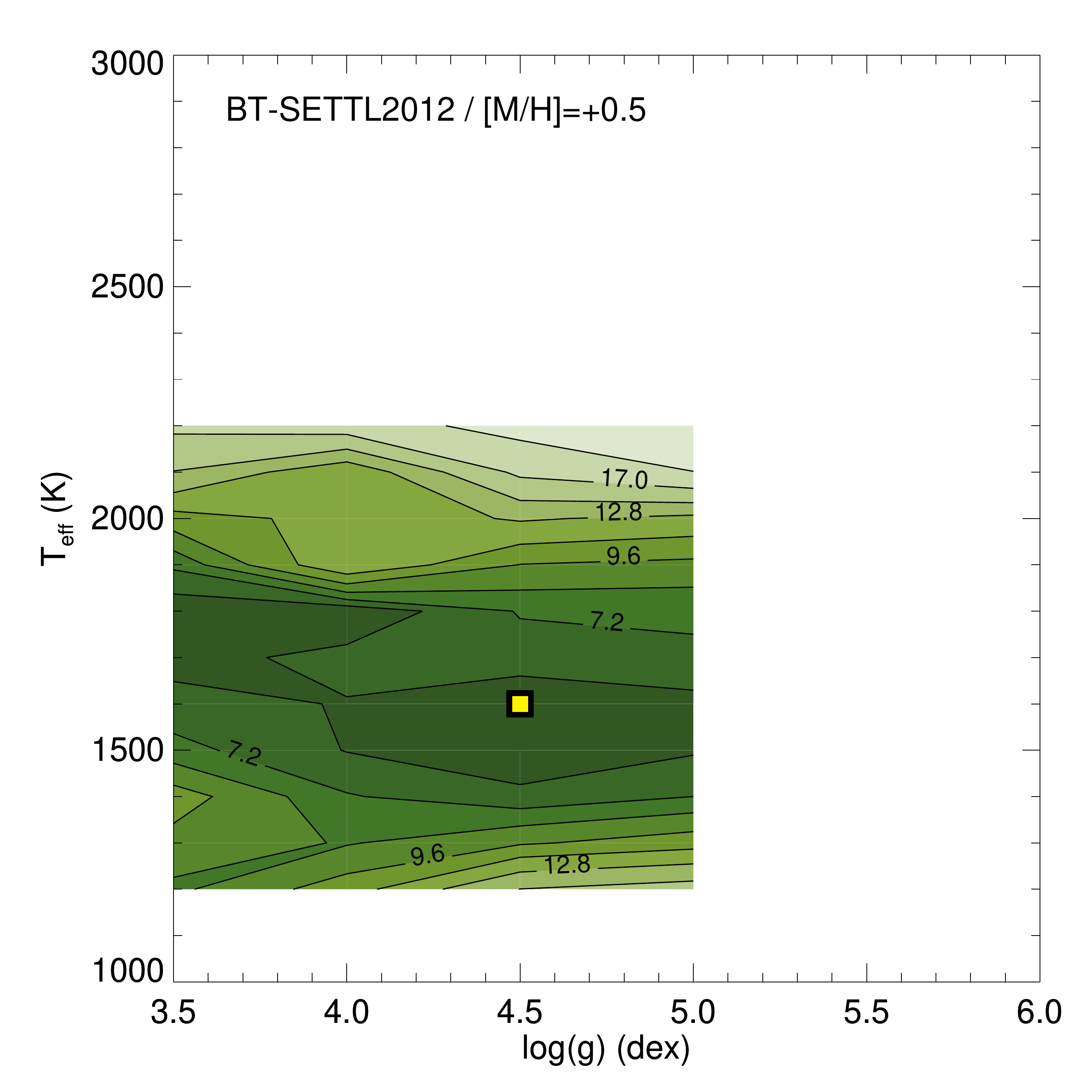} \\
   \includegraphics[width=5.8cm]{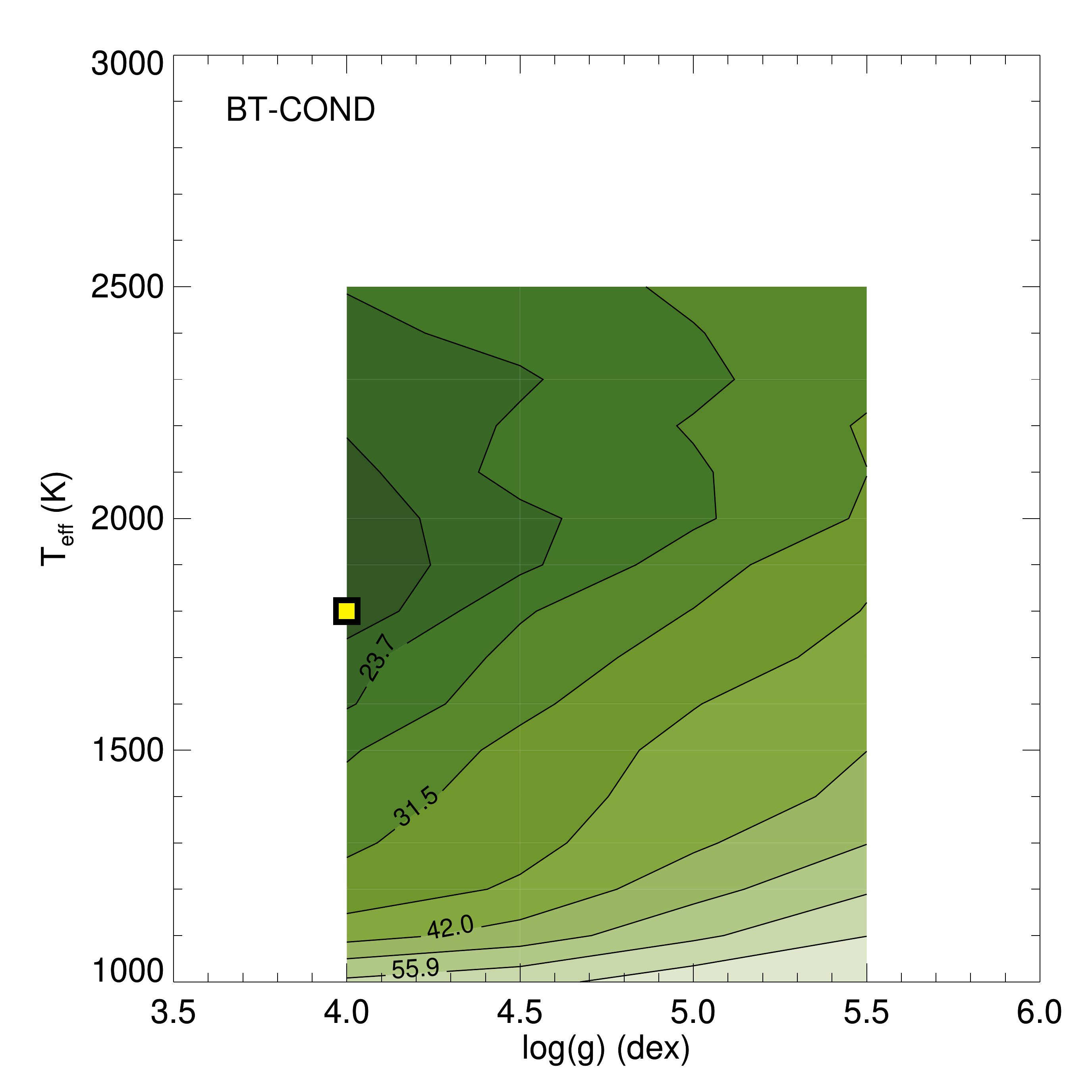} &
   \includegraphics[width=5.8cm]{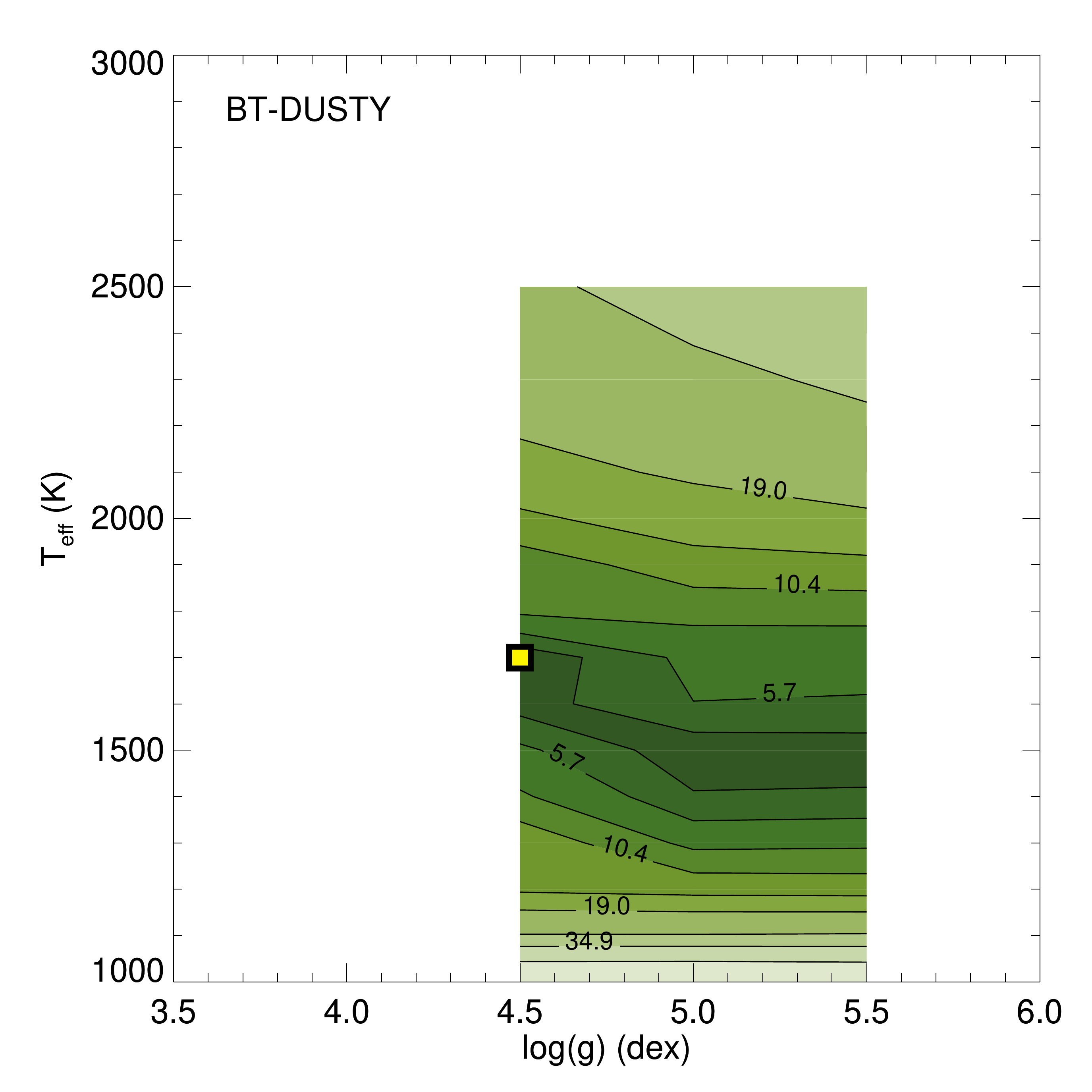} &
   \includegraphics[width=5.8cm]{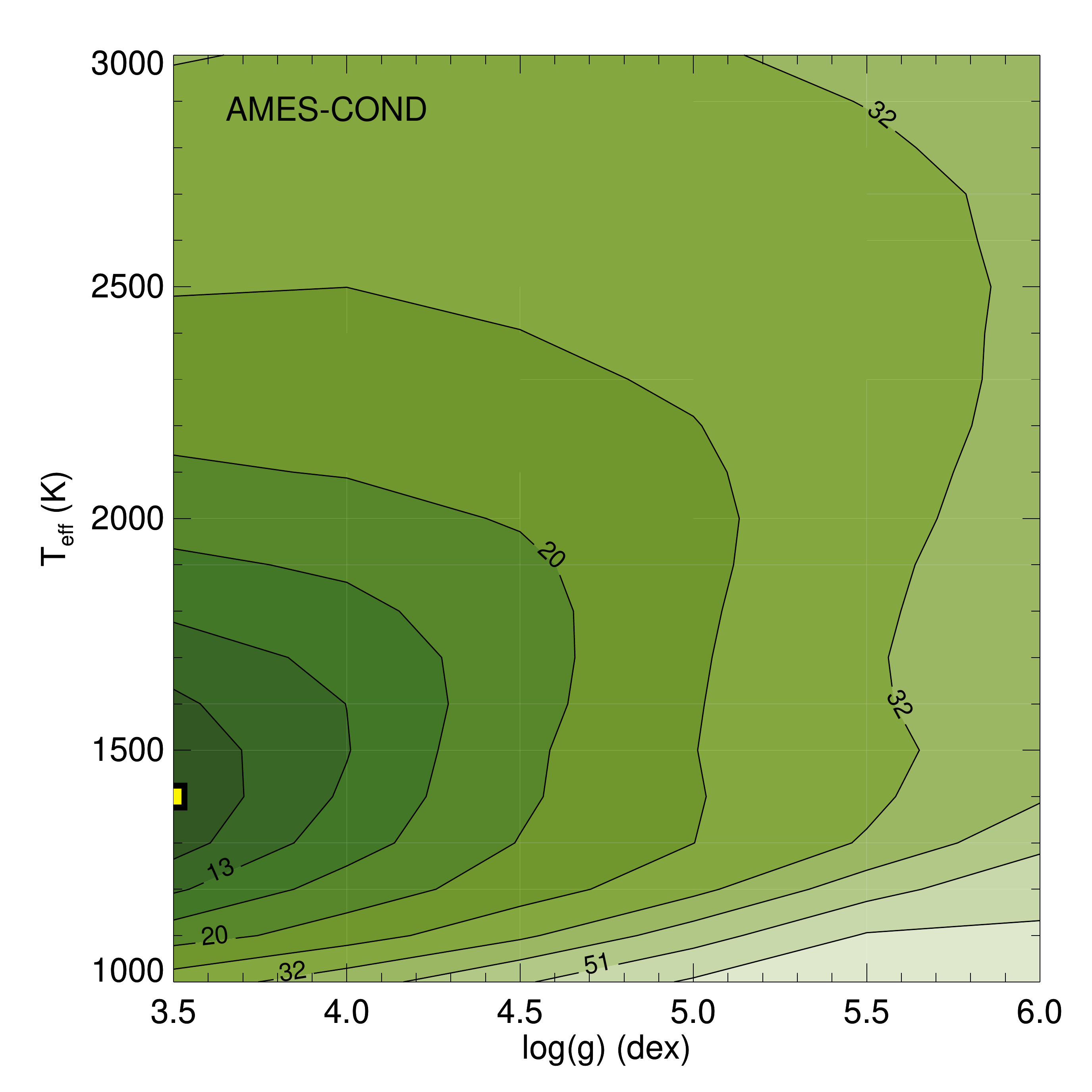} \\
  &  &  \includegraphics[width=5.8cm]{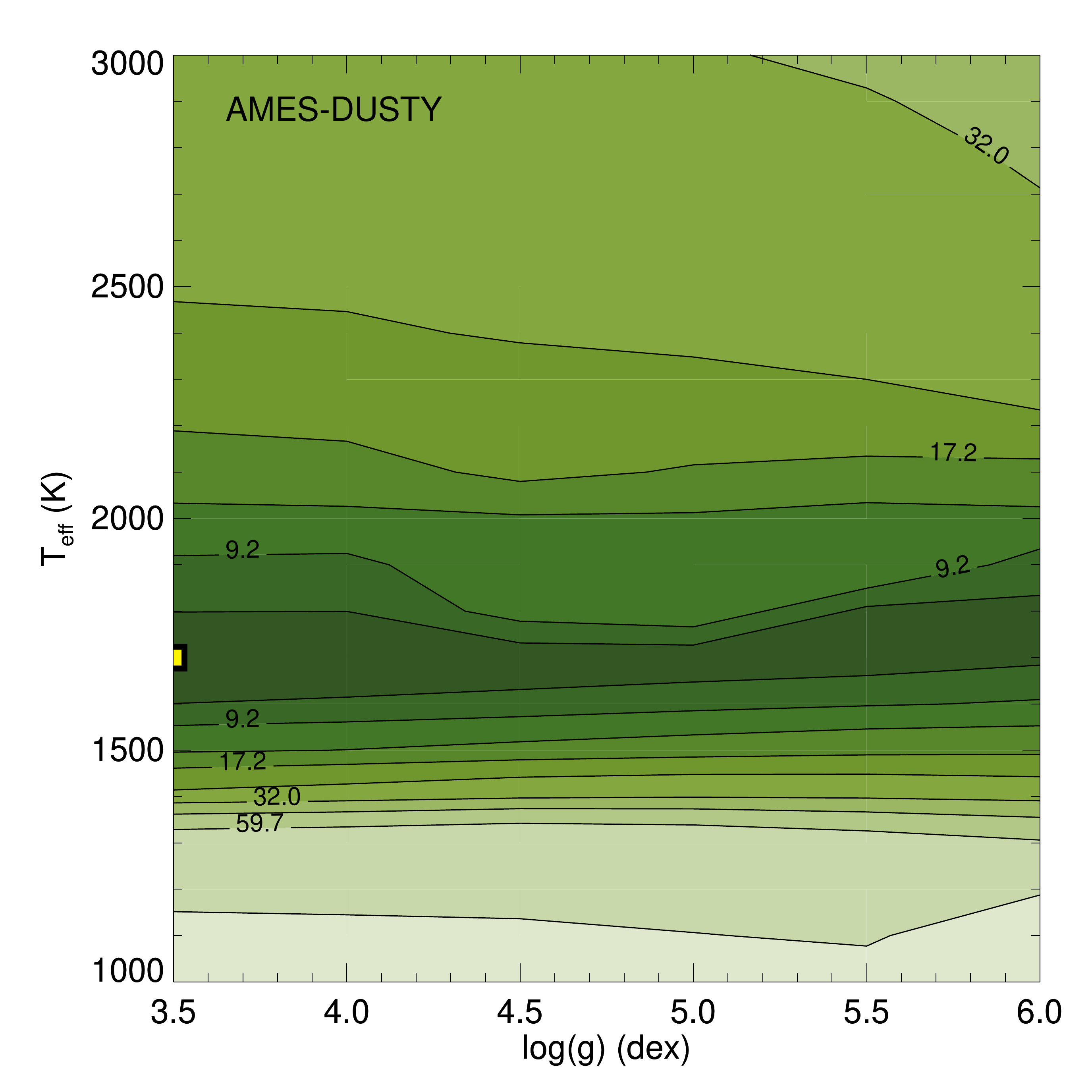}   \\
   
   \end{tabular}
      \caption{$\chi^{2}$ maps obtained when comparing the spectral energy distribution of the planet to synthetic fluxes derived from atmospheric models for given log g and $\mathrm{T_{eff}}$. Minima are indicated by yellow squares.}
         \label{Fig:Figrefchi2}
   \end{figure*}
}

	\subsubsection{BT-Settl, BT-Cond, and BT-Dusty models}
	\label{newmod}
Contrary to AMES-Dusty/Cond atmospheric models, BT-Settl models \citep{2003IAUS..211..325A} use a cloud model to predict the number density and size distribution of dust grains across the 1D atmosphere. These models use the \cite{1978Icar...36....1R}  cloud model which compares layer-by-layer the timescales of the main processes (mixing, sedimentation, condensation, coalescence and coagulation). The mixing timescales are derived from Radiation HydroDynamic (RHD) simulations \citep{2010A&A...513A..19F}. For the 2010 pre-release of the BT-Settl models \citep{2011ASPC..448...91A} (hereafter BT-Settl 2010) the cloud model was improved by a dynamical determination of the supersaturation --- the ratio of the saturation vapor pressure to the local gas pressure $P_{sv}/P_g(t)$. In the 2012 pre-release of the BT-Settl models \citep[][hereafter BT-Settl 2012]{2012RSPTA.370.2765A,2012arXiv1206.1021A} the cloud model was further improved by the implementation of a grain size-dependant forward scattering, and by accounting for nucleation based on cosmic rays studies \citep{Tanaka2005}.  This latter change allowed the cloud model to limit its refractory element depletion and form relatively more dust grains in higher atmospheric layers.

The cloud model is solved from the innermost to the outermost atmospheric layer, depleting the gas composition due to sedimentation gradually from the bottom to the top of the atmosphere. This model obtains both the density distribution of grains and the grain average size per layer, and includes 55 types of grains (including metal, ceramic, salt, and ice crystals). The BT-Settl models do not assume a seed composition since the cloud model modifies the equilibrium chemistry iteratively at each atmospheric layers. The resulting number density of dust grains are not in phase equilibrium, even if the equilibrium chemistry is used to adjust layer-by-layer the gas phase abundances. 

The BT-Settl models are computed with the \texttt{PHOENIX} stellar atmosphere code using the spherical radiate transfer. The code use up-to-date molecular line list of the water-vapor \citep[BT2,][]{2006MNRAS.368.1087B}, $\mathrm{CH_{4}}$ \citep{2005MSAIS...7..157H}, CIA of $\mathrm{H_{2}}$ \citep{2002A&A...390..779B},  TiO, VO, MgH, CrH, FeH, and CaH \citep{1998A&A...337..495P, 2003ApJ...582.1059W, 2003ApJ...594..651D}, and new alkali-line profiles \citep{2007A&A...474L..21A}.  Non-equilibrium chemistry for CO, $\mathrm{CH_{4}}$,  CO$_2$, $\mathrm{N_{2}}$ and $\mathrm{NH_{3}}$ is also included based on the RHD simulation results \citep{2010A&A...513A..19F}.   Finally, the models relies on the revised solar-abundances (\cite{2009CoAst.158..151G} for the BT-Settl 2010 models and \cite{2011SoPh..268..255C} for the BT-Settl 2012 models) which translates to less pronounced water-band absorptions and  redder colors  in the near-infrared than previous version of the models \citep[e.g. those used in ][]{2007A&A...474L..21A, 2010A&A...512A..52B}.

We demonstrate in Bonnefoy et al. (2012 submitted) that the BT-Settl 2010 models represents self-consistently the optical and near-infrared spectra together with the Spitzer photometry (3.6-8 $\mu$m) of the young (1-3 Myr) M9.5 dwarf KPNO-Tau-4 \citep{2002ApJ...580..317B}.  They should then be appropriate for $\beta$ Pictoris b given the spectral type and age range of the planet. We compared $\beta$ Pictoris b photometry with the models  for $\mathrm{1000 K \leq T_{eff}  \leq 3000 K}$ an $\mathrm{3.5 \leq log\:g  \leq 5.5}$.  We plot the best-fit model fluxes and corresponding synthetic spectra in Figure \ref{Fig:Fig5}. Our fitting procedure gives a close ($\pm100$ K) effective temperatures than the AMES-Dusty models. The $\chi^{2}$ is reduced  compared to previous models, given the better agreement of the predicted J band flux with the one of the planet (Figure \ref{Fig:Figrefchi2} amd \ref{Fig:Fig5}).

 
BT-Settl 2012 models at solar metallicity ([M/H]=0) with $\mathrm{1000\:K \leq T_{eff} \leq 3000\:K}$ and $\mathrm{3.5 \leq log\:g \leq 5.5}$ were first considered here. We also used a sub-grid of metal-enriched models  ([M/H]=+0.5) for $\mathrm{1200\:K \leq T_{eff} \leq 2000\:K}$ and  $\mathrm{3.5 \leq log\:g \leq 5.0}$. Revised versions of the Cond and Dusty models (hereafter BT-Cond/Dusty 2012) have been calculated using the same opacities and solar abundances as for the BT-Settl 2012 models for comparison.
BT-Cond 2012 and BT-Dusty 2012 models were only computed on restraints grid of parameters at the time of the analysis. Consequently, the analysis with the BT-Cond 2012 models  was limited to $1000\:K \leq \mathrm{T_{eff} \leq 2500 K}$, $\mathrm{4.0 \leq log\:g \leq 5.5}$, and [M/H]=0 . The BT-Dusty 2012 model grid was also restrained to the same temperature range, [M/H]=0,  and log g $\mathrm{\geq 4.5\:dex}$. Log g = 4.5 falls above the surface gravity expected for $\beta$ Pictoris b (see Section \ref{subsec:Lyon}) and found using BT-Settl 2010 and AMES models. We then only used BT-Dusty 2012 models to provide an additional check of the atmospheric parameters found using other models. 

Parameters of best fitted 2012 models  are reported in Table  \ref{atmoparPHOENIX}. We also overlaid on Figure \ref{Fig:Fig5} the synthetic spectra and fluxes of the BT-Cond 2012 and BT-Settl 2012 models. 
 The analysis confirm trends seen with the other models. $\chi^{2}$ maps all show a clear minimum for $\mathrm{T_{eff}}$=1600-1800 K (Figure \ref{Fig:Figrefchi2}). Best fits are found for reduced surface gravities compared to those determined for mature field dwarfs \citep[$\mathrm{4.5 \lesssim log\:g \lesssim 6}$, see][]{2009A&A...503..639T}.  We note however that the $\chi^{2}$ map corresponding to the BT-Settl 2012 models still shows a weaker contraint on log g than on $\mathrm{T_{eff}}$.  BT-Settl 2012 models reproduces the planet photometry better. Metallicity changes $\mathrm{T_{eff}}$ estimate by only 100 K. The BT-Cond 2012 model predict bluer colors than the planet. BT-Dusty 2012 models give the lowest $\chi^{2}$ despite the limited interval of log g of the model. The fit with BT-Settl 2012 models have a higher $\chi^{2}$ than obtained from the BT-Settl 2010 models. We discovered at  visual inspection that the fit is indeed degraded in the $J$ band.

    \subsubsection{DRIFT-PHOENIX models}
    \label{subsubsec:DF}
   The DRIFT-PHOENIX models couple the \texttt{PHOENIX}  code to a non-equilibrium cloud model \texttt{DRIFT} \citep{2003A&A...399..297W, 2004A&A...414..335W, 2006A&A...455..325H, 2008ApJ...675L.105H} which account for the nucleation, seed formation, growth, evaporation, gravitational settling, and advection of a set of grains. The general differences between these models and the 2008 version of the BT-Settl models are given in \cite{2008MNRAS.391.1854H}. We note that the models use older reference solar abundances \citep[][; also used in AMES-Dusty/Cond models]{1993A&A...271..587G}  than those of the 2010-2012 releases of the BT-Settl models.
 
The grid of synthetic spectra extends from 1500 to  3000 K with 3.0 $\leq$log g$\leq$6.0. The grid was computed for three different metallicities (M/H=-0.5, 0, and +0.5 dex). We showed in Bonnefoy et al. 2013 (submitted) that these models give a correct representation of the near-infared spectra and SED of young M and early-L dwarfs.
 	
Results of the fit are also reported in  Table  \ref{atmoparPHOENIX} and shown in Figure \ref{Fig:Fig5}.  The fit confirms the effective temperature and low surface gravity estimates derived with the other models. $\chi^{2}$ maps indicate fits are sensitive to the effective temperature and do not enable to get an accurate constraint on the surface gravity. The metallicity do not bias the determination of the effective temperature. Models with solar abundances still best match the planet photometry.\\
        
We conclude that the PHOENIX-based models used above predict $\mathrm{T_{eff}=1700 \pm 100\:K}$  for $\beta$ Pictoris b. The SED seem to be systematicaly best reproduced by the models  with  $\mathrm{log\:g=4.0\pm0.5\:dex}$. We note however that  this parameter is less constrained  than the effective temperature. The analysis also demonstrate that once dust is present in the cloud models, our fit provides close effective temperature and surface gravity estimates despite the different underlying physics used in the models. We adopt a radius of $\mathrm{R=1.4\pm0.2\:R_{Jup}}$ for the planet based on the scaling factors between the DRIFT-PHOENIX, BT-Settl, and BT-COND models and the observations. These temperature and radius  correspond to a luminosity $\mathrm{log_{10}(L/L_{\odot})=-3.83\pm0.24}$ dex (Stephan-Boltzmann law). This value is consistent with the empirical luminosity derived in Section \ref{subsub:cmds}.  We use the later for the following analysis.

\subsection{Orbital parameters}
\label{orbitpar}			
We combined our new astrometric data point to those reported in
\cite{2012A&A...542A..41C} and fit orbits using Levenberg- Marquardt
algorithm (LSLM) and Markov-Chain Monte-Carlo (MCMC) simulations as
described in the aforementioned study, but yet with an improved algorithm.
The results of MCMC simulations are reported in Figure \ref{of1}, superimposed with new simulations
performed with the same data set without the new January 2012
astrometric point. The improvement of the probability curves  with respect to
Chauvin et al. (2012) is noticeable, even in the simulations using
the old data set. This is due to the improvement of the MCMC sampling
algorithm. This way we can clearly see what the new point bears.

The main result from \cite{2012A&A...542A..41C} are nevertheless confirmed.
We confirm that the semi-major axis falls in the 8--10 AU range with a
probability higher than 80\%. Subsequently the orbital period ranges
between 17 and 25 years with the same probability. Meanwhile, the planet
eccentricity is $\la 0.15$ with similar probability. Apart from the 2003 point, our astrometric
points are all located on one side of the star, and span over a rather
limited amount of time. Therefore, eccentric orbits with large periods
and currently passing at periastron can not be
excluded. Simultaneously, it is not yet possible to state whether the
eccentricity is zero of not. Consequently, the argument of periastron
is badly constrained, apart from a peak appearing around $\omega=0$.

Conversely, the orbital inclination and the longitude of ascending
node $\Omega$ are well constrained parameters, even if the addition of
the 2012 point causes a small shift in the MCMC distributions. We
confirm the edge-on orientation of the orbit within $\la 4\degr$ with
a probability higher than 90\%. We also confirm that the orbit is more
probably slightly viewed as prograde ($i<90\degr$) than
retrograde. The longitude of ascending node is confined to the range
$145-150\degr$ with similar probability. Due to the edge-on
orientation of the disk, this parameter is closely related to the PA
of the disk. We confirm here the analysis of \cite{2012A&A...542A..40L} who demonstrate that the orbital plan of $\beta$ Pictoris b
was located in the midplane of the inner disk rather than in that of
the main disk. This supports the model of \cite{1997MNRAS.292..896M} and \cite{2001A&A...370..447A} who
attributed the disk warp to the perturbing action of an inner giant
planet, which thus could well correspond to $\beta$ Pictoris b.

Similarly, we computed the MCMC distribution of the possible transit
passages of the planet (Figure \ref{of1}). We also confirm that the putative transit date of
Nov. 1981 claimed by \cite{2009A&A...497..557L} could still be due
to $\beta$ Pictoris b, as the quoted date still falls in a peak of the
distribution. Meanwhile, our analysis predicts a possible next transit
in 2018, which will deserve to be tentatively observed.

The falling evaporating bodies (FEBs) scenario \citep{1996Icar..120..358B, 2000Icar..143..170B, 2001A&A...366..945B}
 also predicts the presence of a Jovian-sized planet at $\sim
10\,$AU as responsible for the infall of numerous star-grazing
planetesimals (FEBs) via its inner mean-motion resonances. Obviously,
$\beta$ Pictoris b could correspond to this planet. However, a major
requirement of the FEBs model is that the suspected planet is slightly
eccentric ($e\ga 0.05$) with a convenient orientation of its
periastron with respect to the line of sight. Due to the paucity of
our data, it is still not possible to state whether $\beta$ Pictoris b
fulfills this condition. This motivates us to pursue the monitoring of
the orbit. The peak that appears at $\omega\simeq 0\degr$, and that
corresponds to orbits with some eccentricity, is nevertheless
interesting. As noted in \cite{2012A&A...542A..41C}, the orientation of the
planet periastron required by the FEB scenario translates into
$0\la\omega\la 20\degr$. The presence of the peak close to this range
in the $\omega$ distribution is encouraging.

These up-to-date constraints on the orbital parameters of $\beta$ Pictoris b combined to present constraints coming from radial velocity measurments \citep{2012A&A...542A..18L} confirm the upper limits of 12 and 15.5 $\mathrm{M_{Jup}}$ on the  planet mass for semi-major axis of 9 and 10 AU respectively. We compare these dynamical constraints to evolutionary models-dependant masses in the following section.

\onlfig{8}{
\begin{figure*}
   \centering
\begin{center}
\begin{tabular}{cc}
\includegraphics[width=7.2cm]{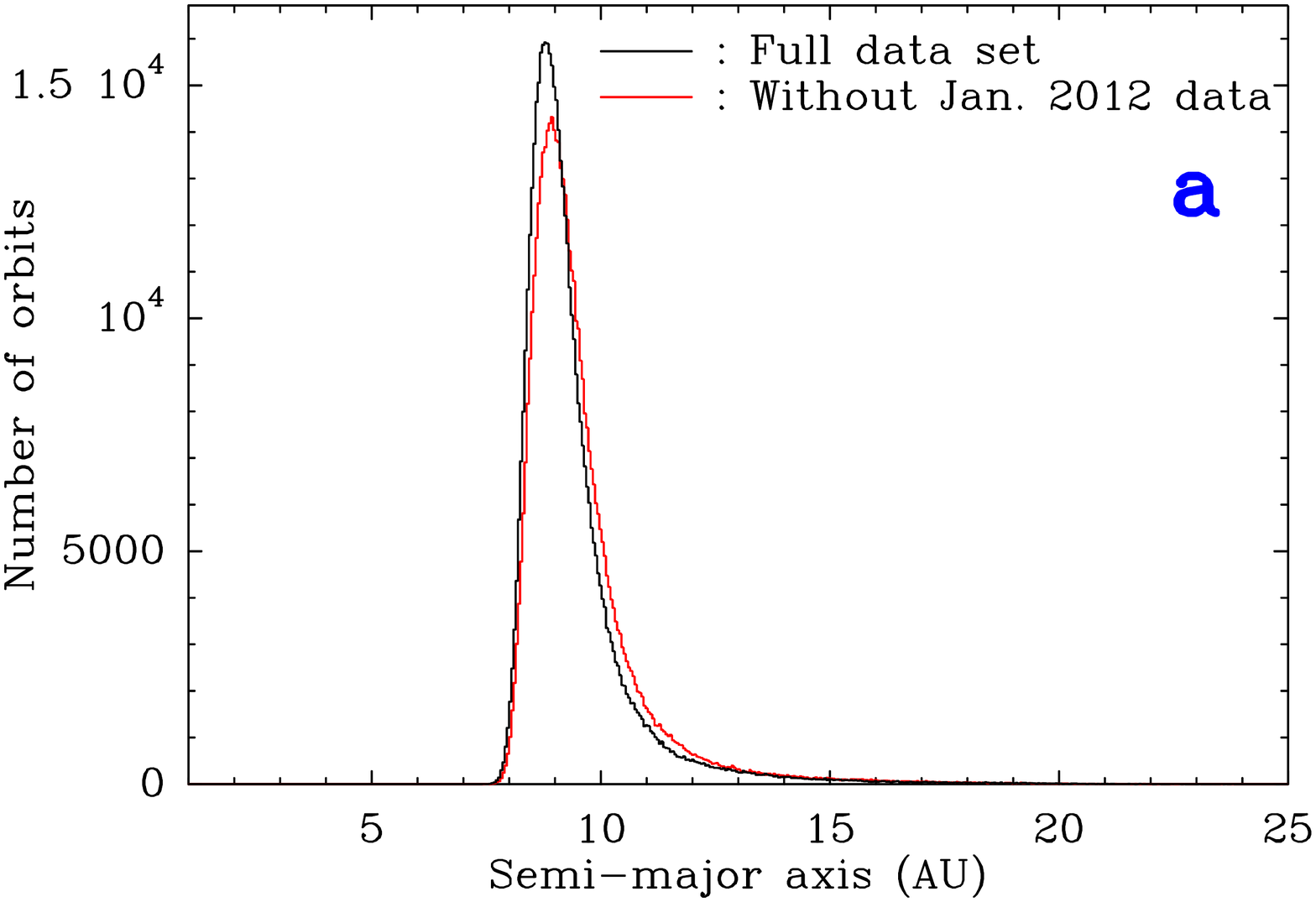} &
\includegraphics[width=7.2cm]{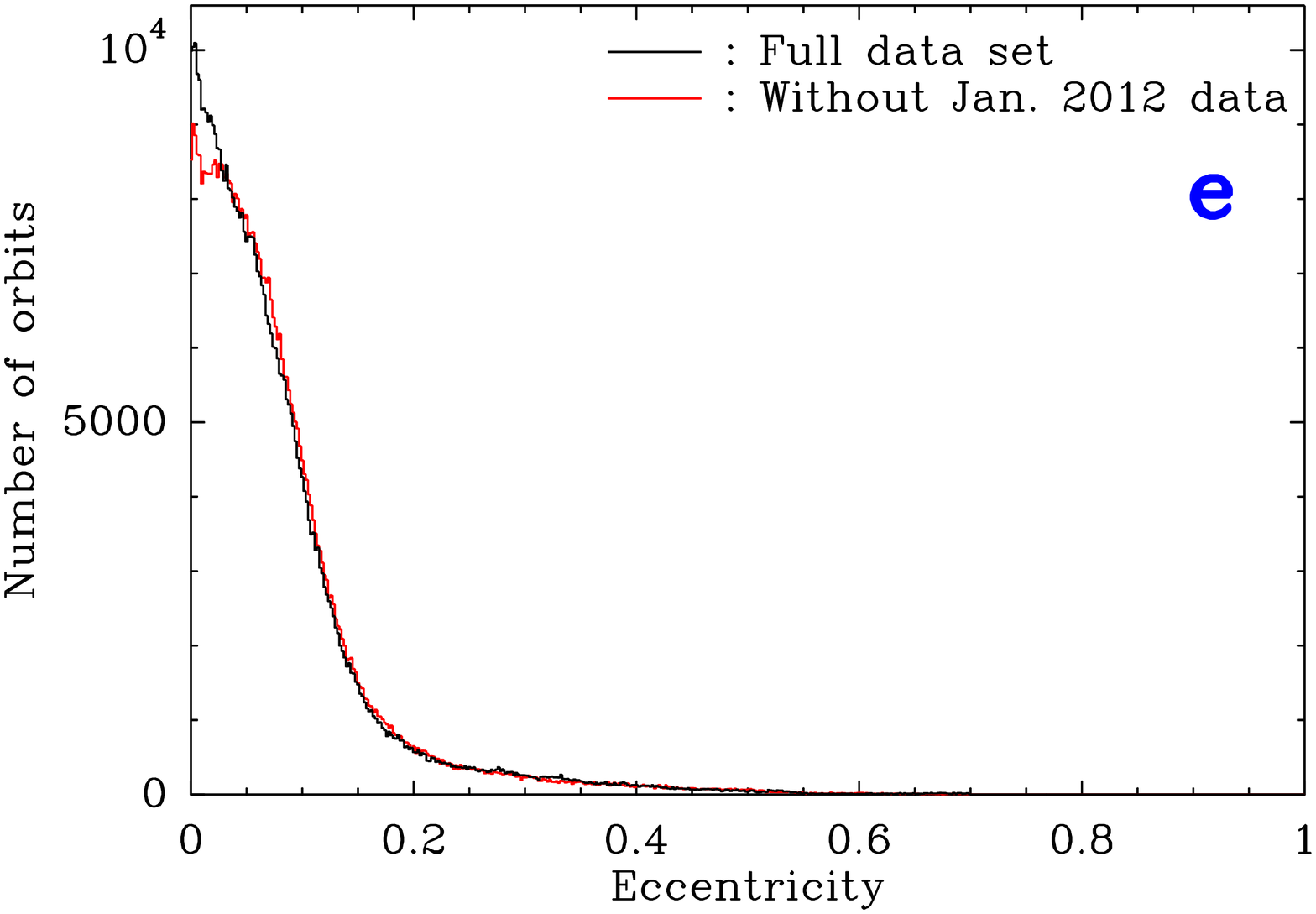} \\
\includegraphics[width=7.2cm]{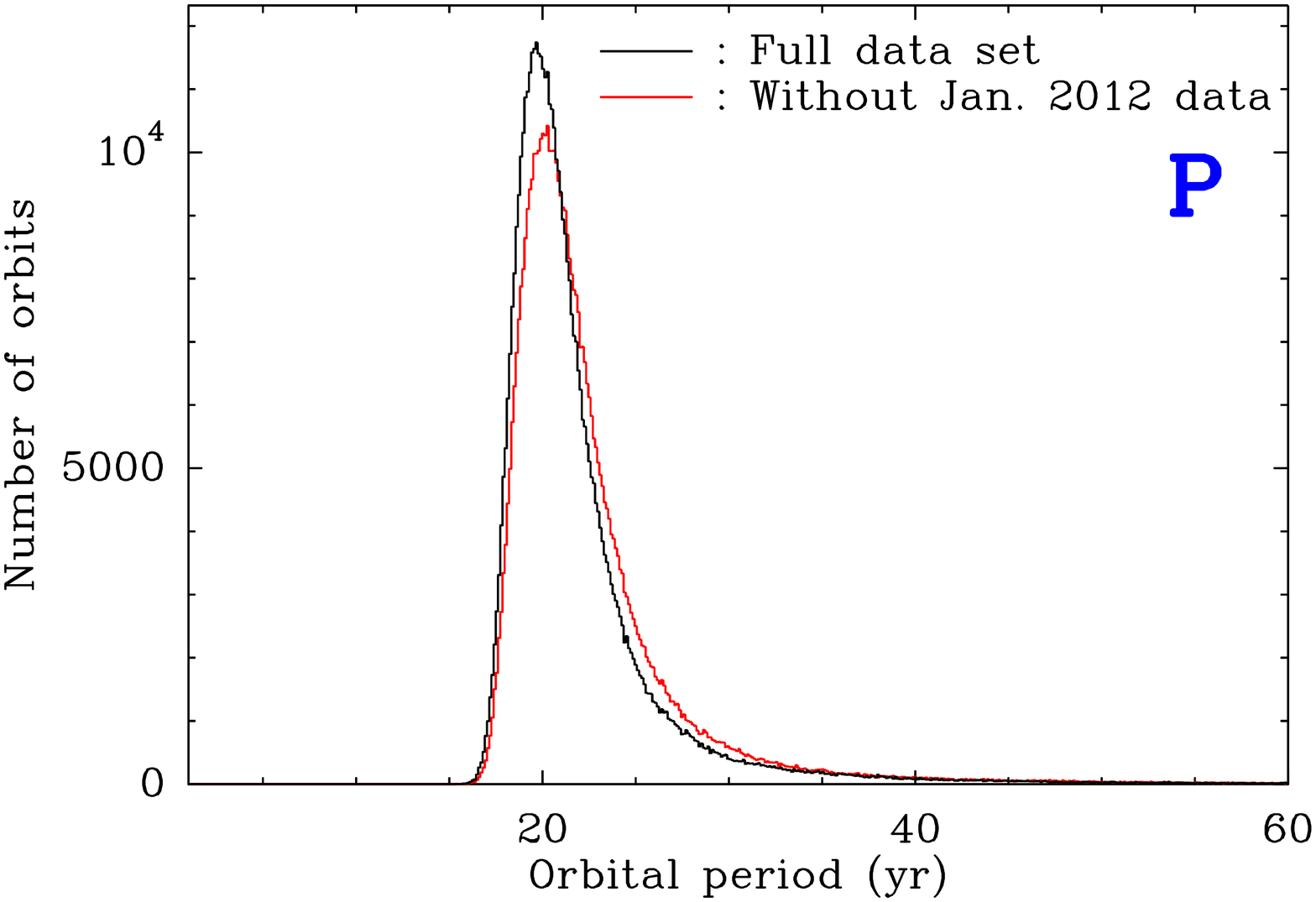} &
\includegraphics[width=7.2cm]{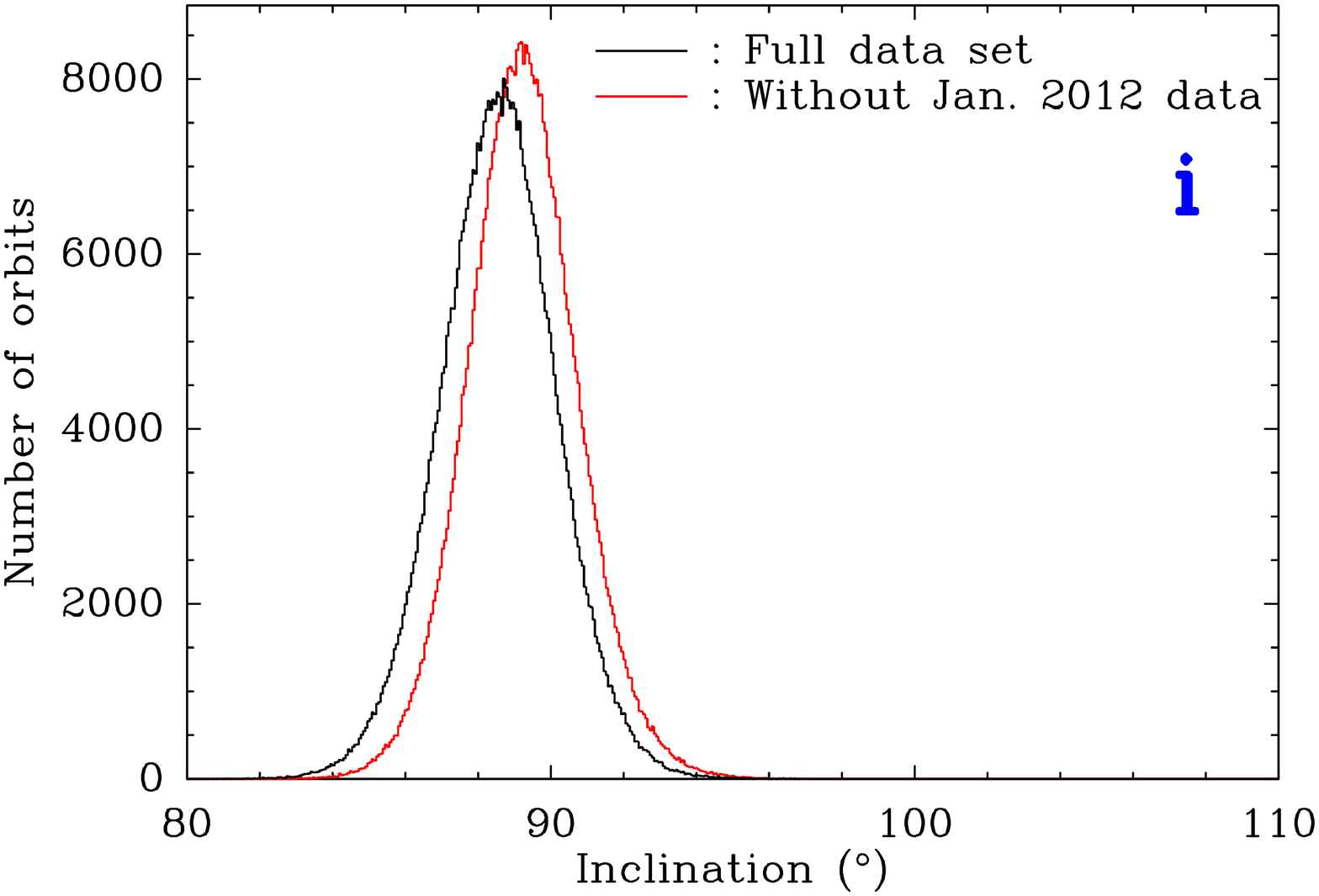} \\
\includegraphics[width=7.2cm]{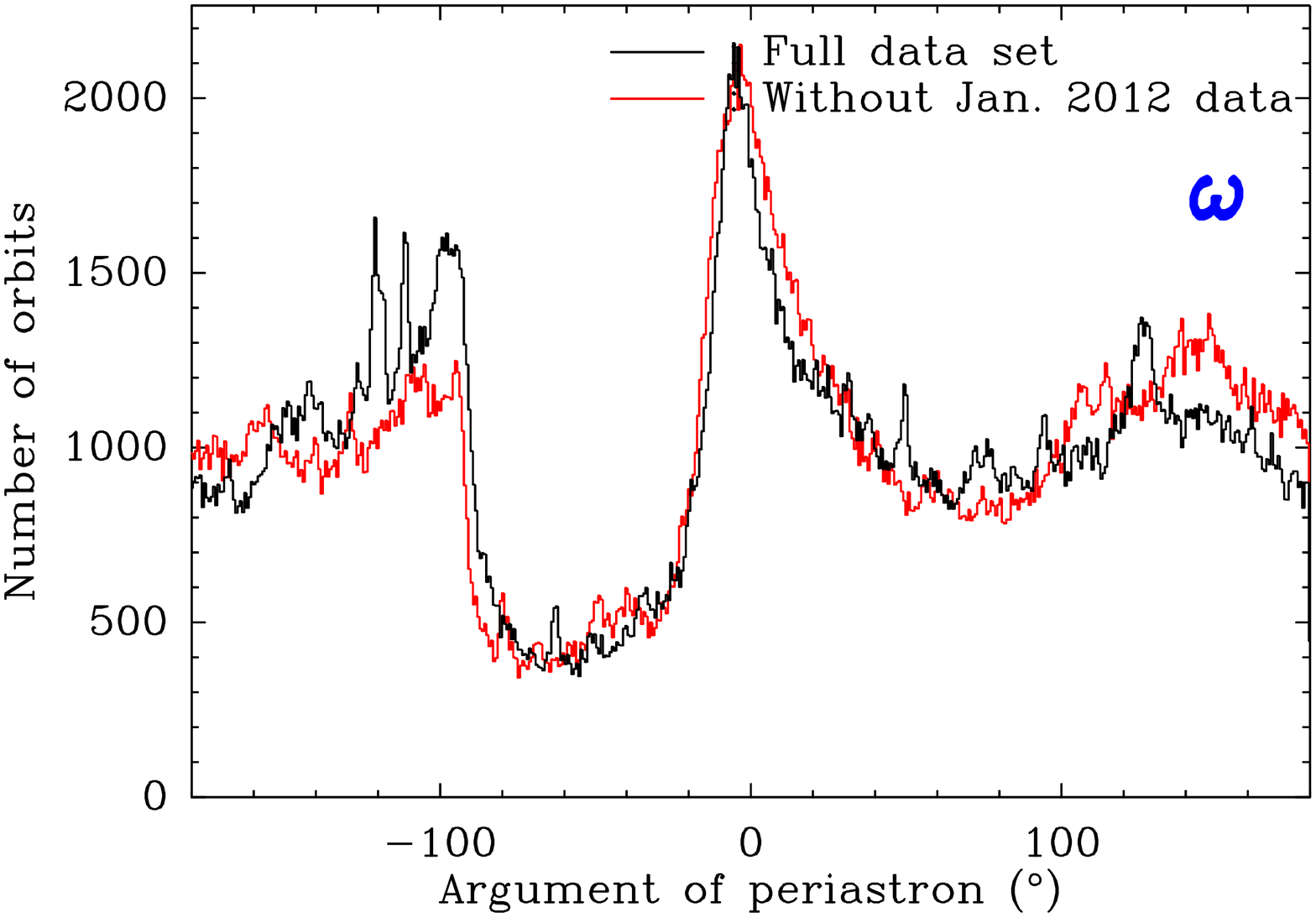} &
\includegraphics[width=7.2cm]{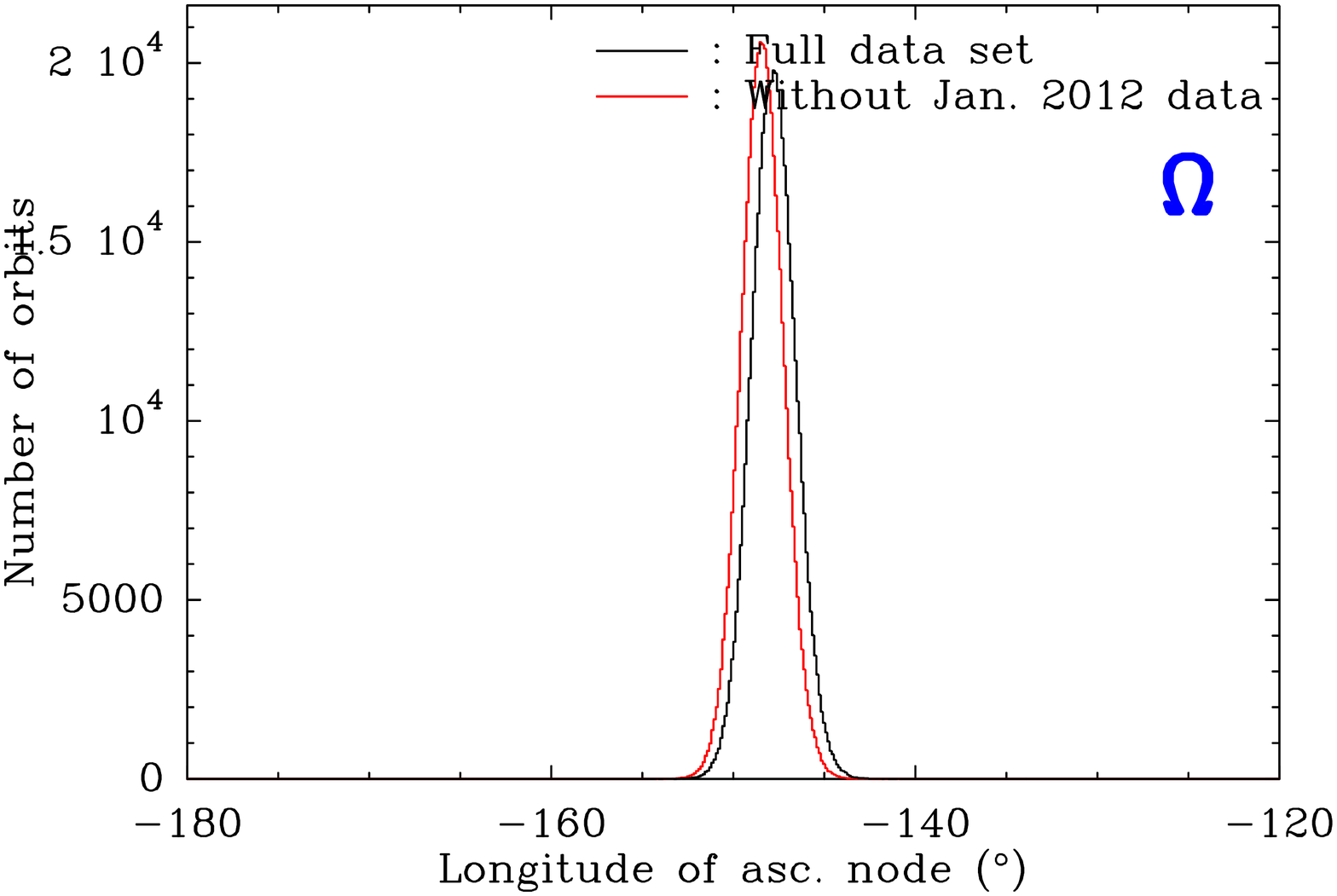} \\
\includegraphics[width=7.2cm]{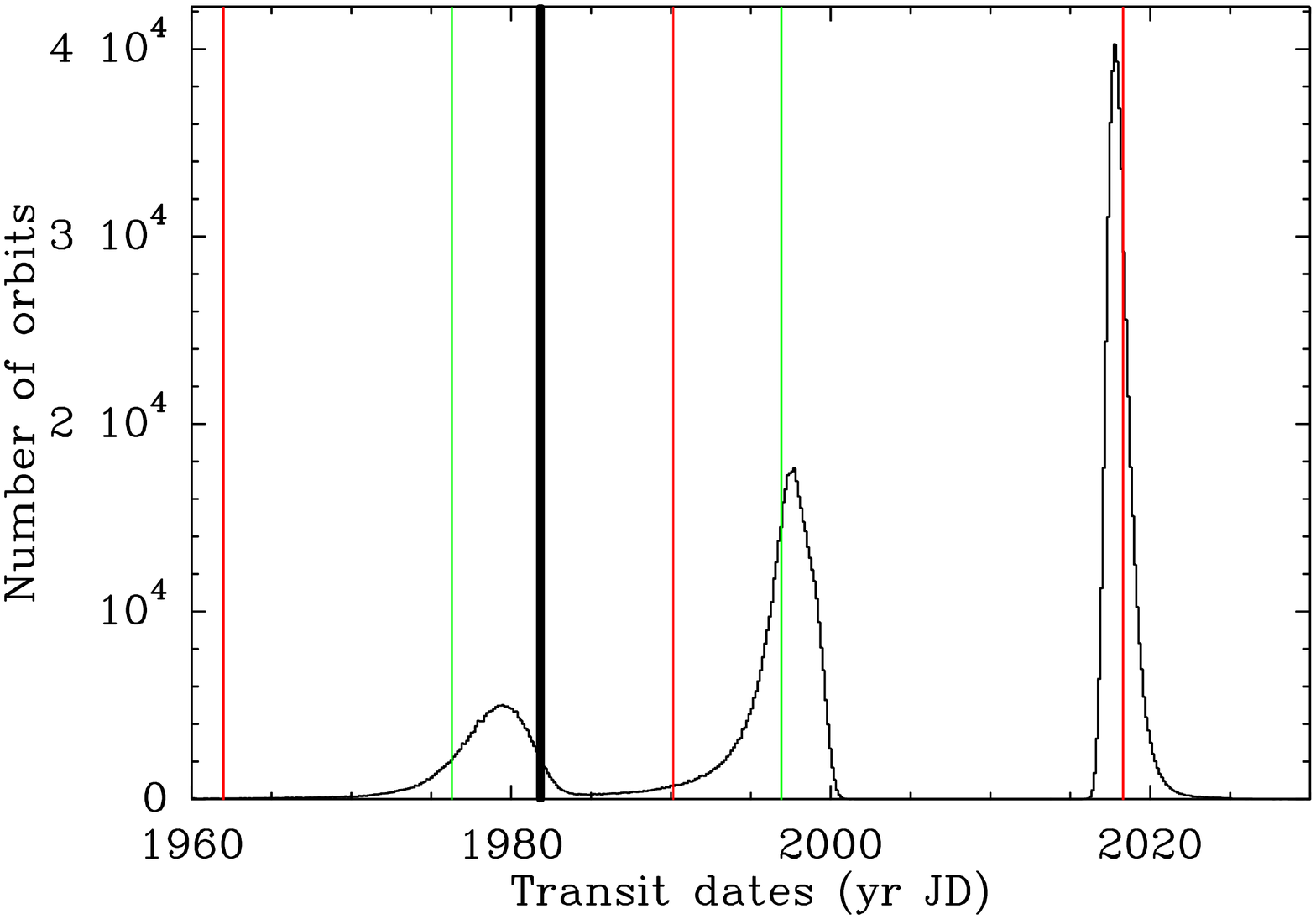} &
\includegraphics[width=7.2cm]{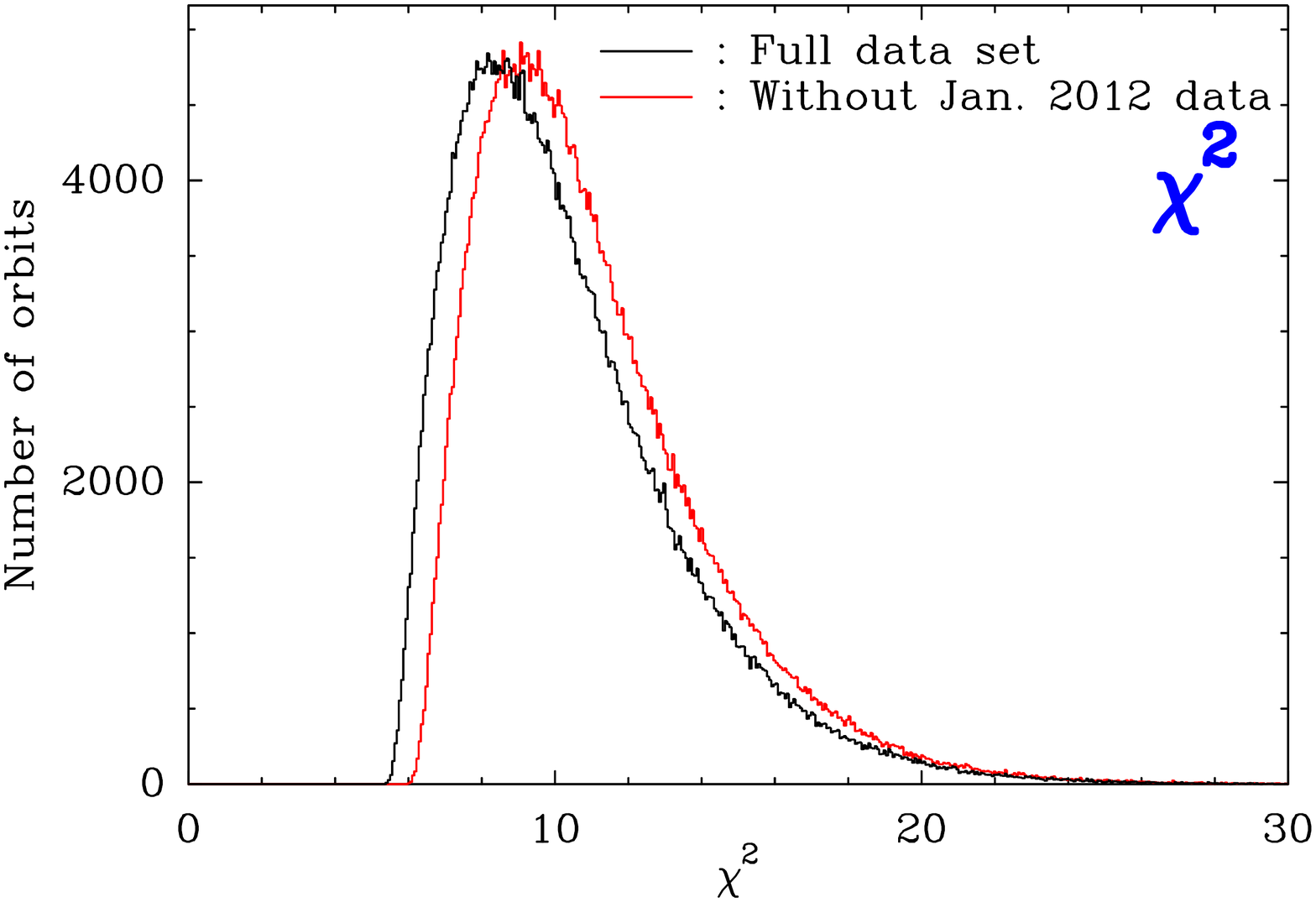} \\
\end{tabular}
\end{center}
\caption{ \tiny
 Distribution of orbital parameters from Markov-Chain Monte-Carlo fit of the set of astrometric data from \cite{2012A&A...542A..41C} (black) and adding the new astrometric data point gathered from the new H-band data (red; see Section \ref{sec:photastro}). Top row: semi-major axis $a$ (left) and $e$ (right). Third row bottom: period $P$ (left) and inclinaison $i$ (right). Second row from bottom: argument of periastrom $\omega$ (left), and longitude of ascending node $\Omega$ (right). Bottom row: transit date (left) and distribution of $\chi^{2}$. Red bars correspond to parameters found from the best LSLM $\chi^{2}$ model. Green bars correspond to highly probable orbital parameters derived from the MCMC approach. The 1981 photometric events is reported as a thick black bar on the bottom-left diagram.}
\label{of1}
\end{figure*}
}

\subsection{New mass estimates}
\label{newmasses}

\begin{table}[t]
\begin{minipage}[ht]{\columnwidth}
\caption{Masses, $\mathrm{T_{eff}}$, surface gravity, and radii estimated from hot-start evolutionary models. FM08 stands for  "Fortney and Marley 2008" models  \citep{2007ApJ...655..541M, 2008ApJ...683.1104F}. SB/cf1s, SB/cf3s, SB/hy1s,  and SB/hy3s corresponds to \cite{2012ApJ...745..174S} evolutionary models with solar-metallicity cloud-free atmospheres, 3x solar-metallicity cloud-free atmospheres, 1x solar-metallicity cloudy atmospheres, and 3x solar-metallicity cloudy atmospheres respectively.}
\label{tab:hot-start}
\centering
\renewcommand{\footnoterule}{}  
\begin{tabular}{llllll}
\hline \hline 
Model		 		&  Input			  		&	 Mass									& 		$\mathrm{T_{eff}}$			&		log g						&	Radius													\\   			
						&								&		($\mathrm{M_{Jup}}$)		&	(K)										&		$\mathrm{log(cm.s^{-2}}$)							&	($\mathrm{R_{Jup}}$)					\\	
\hline
DUSTY				&$\mathrm{M_{J}}$	&$\mathrm{10.4_{-1.3}^{+2.7}}$ & $\mathrm{1720_{-39}^{+65}}$	&$\mathrm{4.04^{+0.13}_{-0.07}}$&$\mathrm{1.53_{-0.09}^{+0.05}}$  \\
DUSTY				&$\mathrm{M_{H}}$	&	$\mathrm{9.2_{-1.1}^{+3.0}}$	&$\mathrm{1615_{-38}^{+53}}$&$\mathrm{4.00^{+0.16}_{-0.05}}$	&$\mathrm{1.50_{-0.09}^{+0.04}}$\\
DUSTY				&$\mathrm{M_{K_{s}}}$	&	$\mathrm{8.4_{-1.1}^{+3.3}}$	&$\mathrm{1537_{-42}^{+79}}$&$\mathrm{3.98^{+0.18}_{-0.07}}$	&$\mathrm{1.48_{-0.09}^{+0.03}}$\\
DUSTY				&$\mathrm{M_{L'}}$	&			$\mathrm{9.8_{-2.5}^{+3.8}}$	&$\mathrm{1674_{-176}^{+177}}$&$\mathrm{4.02^{+0.14}_{-0.09}}$	&$\mathrm{1.52_{-0.10}^{+0.07}}$\\
DUSTY				&$\mathrm{M_{NB}}$ & 		$\mathrm{7.4_{-1.9}^{+5.0}}$	&$\mathrm{1434_{-134}^{+261}}$&$\mathrm{3.94^{+0.23}_{-0.12}}$&	$\mathrm{1.45_{-0.08}^{+0.06}}$	\\
DUSTY				&$\mathrm{M_{M'}}$ & 		$\mathrm{9.0_{-2.5}^{+4.6}}$	&$\mathrm{1635_{-251}^{+309}}$&$\mathrm{4.01^{+0.12}_{-0.15}}$&	$\mathrm{1.51_{-0.06}^{+0.12}}$	\\
DUSTY				&$\mathrm{T_{eff}}$	&		$\mathrm{10.1_{-1.8}^{+3.2}}$	&			n.a.									&$\mathrm{4.03^{+0.14}_{-0.08}}$&	$\mathrm{1.53_{-0.12}^{+0.06}}$	\\
DUSTY				&$\mathrm{L/L_{\odot}}$	&	$\mathrm{9.0_{-1.5}^{+3.4}}$&		$\mathrm{1594_{-77}^{+105}}$		&$\mathrm{4.00^{+0.17}_{-0.08}}$&		$\mathrm{1.50_{-0.10}^{+0.04}}$\\
COND				&$\mathrm{M_{J}}$	&	$\mathrm{7.9_{-1.5}^{+3.8}}$	&	$\mathrm{1490_{-93}^{+124}}$	&	$\mathrm{3.96^{+0.20}_{-0.09}}$	&	$\mathrm{1.47_{-0.11}^{+0.04}}$\\
COND				&$\mathrm{M_{H}}$	&		$\mathrm{8.5_{-1.4}^{+3.5}}$	&$\mathrm{1544_{-68}^{+107}}$&	$\mathrm{3.98^{+0.18}_{-0.08}}$	&	$\mathrm{1.48_{-0.09}^{+0.04}}$\\
COND				&$\mathrm{M_{K_{s}}}$	&$\mathrm{9.7_{-1.3}^{+3.1}}$	&$\mathrm{1663_{-53}^{+84}}$&	$\mathrm{4.02^{+0.15}_{-0.07}}$	&	$\mathrm{1.52_{-0.08}^{+0.03}}$\\
COND				&$\mathrm{M_{L'}}$	&	$\mathrm{11.7_{-2.3}^{+2.6}}$	&$\mathrm{1838_{-131}^{+112}}$&	$\mathrm{4.06^{+0.10}_{-0.08}}$&$\mathrm{1.58_{-0.08}^{+0.07}}$\\
COND				&$\mathrm{M_{NB}}$ &$\mathrm{9.4_{-1.1}^{+3.9}}$&$\mathrm{1636_{-153}^{+178}}$&	$\mathrm{4.01^{+0.15}_{-0.11}}$&	$\mathrm{1.51_{-0.11}^{+0.07}}$\\
COND				&$\mathrm{M_{M'}}$ &$\mathrm{13.2_{-1.8}^{+2.0}}$&$\mathrm{2017_{-131}^{+115}}$&	$\mathrm{4.07^{+0.01}_{-0.05}}$&	$\mathrm{1.70_{-0.08}^{+0.08}}$\\
COND				&$\mathrm{T_{eff}}$	&	$\mathrm{10.1_{-1.8}^{+3.2}}$&			n.a.							&$\mathrm{4.03^{+0.14}_{-0.08}}$&		$\mathrm{1.53_{-0.12}^{+0.06}}$\\
COND				&$\mathrm{L/L_{\odot}}$	&	$\mathrm{9.0_{-1.5}^{+3.4}}$&		$\mathrm{1594_{-77}^{+105}}$		&$\mathrm{4.00^{+0.17}_{-0.08}}$&		$\mathrm{1.50_{-0.10}^{+0.04}}$\\
\hline
FM08/1s				&$\mathrm{M_{J}}$	&			$\mathrm{\geq 7.7}$		&	 $\mathrm{\geq 1333}$	&-- & -- \\
FM08/1s				&$\mathrm{M_{H}}$	&	$\mathrm{\geq 10}$	&		$\mathrm{\geq 1554}$	&-- & -- \\
FM08/1s				&$\mathrm{M_{K_{s}}}$	&	$\mathrm{> 10}$	&		$\mathrm{> 1400}$	&-- & -- \\
FM08/1s				&$\mathrm{M_{L'}}$	&			$\mathrm{> 10}$   &		$\mathrm{> 1400}$	&-- & -- \\
FM08/1s				&$\mathrm{M_{NB}}$	&		$\mathrm{> 10}$   &		$\mathrm{> 1400}$	&-- & -- \\
FM08/1s				&$\mathrm{M_{M'}}$	&			$\mathrm{> 10}$   &		$\mathrm{> 1400}$	&-- & -- \\
FM08				&$\mathrm{T_{eff}}$	&   		$\mathrm{\geq 9.0}$	&				n.a.						&	  --  &  -- \\	
FM08				&$\mathrm{L/L_{\odot}}$	&   		$\mathrm{\geq 8.0}$	&	$\mathrm{\geq 1496}$	&	  --  &-- \\	
FM08/5s				&$\mathrm{M_{J}}$	& $\mathrm{\geq 9.6}$		&	 $\mathrm{\geq 1483}$		&-- & -- \\
FM08/5s				&$\mathrm{M_{H}}$	&		$\mathrm{\geq 10}$ &		$\mathrm{> 1554}$	&-- & -- \\
FM08/5s				&$\mathrm{M_{K_{s}}}$	&		$\mathrm{> 10}$ &		$\mathrm{> 1400}$	&-- & -- \\
FM08/5s				&$\mathrm{M_{L'}}$	&			$\mathrm{> 10}$	&		$\mathrm{> 1400}$  &-- & -- \\
FM08/5s				&$\mathrm{M_{NB}}$	&		$\mathrm{\geq 10}$   &		$\mathrm{\geq 1400}$	&-- & -- \\
FM08/5s				&$\mathrm{M_{M'}}$	&			$\mathrm{> 10}$	&		$\mathrm{> 1400}$  &-- & -- \\
\hline
SB/cf1s			&$\mathrm{M_{J}}$	&	$\mathrm{7.4^{+1.0}_{-1.0}}$		&		--										&	$\mathrm{3.98^{+0.12}_{-0.12}}$	&			$\mathrm{1.41^{+0.1}_{-0.1}}$									\\
SB/cf1s				&$\mathrm{M_{H}}$	&	$9.2^{+0.5}_{-0.5}$		&		--										&	$4.06^{+0.09}_{-0.09}$		&			$1.43^{+0.1}_{-0.1}$												\\
SB/cf1s				&$\mathrm{M_{L'}}$	&		$\mathrm{\geq 10}$										&				--								&		--						&		--					\\
SB/cf3s				&$\mathrm{M_{J}}$	&	$\mathrm{7.8^{+1.0}_{-1.0}}$		&					--							&	$\mathrm{4.00^{+0.12}_{-0.12}}$	&		$\mathrm{1.42^{+0.1}_{-0.1}}$				\\
SB/cf3s				&$\mathrm{M_{H}}$	&		$9.1^{+0.5}_{-0.5}$		&						--						&	$4.06^{+0.09}_{-0.09}$	&		$1.43^{+0.1}_{-0.1}$			\\
SB/cf3s				&$\mathrm{M_{L'}}$	&		$\mathrm{\geq 10}$											&		--										&		--								&			--				\\
SB/hy1s				&$\mathrm{M_{J}}$	&		$\mathrm{\geq 9.5}$			&		--										&	--	&	--													\\
SB/hy1s				&$\mathrm{M_{H}}$	&	$9.5^{+0.5}_{-0.5}$			&			--									&		$4.07^{+0.09}_{-0.09}$								&	  	$1.43^{+0.1}_{-0.1}$													\\
SB/hy1s				&$\mathrm{T_{eff}}$	&   		$\mathrm{\geq 8.5}$	&				n.a.						&	  --  &  -- \\	
SB/hy1s				&$\mathrm{L/L_{\odot}}$	&   		$\mathrm{\geq 8.2}$	&	$\mathrm{\geq 1574}$	&	  --  &-- \\	
SB/hy1s				&$\mathrm{M_{L'}}$	&		$\mathrm{\geq 10}$											&		--										&			--							&		--				\\
SB/hy3s				&$\mathrm{M_{J}}$	&		$\geq 9.9$		&				--								& -- &	--											\\
SB/hy3s				&$\mathrm{M_{H}}$	&		$9.4^{+0.5}_{-0.5}$										&					--							&	$4.08^{+0.09}_{-0.09}$		&		$1.43^{+0.1}_{-0.1}$												\\
SB/hy3s				&$\mathrm{M_{L'}}$	&			$\mathrm{\geq 10}$									&						--						&			--							&	--													\\
\hline
\end{tabular}
\end{minipage}
\end{table}

	   \begin{figure}
   \centering
   \includegraphics[width=\columnwidth]{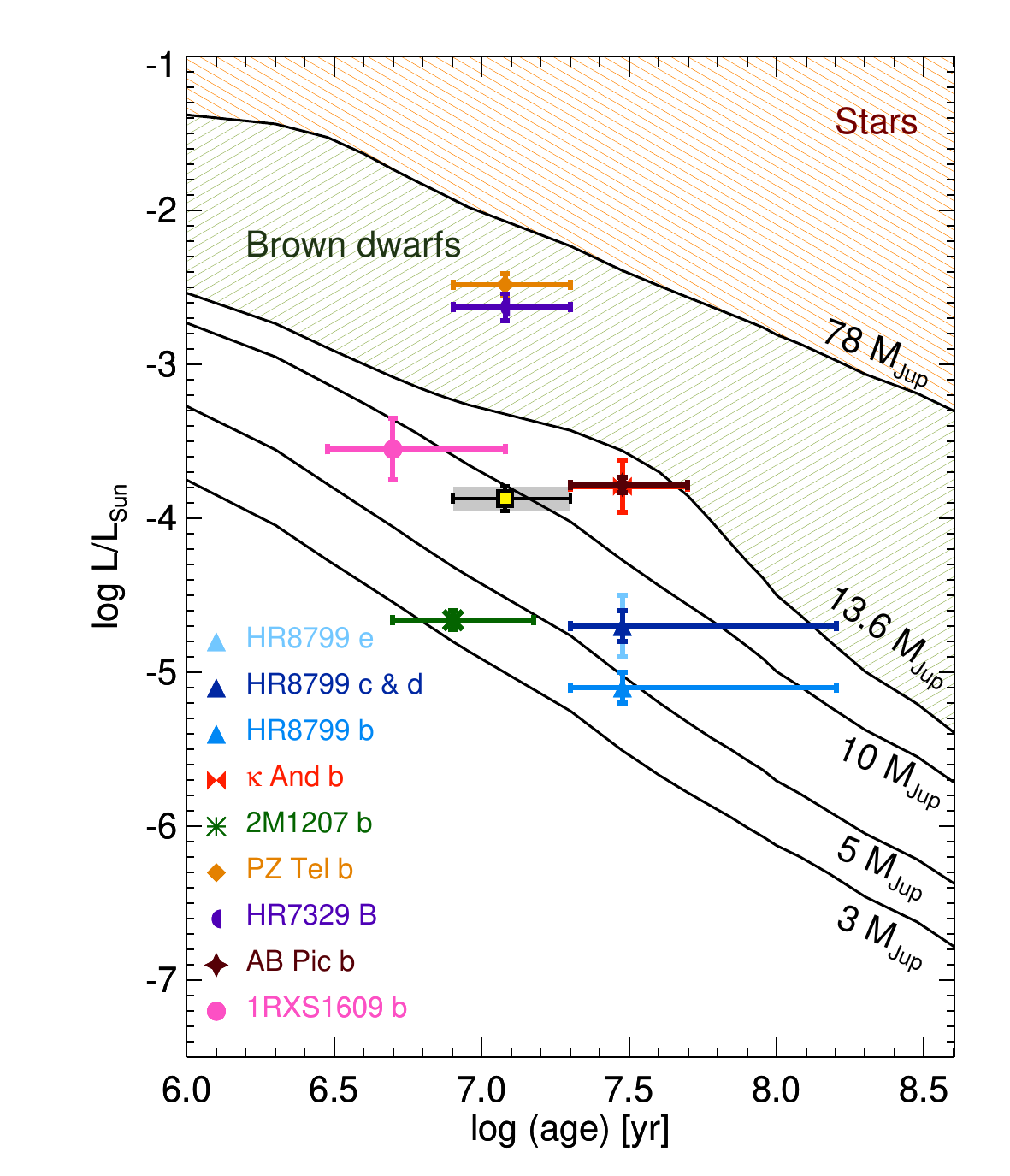}
      \caption{Luminosities of reference young planetary mass companions compared to COND evolutionary tracks for masses of 3, 5, 10, 13, and 78 $\mathrm{M_{Jup}}$. The luminosities of the two brown-dwarfs companions to stars members of the $\beta$ Pictoris moving group (HR7329 B, PZ Tel b) are overlaid for comparison.}
         \label{Fig:Figref}
   \end{figure}

	We used  evolutionary models that couple non-grey radiative-convective atmosphere models to thermal evolution models based on different intial conditions to estimate new physical parameters (mass, $\mathrm{T_{eff}}$, log g, radius) for the planet.

		\subsubsection{Hot-start and cold-start evolutionary tracks}
				\label{subsec:Lyon}
We first considered the DUSTY and COND evolutionary tracks of \cite{2000ApJ...542..464C} and \cite{2003A&A...402..701B} to derive the mass, $\mathrm{T_{eff}}$, log g, and radius of the planet (Table \ref{tab:hot-start}) using the photometry in each  bands, the luminosity, or the temperature determined in Sections \ref{subsub:cmds} and \ref{subsec:specsynth}. Masses derived from J, H, and M' band photometry  agree with those found in previous studies \citep{2010Sci...329...57L,2010A&A...512A..52B, Quanz2010} with the exeption of COND evolutionary models predictions for the M' band. All predicted surface gravities and radii but the one derived from the COND evolutionary models in the M' band are also consistent with those found from the SED fit (see Section \ref{subsec:specsynth}). The problem in the M' band could arise from an incorrect representation of the emergent fluxes of $\beta$ Pictoris b longward 4.0 $\mu$m by the associated AMES-Cond atmospheric models. Predictions based on the measured temperature and luminosity should be less affected by this problem.  We find masses of $\mathrm{10.1^{+3.2}_{-1.8}\: M_{Jup}}$  and $\mathrm{9.0_{-1.5}^{+3.4}\: M_{Jup}}$ for these two input parameters respectively. We plot the position of $\beta$ Pictoris along with those of other young companions in a diagram showing the evolution of the luminosity of objects with planetary, sub-stellar, and stellar masses (Figure \ref{Fig:Figref}). The planet sits above 2M1207b and clearly bellow the two brown dwarf companions discovered around stars members of the $\beta$ Pictoris moving group.

\begin{figure}
   \centering
   \includegraphics[width=\columnwidth]{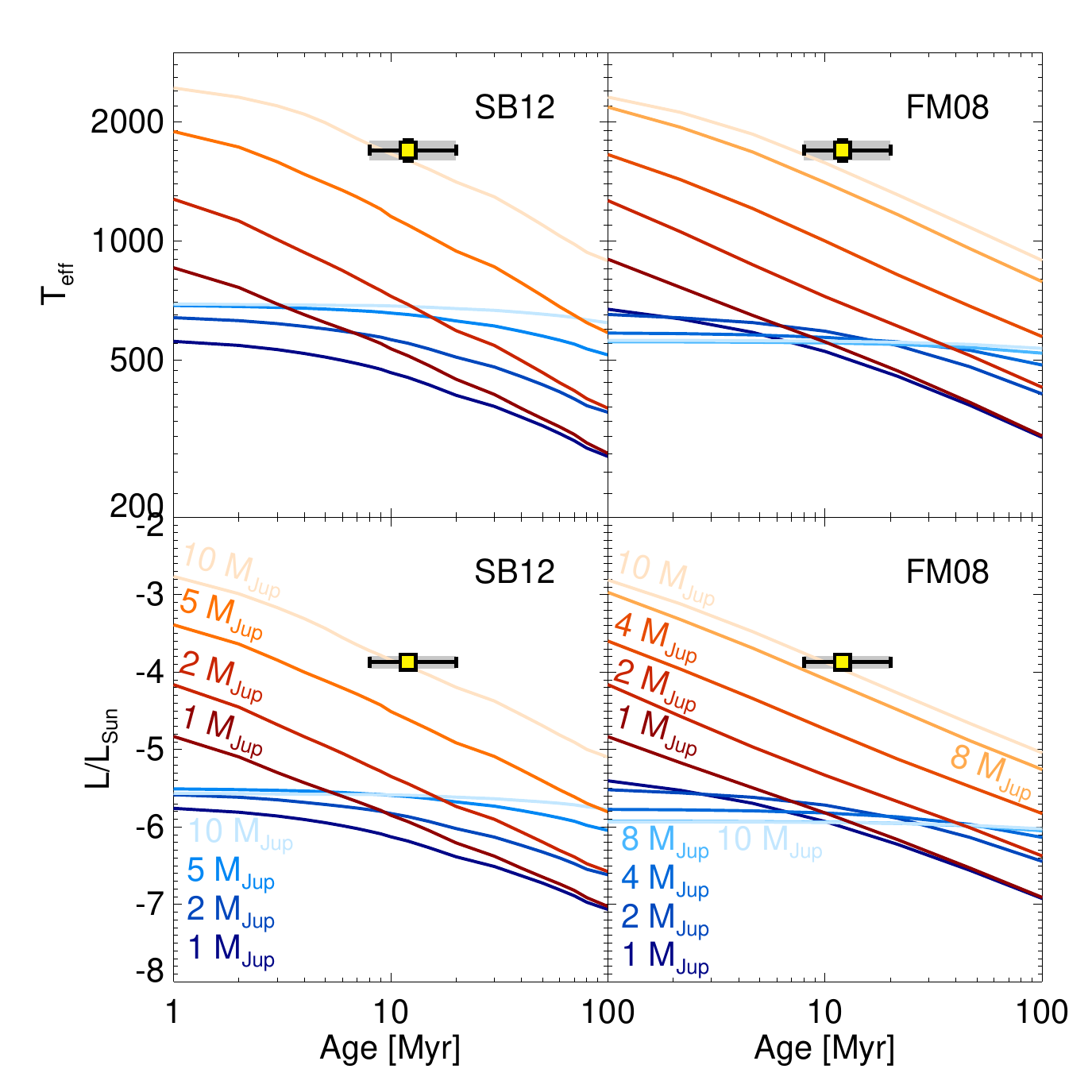}
      \caption{Luminosity and effective temperatures of $\beta$ Pictoris b compared to those predicted by ``cold-start" (blue curves) and ``hot-start" (red curves) models of \cite{2008ApJ...683.1104F} --FM08 -- and \cite{2012ApJ...745..174S} -- SB12 -- at solar metallicity for different masses and ages.}
         \label{Fig:FigrefHC}
   \end{figure}

Other ``hot-start" models have been proposed by \cite{2007ApJ...655..541M} for 1-10 $\mathrm{M_{Jup}}$ planets. \cite{2008ApJ...683.1104F} couple an upgrated version of these models to cloudy  atmospheric models with 1 and 5 times solar abundances. Models with enriched atmospheres were created to account for the possible enhancement of metal content of exoplanets formed in disks, like  Jupiter and Saturn \citep[see ][and ref. therein]{2008ApJ...683.1104F}. We report evolutionary models predictions (generated for the NaCo passbands) for these two choices of boundary conditions in Table \ref{tab:hot-start} and show the location of $\beta$ Pictoris b luminosity and $T_{eff}$ with respect to tracks in Figure \ref{Fig:FigrefHC}. Predictions for the highest magnitude values fall  above the mass range covered by the model (predictions up to 0.8 mag fainter in the $K_{s}$ band at solar metallicity, and up to 1 mag fainter in the M' band at 5 times solar metallicity). For this reason, we are only able to give lower limits on the mass and effective temperature and can not give secure lower limits for radii and surface gravities. Nevertheless, the lower limits on masses and $\mathrm{T_{eff}}$ agree  with those derived from DUSTY/COND evolutionary models.
 
We also use ``hot-start" evolutionary models of  \cite{2012ApJ...745..174S} (hereafter SB12) and report predictions in Table  \ref{tab:hot-start} for the J, H, and L' band filters. SB12 models consider four atmospheric models (Burrows et al. 2011) as boundary conditions of their interior models:  cloud-free models at solar-metallicity (hereafter cf1s), three time solar-metallicity (cf3s), models with hydrid dust clouds at solar metallicity (hy1s) and models with hydrid clouds with three time solar metallicity (hy3s).  We retrieve predictions in agreement with estimates from the two other hot-start models. We also derived temperature and/or mass  predictions from the effective temperature and the luminosity of the planet for the hy1s models\footnote{Predictions come from Figure 5 and Table 1 of SB12. The temperature for the other boundary conditions are not given in the paper and do not enable to make a similar comparison.}.

 We also  compared predictions of the so-called ``cold-start" models proposed by \cite{2007ApJ...655..541M} and SB12 to  the new $\beta$ Pictoris b photometry, temperature and luminosity (see Figure \ref{Fig:FigrefHC}). We confirm that only masses (and lower limits) derived from ``hot-start"  models agree  with the independent upper limit of 15.5 $\mathrm{M_{Jup}}$ of the mass of $\beta$ Pictoris b given the mass range covered by the models.	\\

   \begin{figure}
   \centering
   \includegraphics[width=\columnwidth]{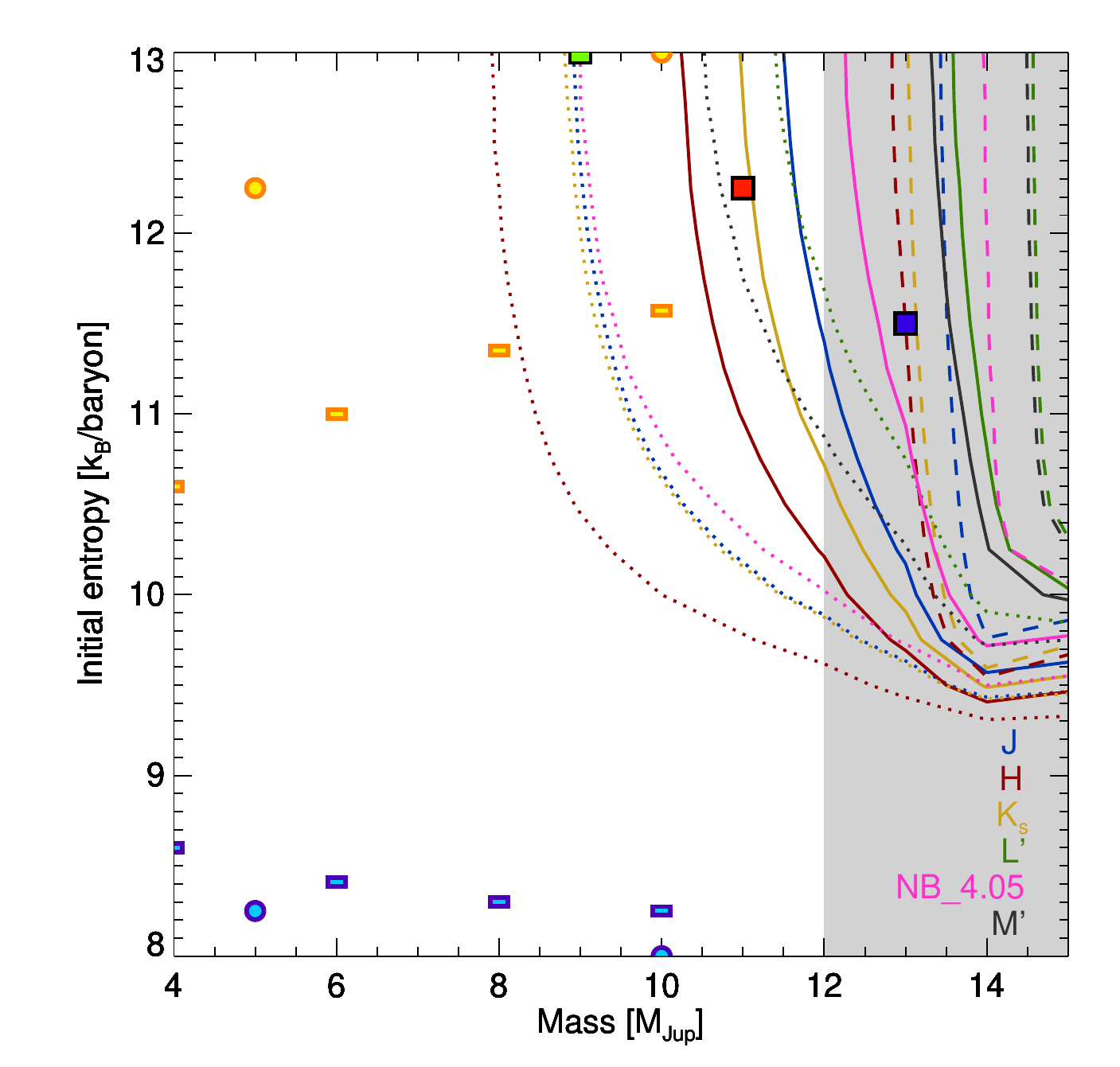}
      \caption{Masses and initial entropies for $\beta$ Pictoris b (solid lines) for which  warm-start evolutionary models of \cite{2012ApJ...745..174S} with cloudy atmospheres and solar metallicity (hy1s) at 12 Myr predict the measured absolute magnitudes of the planet. The dotted and dashed lines brakett the range of possible initial entropies and masses of the planet considering the error on the measured photometry and on the age (8-20 Myr) of the system. Green, red, and blue squares correspond to masses and initial entropies for which the predicted photometry best fit simultaneously the companion one for ages of 8, 12, and 20 Myr respectively. Masses excluded from present radial velocity measurements and assuming a planetary orbit with a semi-major axis  of 9 AU are reported in grey. Yellow and blue dots represents the initial entropies considered in the SB12 hot and cold star models respectively. Yellow and blue rectangles represents the entropies at 1 Myr in the \cite{2007ApJ...655..541M} hot and cold star models.}
         \label{Fig:Fig6}
   \end{figure}

		\subsubsection{``Warm-start" models}
\cite{2012ApJ...745..174S} also propose to explore intermediate cases between cold-start and hot-start models where only a fraction of the gravitational potential energy liberated at the accretion shock do not enter the initial energy budget of the planet. These ``warm-start" models do not aim at modeling self-consistently the formation history but rather propose a continuum of evolutionary tracks corresponding to a range of initial entropies  ($\mathrm{S_{init}}$; from 8 Boltzmann constants per baryon -- $\mathrm{k_{B}}$/baryon -- and up to 13 $\mathrm{k_{B}}$; with 0.25$\mathrm{k_{B}}$/baryon increments). 

We computed the predicted absolute NaCo magnitudes of 1-15 $M_{Jup}$ planets using the predicted ``warm-start" low-resolution (R$\sim200$) spectra\footnote{Spectra can be downloaded at \url{http://www.astro.princeton.edu/~burrows/warmstart/spectra.tar.gz}} of planets placed at 10 pc, the instrument passbands, and the same flux-calibrated spectrum of Vega \citep{2007ASPC..364..315B} used to generate NaCo magnitudes of the LYON tracks. As expected, predicted absolute magnitudes increase with the object mass and $\mathrm{S_{init}}$. We report in Figure \ref{Fig:Fig6} the initial entropies and masses for which $\beta$ Pictoris photometry matches the prediction of hy3s models for an age of 12 Myr (solid lines), and using the extreme values of our photometry for ages of 8 and  20 Myr (dashed lines). For each considered filter, the allowed range of $\mathrm{S_{init}}$ and mass of the planet is contained between the dotted and dashed lines. This range can be compared on the figure to the interval of masses excluded from independent constraints on the mass of the planet and shaded on the figure. We olverlay $\mathrm{S_{init}}$  considered for the hot-start (yellow/orange symbols) and cold-start (blue/violet symbols) models of \cite{2007ApJ...655..541M} and \cite{2008ApJ...683.1104F} (rectangles; entropy at 1 Myr) and SB12 (big dots). 

We also compare our combined set of photometric points to  model predictions using a $\chi^{2}$ minimization (which do not require adjusting a dillution factor). Results of the minimization are reported for  8 Myr, 12 Myr, and 20 Myr predictions and for the four  boundary conditions in Table \ref{tab:warm-start}.  We report in Figure \ref{Fig:Fig6} the corresponding entropies and masses. 

\begin{table}[t]
\begin{minipage}[ht]{\columnwidth}
\caption{Best fitted photometric predictions of the warm-start evolutionary models}
\label{tab:warm-start}
\centering
\renewcommand{\footnoterule}{}  
\begin{tabular}{lllll}
\hline \hline 
Atmospheric model			 &  	 Age     &			Mass								& 		$\mathrm{S_{init}}$				&	$\mathrm{\chi^{2}}$	\\
										 &		(Myr)		&			($\mathrm{M_{Jup}}$)	&	($\mathrm{k_{B}}$/baryon)   	&									\\		
\hline
Cloud free - 1x solar	 &			8		&			10								&		11.50									&			4.64					\\							
Cloud free - 3x solar	 &			8		&			9								&		13.00									&			3.40					\\							
Hybrid cloud - 1x solar		 &			8		&			9								&		13.00									&			1.59					\\							
Hybrid cloud - 3x solar		 &			8		&			9								&		12.00									&			1.59					\\							
Cloud free - 1x solar	 &			12	&			12								&		11.00									&			4.86					\\							
Cloud free - 3x solar	 &			12	&			11								&		13.00									&			3.49					\\							
Hybrid cloud - 1x solar		 &			12	&			11								&		12.25									&			1.67					\\							
Hybrid cloud - 3x solar		 &			12	&			11								&		11.75									&			1.69					\\							
Cloud free - 1x solar	 &			20	&			13								&		13.00									&			4.94					\\							
Cloud free - 3x solar	 &			20	&			13								&		11.50									&			3.62					\\							
Hybrid cloud - 1x solar		 &			20	&			13								&		11.50									&			1.75					\\							
Hybrid cloud - 3x solar		 &			20	&			13								&		11.00									&			1.76					\\							
\hline
\end{tabular}
\end{minipage}
\end{table}


We reach the following conclusions:
\begin{itemize}
\item Single band photometry constrains the planet mass to   $\geq6\mathrm{\:M_{Jup}}$. The H-band, L'-band, and M'-band magnitudes of $\beta$ Pictoris (which tend to be less dependant of the choice of the atmospheric model according to Figure 8 of SB12), give the same constraints on the initial entropies ($\mathrm{S_{init} \geq}$9.3, 9.6, and 9.7 $\mathrm{k_{B}}$/baryon respectively) for the four boundary conditions. The constraints rise to $\mathrm{S_{init} \geq10.0\:k_{B}}$/baryon considering all the bands, the independent mass constraints,  and an age of 12 $\mathrm{M_{Jup}}$.
\item Cloud-free models fail to give coherent predicted masses and initial entropies for the different photometric bands. Cloudy models hy1s and hy3s reduce the dispersion. They predict masses greater than  8 $\mathrm{M_{Jup}}$ for the planet and indicate $\mathrm{S_{init} \geq9.3}\mathrm{\:k_{B}}$/baryon.  Comparatively, cf1s and cf3s models restraint the possible range of initial entropies to $\mathrm{\geq}$ 9.1 $\mathrm{k_{B}}$/baryon.  If we consider the independent limit of 12 $\mathrm{M_{Jup}}$ on the mass of the planet,  $\mathrm{S_{init}}$ needs to be larger than 9.2 and 9.3 $\mathrm{k_{B}}$/baryon for the cf1s and cf3s boundary conditions. The constraint rises to $\mathrm{S_{init} \geq 9.6}$ $\mathrm{k_{B}}$/baryon for hy1s and hy3s atmospheres. These initial entropies all appear larger than those considered for the cold-start models of FM08 and SB12 but lower than those used for the hot-start models. Masses predicted for the lowest magnitude values of the planet at an age of 20 Myr fall in the range proscribed by radial velocity measurements if the planet semi-major axis is 9 AU. 
\item Results from the combined fit of our photometry confirm the latter conclusions. They predict masses between 9 and 13 $\mathrm{M_{Jup}}$ and initial entropies ($\mathrm{\geq 11\:k_{B}/baryon}$) that better corresponds to those of hot-start evolutionary tracks. Masses predicted for a 20 Myr old system fall in the  range of masses excluded by dynamical constraints if $\beta$ Pictoris b semi-major axis is $\leq$9 AU. The $\chi^{2}$ of the fit is  reduced for  cloudy atmospheres with solar abundances. This is in agreement with conclusions derived  from atmospheric models fits in Section \ref{subsec:specsynth}.
\end{itemize}

 The present analysis suggests that initial conditions are intermediate between those adopted for cold-start and hot-start models of SB12 and FM08, and more likely lies closer to those adopted for hot-start models. We discuss the implications of these conclusions in the Section \ref{subsec:formBpic}.

\section{Discussion}
\label{Discussion}
\subsection{Comparison of the empirical and model-dependent analysis}
\label{subsec:cloudatmp}
Comparison of $\beta$ Pictoris b photometry  to empirical objects and atmospheric/evolutionary models converge toward the picture of an exoplanet with a cloudy atmosphere. The effective temperature derived in Section \ref{subsec:specsynth} is compatible with those derived for other young L-type companions such as AB Pic b ($\mathrm{T_{eff}=1700\pm100}$ K, Bonnefoy et al 2013, submitted), CD-35 2722 B \citep[$\mathrm{T_{eff}=1700-1900}$K,][]{2011ApJ...729..139W}, or 1RXSJ1609-2105b \citep[$\mathrm{T_{eff}=1800^{+200}_{-100}}$K,][]{2008ApJ...689L.153L} using similar PHOENIX-based atmospheric models. We also recently derived $\mathrm{T_{eff}=1700-1900K}$ for a sample of young M9.5-L0 dwarfs comparing their near-infrared spectra to BT-SETTL2010 models  (Bonnefoy et al 2013, submitted). We note however some exeptions like the young L0 companion GSC 06214−00210 b \citep[][]{2011ApJ...743..148B} that has a significant higher estimated temperature ($\mathrm{T_{eff}=2700\pm200}$K). Nevertheless, we believe that this exeption most certainly arises from the non-homogeneous classification of these companions\footnote{Some of the companions have been classified based on the comparison to mature field dwarfs spectra in the near-infrared. A classification scheme which account for the different ages of the objects \citep{2005ARA&A..43..195K, 2009AJ....137.3345C} might be more appropriate.} and that the present comparison rather validate the self-consistency of the analysis of $\beta$ Pictoris b near-infrared SED. We also note that the spectroscopic $\mathrm{T_{eff}}$ of the planet corresponds to a spectral type in the range L3-L4.5 using the conversion scale of \cite{2009ApJ...702..154S} valid for field dwarfs. Despite the scale was generated using another family of atmospheric models \citep[e.g.][]{2008ApJ...689.1327S}, this spectral type range is  consistent with that infered from the location of $\beta$ Pictoris with respect to  field dwarfs  in J-H vs $\mathrm{H-K_{s}}$ and color-magnitudes diagrams. 

If we choose to adopt a spectral type $\mathrm{L2_{\gamma}\pm2}$ for $\beta$ Pictoris b, the planet appears to have redder $\mathrm{K_{s}-L'}$ colors  with respect to early-L field dwarfs in Figure \ref{Fig:Fig2}.  SB12 evolutionary models and all but the DRIFT-PHOENIX models (Figure \ref{tab:hot-start}) suggest the planet could be overluminous in the L' band. Non-equilibrium chemistry $CO \leftrightarrows CH_{4}$  induced  by the reduced surface gravity of the object could be invoked \citep{2011ApJ...735L..39B}. But we would expect the planet to have bluer L'-M' colors than those of mature dwarfs (by decreasing the strength of the $CH_{4}$ shortward 4 $\mu$m and increasing of the CO one longward 4 $\mu$m). BT-Settl (which account for non-equilibrium chemistry) and DRIFT-PHOENIX models predict that the $\mathrm{K_{s}-L'}$ color is mostly independent of the surface gravity of the object. DRIFT-PHOENIX also suggest that metallicity is not able to explain the planet color. The same is true for the other colors represented in Figure \ref{Fig:Fig2}.  To conclude, one would expect to find a similar location for $\beta$ Pictoris b and  the other young L-type companions in the diagrams. The problem could more simply arise from an error in the neutral density factor estimates.  Adopting the default value from the NaCo User Manual ($\sim$1.8\%; and used for all the studies of $\beta$ Pictoris b so far) would bring back $\beta$ Pictoris b colors in better agreement with those of L dwarfs and young companions.  But we note that our transmission value relies on three different measurements and are not affected by any residual flux level from background subtraction. The choice of the neutral density transmission do not bias our best fitted effective temperatures. \\

\begin{table}[t]
\begin{minipage}[ht]{\columnwidth}
\caption{Properties of the $\beta$ Pictoris system}
\label{tab:sum}
\centering
\renewcommand{\footnoterule}{}  
\begin{tabular}{llll}
\hline \hline 
Parameter			 &  	$\beta$ Pictoris A    &			$\beta$ Pictoris b	&		References \\
\hline 
d	(pc)										&		$19.44\pm0.05$		&		--		&	2		\\					
Age	(Myr)										&		$12^{+8}_{-4}$	&	--			&		3	\\					
J	(mag)									&		$3.524\pm0.013$		&	$\mathrm{14.0\pm0.3}$		&	1, 4			\\
H	(mag)											&	$3.491\pm0.009$	&	$13.5\pm0.2$			&	1, 4		\\
$\mathrm{K_{s}}$ (mag)						& 	$3.451\pm0.009$	&		$12.6\pm0.1$	&	4, 5			\\
L'		(mag)										&	$3.454\pm0.003$\tablefoottext{a}	&	$\mathrm{11.0\pm0.2}$		&	1,  5, 6			\\
NB\_4.05	(mag)								&		--		&	$11.20\pm0.23$		&	 7		\\
M'		(mag)										&	$3.458\pm0.009$\tablefoottext{a}	&	$11.0\pm0.3$		&	 1, 6			\\
$\mathrm{M_{J}}$(mag)					&		$2.081\pm0.014$		&	$\mathrm{12.6\pm0.3}$		&	1			\\
$\mathrm{M_{H}}$	(mag)				&		$2.048\pm0.011$		&		$12.0\pm0.2$		&	1		\\
$\mathrm{M_{K_{s}}}$ (mag)						&	$2.007\pm0.011$		&	$11.2\pm0.1$		&		5		\\
$\mathrm{M_{L'}}$		(mag)				&	$2.011\pm0.006$		&		$\mathrm{9.5\pm0.2}$		&		1, 5		\\
$\mathrm{M_{NB\_4.05}}$	(mag)					&		--		&		$9.76\pm0.24$	&		 7		\\
$\mathrm{M_{M'}}$	(mag)					&		$\mathrm{2.014\pm0.011}$	&		$\mathrm{9.5\pm0.3}$	&		1		\\
Spectral type								&		A6V		&	$\mathrm{L2_{\gamma}\pm2:}$		&		1, 8			\\
$\mathrm{T_{eff}}$	(K)				&		8052, 8036		&	$1700\pm100$		&	1, 8, 9			\\
log g		(dex)								&		4.15, 4.21		&	$4.0\pm0.5$			&	1, 8, 9		\\
M/H			(dex)								&		+0.05, +0.11		&		0.0 ?		&		1, 8, 9	\\
$\mathrm{log_{10}(L/L_{\odot}}$)		&		$0.91\pm0.01$\tablefoottext{b}					&		$\mathrm{-3.87\pm0.08}$ &  1 \\
Mass		($\mathrm{M_{\odot}}$)									&		$1.70\pm0.05$\tablefoottext{b}			&	$0.010^{+0.003}_{-0.002}$\tablefoottext{c}		&		1	\\
\hline
\end{tabular}
\end{minipage}
\tablefoot{[1] - this work,  [2] - \cite{2007A&A...474..653V}, [3] - \cite{2001ApJ...562L..87Z}, [4] - \cite{1996yCat..41190547V}, [5] - \cite{2011A&A...528L..15B}, [6] - \cite{1991A&AS...91..409B}, [7] - \cite{2010ApJ...722L..49Q}, [8] - \cite{2006AJ....132..161G}, [9] - \cite{2008A&A...490..297S}}
\tablefoottext{a}{ESO L and M-band magnitude.} \\
\tablefoottext{b}{Using the evolutionary tracks of \cite{2012A&A...537A.146E} with and without rotation.}\\
\tablefoottext{c}{Using COND and DUSTY ``hot-start" evolutionary models.\\} 
\end{table}

\subsection{$\beta$ Pictoris b abundances}

 \cite{2006AJ....132..161G} and \cite{2008A&A...490..297S} estimate solar metallicity (M/H=+0.05 and +0.11 respectively) for $\beta$ Pictoris A comparing the optical spectra of the star to atmospheric models.   The abundances of $\beta$ Pictoris A are consistent with measurements found for other young nearby associations members \citep{2009A&A...501..965V, 2012arXiv1209.2591B}. 
BT-Settl 2012, DRIFT-PHOENIX, and SB12 models at solar metallicity best match the SED of $\beta$ Pictoris b.   A careful visual check of the fits reveals that differences in magnitudes between the best fitting SB12 models (and associated $\chi^{2}$ values) with 1x and 3x solar-metallicities (or M/H=0 and +0.5 dex respectely) are negligible.  The planet would need to be fainter of 0.45 magnitude in the J band and brighter than 0.25 magnitudes in the $\mathrm{K_{s}}$ to be best fitted by the 3x solar-metallicity DRIFT-PHOENIX models.  The J-band of best fitted metal-enriched BT-Settl2012 models is 1 magnitude fainter than that of $\beta$ Pictoris b. The higher departure of BT-Settl2012 models flux in the J band at [M/H]=+0.5 compared to the DRIFT-PHOENIX might come from the use of different reference solar abundances. The reference solar abundances of \cite{2011SoPh..268..255C} should tend to make BT-Settl2012 models under-metallic  with respect to DRIFT-PHOENIX for the same [M/H] value.  The comparison of  low-resolution spectra of $\beta$ Pictoris b  to synthetic spectra further exploring the impact of metallicity  (C/O ratio for instance) would be needed to firmly estimate this atmospheric parameter. 


\subsection{The formation of $\beta$ Pictoris b}
\label{subsec:formBpic}
It is thought that disk instability leads gas to retain much more of its initial entropy than core-accretion, leading to a planet with ``hot-start"  conditions. New 3D simulations of the collapse of disk clumps confirm this fact \citep{2012MNRAS.427.1725G}. Constraints on the initial entropies of $\beta$ Pictoris b would then suggest a formation by disk-instability. \cite{2011ApJ...736...89J} have shown that disk-instability is not the dominant formation scenario of wide-orbit companions around stars more massive than $\beta$ Pictoris (and then more favorable to disk-instability) using models which can predict within which boundaries in mass/semimajor axis space fragments can form from disk instabilities (see \cite{2011ApJ...736...89J} and \cite{2010arXiv1012.5281M} for a description of the models). Rameau et al 2013 (submitted) shows that  the same models  applied to the case of $\beta$ Pictoris  can only form self-gravitating clumps with masses $\mathrm{\geq 3 M_{Jup}}$ at distances greater than $\sim$45 AU from the star.  Additional dynamical process such as tidal interaction with the disk \citep{2011MNRAS.416.1971B, 2011ApJ...737L..42M, 2012ApJ...746..110Z} and/or planet-planet interactions \citep{2012ApJ...754...57B} would then be needed to explain the present location of the planet.

Nevertheless, \cite{2012A&A...547A.111M} argue that the link between cold/hot start evolutionary models, and more generally the link between initial entropy and the the formation scenario (e.g. core-accretion vs disk instability) can not presently be  established given our limited knowledge of the way the material is accreted, and  in particular at the accretion shock \citep{2007ApJ...655..541M, 2011A&A...530A..13C}. For example, core-accretion with a subcritical shock (no radiative losses) also leads to a ``hot start". Constraints on the initial entropy could nevertheless be compared to those of planets similar to $\beta$ Pictoris b (discovered in disks, with dynamical constraints) that will be unearthed by the planet imagers (LBTI/LMIRCam, Subaru/SCExAO, VLT/SPHERE, Gemini/GPI) in the incoming years.

   \begin{figure}
   \centering
   \includegraphics[width=\columnwidth]{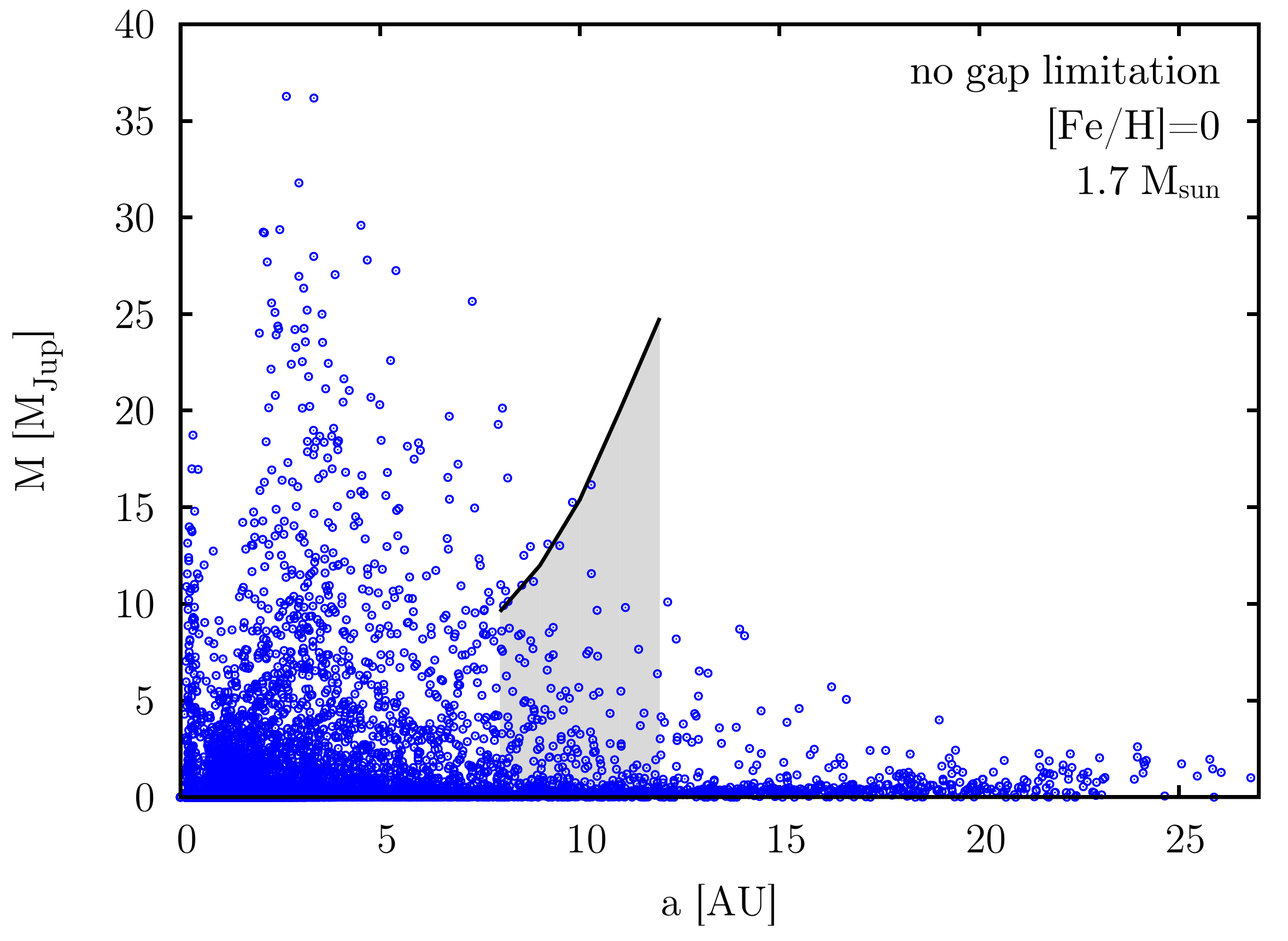}
      \caption{Population synthesis of planets formed one-by-one by core accretion around a 1.7 $\mathrm{M_{\odot}}$. We overlaid limits on the mass coming from radial velocity measurments (black solid line).}
         \label{Fig:FigPopsynth}
   \end{figure}

Could $\beta$ Pictoris b be formed by core-accretion then? \cite{2008ApJ...673..502K} suggest that cores can form at 8-10 AU around a star as massive as $\beta$ Pictoris. We used here the framework of \cite{2012A&A...547A.111M} for planet population synthesis to deepen this hint. It handles the formation of single planets in disks around solar-mass stars and account for disk evolution and migration processes (type I, type II). It can be used to generate population of planets with individual properties that can be compared to the observations. We modified the code in order to simulate the formation of planets around a 1.7 $M_{\odot}$ star at solar-metallicity.  The disk mass distribution was shifted by a factor $\mathrm{M_{star}^{1.2}}$ with respect to the \cite{2012A&A...547A.111M} models in order to get an accretion rate on the star which varies approximatively as $\mathrm{M_{star}^{2}}$ \citep[with $\mathrm{M_{star}}$ the stellar mass; see][]{2011A&A...526A..63A}. We show in Figure \ref{Fig:FigPopsynth} the resulting populations of 50 000 planets. Synthetic planets as massive as  $\beta$ Pictoris b (considering masses from the hot-start models or  dynamical constraints on the mass) can be found at 12 Myr around such a massive star up to $\sim$14 AU.  Generally, this population synthesis shows that massive planets such as $\beta$ Pictoris  b represent the boundary of the population of planets for the considered semi-major axis. However, we want to stress that these simulations are defined here for a given set of model parameters and that several assumptions are made. An important assumptions  in this context is that the gas accretion rate is not reduced due to gap formation \citep{1999ApJ...526.1001L} because of the eccentric instability \citep{2006A&A...447..369K}. We did not included in these simulations disk heating by the irradiation of the host star (viscous heating only). Next, one planet embryo is injected per disk. This situation might not be realistic for $\beta$ Pictoris given observational clues for additional -- but lower mass --  planets in the system \citep[e.g.][]{2004Natur.431..660O}. We also took here a core-accretion rate  of \cite{1996Icar..124...62P} that might be too much optimistic \citep{2013A&A...549A..44F}, meaning that we  assume that massive cores form relatively quickly. We also assume that massive cores can be formed. In the range of semi-major axis of $\beta$ Pictoris b (8-10 AU), massive planets ($\mathrm{> 9\:M_{Jup}}$) have predicted core masses of 160 to 230 $M_{\oplus}$. Nevertheless, recent detections of dense transiting companions suggest that this hypothesis is relevant \citep{2008A&A...491..889D, 2012A&A...538A.145D}.   Simulations studying the impact on the choice of free parameters (and hypothesis made in the models) and accounting for multiple planets formation per disk would now be needed to better understand the formation of the planet. A forthcoming study (Molli\`{e}re, in preparation) coupling the formation and evolution (luminosity, $\mathrm{T_{eff}}$) of synthetic planets will also enable to better understand if core-accretion can form high entropy objects at the planet separation.

\section{Conclusions}
We present the first resolved J, H, and M' band images of the exoplanet $\beta$ Pictoris b. We extract fluxes of the planet at these wavelengths and combined them to published observations at 2.18, 3.8, and 4.05 $\mu$m in order to build the spectral energy distribution of the planet from 1 to 5 $\mu$m.

 We find that the planet has similar colors and flux to young and old field L0-L4 dwarfs, and in particular to the $\sim$30 Myr old  planet/brown-dwarf companion recently discovered around the massive star $\kappa$ Andromedae. We use these broad spectral type estimates and bollometric corrections defined for young objects to derive  $\mathrm{Log_{10}(L/L_{\sun})=-3.87\pm0.08}$ for the planet. Comparison of the planet SED to atmosphere models shows  that the atmosphere contains dust and gives a more robust estimate of its effective temperature ($\mathrm{T_{eff}=1700\pm100}$K). Models also suggest that the planet has a reduced surface gravity  ($\mathrm{log\:g =4.0 \pm 0.5}$) characteristic of young objects.  

We  measure the position of the planet in our new data and combine it with existing astrometry to refine the distribution of orbital parameters. We confirm that the semi-major axis lies in the 8-10  AU range and the extentricity is lower or equal to 0.15 (80\% probability). This validates the independent upper limit on the mass of the companion of 12 (for an orbit at 9 AU) and 15.5 $\mathrm{M_{Jup}}$ (10 AU) derived from radial velocity. 

We compare this independent mass determination to predictions of three classes of evolutionary tracks based on different initial conditions.  We confirm that masses predicted by  \cite{2007ApJ...655..541M} and \cite{2012ApJ...745..174S} ``cold-start" evolutionary models for the estimated $\mathrm{T_{eff}}$ and photometry of $\beta$ Pictoris b  do not comply with dynamical constraints. Instead, we find that classical ``hot-start" models successfully predict the temperature and luminosity of the object for masses of $\mathrm{10^{+3}_{-2}M_{Jup}}$ and  $\mathrm{9^{+3}_{-2}\:M_{Jup}}$ respectively. ``Warm-start" models, which deal with the intermediate cases between ``cold" and ``hot-starts" predict masses $\geq$6 $\mathrm{M_{Jup}}$. These models also constrain the initial entropy of the planet to $\mathrm{S_{init} \geq 9.3 k_{b}/baryon}$. Initial entropy values are then at least intermediate between those considered for hot-start and cold-start models, and most likely closer to those of hot-start models\footnote{At the time of acceptance of this study, Marleau \& Cumming  (2013, in prep.) find close simultaneous constraints on the mass and initial entropy of $\beta$ Pictoris b by comparing the planet's bolometric luminosity to their custom ``warm-start" models.}. We argue that these values can not presently  be used to identify the formation scenario of the planet. We generate a population of planets formed by core-accretion around a star as massive as $\beta$ Pictoris. Some planets of the population have masses and semi-major axis which agrees with current estimates for $\beta$ Pictoris b. We show in addition that synthetic analogs to $\beta$ Pictoris b are at the edge of the core-accretion capabilities. A more complete version of these models better adapted to the specificities of the $\beta$ Pictoris system and which explores several free parameters set in the models would now be needed to better understand the formation of the system.

$\beta$ Pictoris b will likely be used as a reference target for the science verification of the planet imagers SPHERE and GPI next year. These two facilities  should pursue the monitoring of the planet orbital motion, at a time when it should be close to periastron. The intruments should also easily provide the first low-resolution (R $\lesssim$400) near-infrared (1-2.5 $\mu$m) spectrum of the planet. Spectral features, such as the pseudo-continuum shape from 1.5 to 1.8 $\mu$m,  will enable to firmly confirm the intermediate surface gravity of the companion. We could also expect to estimate the metallicity (possibly from the shape of the K-band pseudo-continuum) of the planet.
 
 \begin{appendix} 
 \section{Transmission of neutral density filters}
We measured the transmission of the two neutral densities (ND\_short, ND\_long)  installed in NaCo for the J, H, $\mathrm{K_{s}}$, L', and M' filters  using datasets gathered from the ESO archives where bright targets were observed consecutively with and without a neutral density (see Table \ref{tab:transmobslog}). Each dataset was reduced with the ESO \textit{Eclipse} package \citep{1997Msngr..87...19D}.  \textit{Eclipse} carried out sky and/or dark subtraction, the division by the flat field, bad pixel flagging and interpolation. The scripts finally shifted and added dithered images to produce a post-cosmetic frame of the sources. We performed aperture photometry on the sources for each setup and compared them to retrieve the transmission factor (different integration radii were considered depending on the dataset).  We repeated the procedure for different targets for a given filters in order to estimate an error on the transmission factor. Values are reported in Table \ref{tab:transmobs}.  We considered in addition the transmission factor reported in \citep{2008A&A...482..939B} for the ND\_short neutral density for the computation of the error in the $K_{s}$ band. To conclude, we only got a single measurment of the ND\_long neutral density  transmission in the M' band. We considered a higher error for this transmission based on the impact of the level of fluctuations of the background around the target.

 \label{App:transm}
 \begin{table*}[t]
\begin{minipage}[ht]{\linewidth}
\caption{Observations log for the calibration of the neutral density transmission.}
\label{tab:transmobslog}
\begin{center}
\renewcommand{\footnoterule}{}  
\begin{tabular}{llllllllll}
\hline \hline 
Date					&		Target		&	Camera	&		Filter		& Neutral density		&		DIT		&		NDIT			&		Nexp		&		EC Mean		&		$\tau_{0}$ mean	\\
						&						&					&						&								&		(s)			&						&						&		(\%)				&q		(ms)					\\
\hline
07/04/2004		&	HIP59502		&	S13			&			J			&		ND\_short &		1.500		&			25		&		6		&		30.8  &   6.3 		\\		
07/04/2004		&	HIP59502		&	S13			&			J			&		\dots		&		0.350		&			35		&		6		&		18.9  &   3.6 		\\		
07/04/2004		&	HIP61265		&	S13			&			J			&		ND\_short &		2.500		&			15		&		6		&		39.0	&	7.0 		\\
07/04/2004		&	HIP61265		&	S13			&			J			&		\dots		&		0.350		&			35		&		6		&		14.9	&	2.6 		\\
09/06/2004		&	HIP81949		&	S13			&			J			&		ND\_short &		2.200		&			20		&		6		&		33.6	&	4.0		\\
09/06/2004		&	HIP81949		&	S13			&			J			&		\dots		&		0.350		&			35		&		6		&		29.2	&	2.7		\\
26/06/2004		&	HIP81949		&	S13			&			J			&		ND\_short &		2.200		&			20		&		6		&		38.7	& 2.1		\\
26/06/2004		&	HIP81949		&	S13			&			J			&		\dots		&		0.350		&			35		&		6		&		17.6	&	1.2		\\
28/11/2005		&	AB Dor B			&	S13			&			J			&		ND\_short	& 10.000		&			3		&		6		&		35.9		&		5.3  \\
28/11/2005		&	AB Dor B			&	S13			&			J			&		\dots			&	0.347		&			100		&		6		&		41.1		&		5.7	\\									
\hline
09/06/2004		&	HIP81949			&	S13		&		H			&	ND\_short	& 2.200		&		20		&		6		&		32.0		&		3.0		\\
09/06/2004		&	HIP81949			&	S13		&		H			&		\dots			&		0.350		&		 35			&		6		&		28.3	&		2.5		\\	
20/06/2004		&	GSPC S273-E	&	S13		&		H			&	ND\_short	& 10.000		&		5			&		6		&		22.2  &		2.4		\\
20/06/2004		&	GSPC S273-E	&	S13		&		H			&		\dots 			&		2.000		&			15		&		6		&		13.0		&		1.5		\\
28/11/2005		&	AB Dor B			&	S13			&			H			&		ND\_short	& 6.000		&			2		&		6		&		23.6		&		1.9  \\
28/11/2005		&	AB Dor	B		&	S13			&			H			&		\dots			&	0.347		&			100		&		6		&		29.6		&		2.3	\\									
\hline
09/06/2004		&	HIP81949		&		S13		&			$\mathrm{K_{s}}$	&		ND\_short	&  5.000		&		20		&		6		&		32.1		&		3.3		\\
09/06/2004		&	HIP81949		&		S13		&			$\mathrm{K_{s}}$	&		\dots			&	0.350		&		35		&		6		&		26.2		&		1.9		\\
28/11/2005		&	AB Dor 	B		&	S13			&			$\mathrm{K_{s}}$	&		ND\_short	& 2.200		&			3		&		6		&		27.8		&		3.2  \\
28/11/2005		&	AB Dor	B		&	S13			&			$\mathrm{K_{s}}$	&		\dots			&	0.347		&			100		&		6		&		43.5		&		4.7	\\									
\hline
25/12/2009		&	GSC 08056-00482		&		L27		&		L'		& ND\_long	&  0.200			&		100			&		5		&		28.9		&		0.9		\\
25/12/2009		&	GSC 08056-00482		&		L27		&		L'		&		\dots		&		0.200		&		150		&		5  &  42.3		&		2.9		\\
27/12/2009		&	WW PsA						&		L27		&		L'		&		ND\_long	&  0.200			&		100			&	5		&		25.3		&		3.4		\\
27/12/2009		&	WW PsA						&		L27		&		L'		&		\dots		&		0.200		&		150			&		5		&		30.2		&		12.9		\\
28/12/2009		&	 TX PsA							&		L27		&		L'		& ND\_long	&  0.200			&		100			&		4		&	36.4		&		2.0		\\
28/12/2009		&	 TX PsA							&		L27		&		L'		&		\dots		&	0.200		&		150		&		4		&		41.8		&		5.7		\\
\hline
03/03/2007		&	HD158882	&	L27			&			M'			&		ND\_long	& 0.056		&			70		&		4		&		n.a.		&		n.a.  \\			
03/03/2007		&	HD158882	&	L27			&			M'			&		\dots	& 0.056		&			70		&		4		&		n.a.		&		n.a.  \\			
\hline
\end{tabular}
\end{center}
\end{minipage}
\end{table*}

\begin{table*}[t]
\begin{minipage}[ht]{\linewidth}
\caption{Transmission of the neutral density filters}
\label{tab:transmobs}
\begin{center}
\renewcommand{\footnoterule}{}  
\begin{tabular}{lll}
\hline \hline 
Neutral density		&		Filter					&		Transmission (\%)	\\
\hline
ND\_short			&		J							&		$1.42\pm0.04$	 	\\
ND\_short			&		H							&		$1.23\pm0.05$	    \\
ND\_short			&		$\mathrm{K_{s}}$&		$1.13\pm0.04$	   	\\	
ND\_long				&		L'							&		$2.21\pm0.07$ 	 	\\
ND\_long				&		M'							&		$2.33\pm0.10$ 	 	\\
\hline
\end{tabular}
\end{center}
\end{minipage}
\end{table*}

\section{Details on comparison objects}
We report in Table \ref{tab:compobj}  the properties of the young companions and binaries used as comparison objects in Figures \ref{Fig:Fig2} to \ref{Fig:Fig4bis}.
 
\begin{table*}[t]
\begin{minipage}[ht]{\linewidth}
\caption{Properties of young companions and binaries used as comparison objects.}
\label{tab:compobj}
\begin{center}
\renewcommand{\footnoterule}{}  
\begin{tabular}{lllllllll}
\hline \hline 
Name		&			Distance			&			Association			&		Age			&		Spectral type A			&	Spectral type b&		Separation\tablefootmark{a}			&	 Mass		&			References \\
				&			(pc)					&								&			(Myr)				&				&				&			(AU)			  &		($\mathrm{M_{Jup}}$) & 	\\
\hline 				
HIP 78530B		&		$156.7\pm13.0$			&		Up. Sco.		&		5		&		B9V		&		$\mathrm{M8\pm1}$	&		$710\pm60$	&  $22\pm4$	  &  1 \\
GSC 06214-00210b		&		$145\pm14$	&	Up. Sco.		&	5	&		K7		&		$\mathrm{L0\pm1}$		&		$320\pm30$		&		 $14\pm2$		&	2, 3 \\
USCO CTIO 108B		&		$145\pm14$		&	Up. Sco.		&	5	& M7		&		M9.5		&		$670\pm64$		&		$14^{+2}_{-8}$		&	4 \\	
USCO CTIO 108A		&		$145\pm14$		&	Up. Sco.		&	5	& M7		&		M9.5		&		$670\pm64$		&		$60\pm20$		&	4 \\	
HR8799b						&		$39.4\pm1.0$		&		Columba		&	30	&		A5V		&	L-T		&		$67.5-–70.8$		&		$<7$		&		5,6,7,8 \\	
HR8799c						&		$39.4\pm1.0$		&		Columba		&	30	&		A5V		&	L-T		&		$42.1–-44.4$		&		$<10$		&		5,6,7,8 \\	
HR8799d						&		$39.4\pm1.0$		&		Columba		&	30	&		A5V		&	L-T		&		$26.4–-28.1$		&		$<10$		&		5,6,7,8 \\	
HR8799e						&		$39.4\pm1.0$		&		Columba		&	30	&		A5V		&	L-T		&		$14.5\pm0.5$		&		$<10$		&		6,7,9 \\	
2M1207A						&		$52.4\pm1.1$		&		TW Hydrae	&	8		&		M8 pec  & $\mathrm{\gtrsim L5_{\gamma}}$\tablefootmark{b}		&	$40.8\pm0.9$   &  $24\pm6$  &  10, 11 \\
2M1207b						&		$52.4\pm1.1$		&		TW Hydrae	&	8		&		M8 pec  & $\mathrm{\gtrsim L5_{\gamma}}$\tablefootmark{b}		&	$40.8\pm0.9$   &  $8\pm2$  &  10,11,12, 13 \\
SDSS J2249+0044A					&		$54\pm16$		&		Unknown	&	$\sim$100	&		$\mathrm{L3\pm0.5}$		&		$\mathrm{L5\pm1}$  & $17\pm5$	&		$29\pm6$	&	14 \\
SDSS J2249+0044B					&		$54\pm16$		&		Unknown	&	$\sim$100	&		$\mathrm{L3\pm0.5}$		&		$\mathrm{L5\pm1}$  & $17\pm5$	&		$22^{+6}_{-9}$	&	14 \\
AB Pic b					&$45.5^{+1.8}_{-1.7}$&		Carina	&	30		&		K1Ve & $\mathrm{L0 \pm 1}$	&	$248\pm10$ & 	$13-14$	&  15, 16, 17 \\			
1RXSJ1609-2105b	&	$145 \pm 20$	& Up. Sco. & 5	&	$\mathrm{K7V}$	&	$\mathrm{L4^{+1}_{-2}}$  &  $\sim$330  &  $8^{+4}_{-2}$  &  18, 19 \\
CD-35 2722 B &	$21.3\pm1.4$	&	AB Dor	&	$\sim$100 & M1Ve	&	$\mathrm{L4\pm1}$  &  $67\pm4$  & $31\pm8$ & 20 \\	
PZ Tel B		&		$51.5\pm2.6$ & $\beta$ Pic & 12	&	K0	&	$\mathrm{M7\pm2}$	&	$17.9\pm0.9$	&	$36\pm6$	&	21, 22 \\	
HR7329	B	&	$47.7\pm1.5$	&	 $\beta$ Pic & 12	&	A0  & M7.5  & $200\pm7$  &  $26\pm4$		&	23, 24 \\
TWA 5B		&		$44\pm4$	&	TW Hydrae	&  8		&		M2Ve	&	M8-M8.5	&		$\sim$98 & $\sim$20		&	25	\\
$\kappa$ Andromedae b & $51.6\pm0.5$ & Columba & 30 & B9IV &  L2-L8	&	$55\pm2$ & $12.8^{+2}_{-1}$		&	26	\\
\hline
\end{tabular}
\end{center}
\end{minipage}
\tablefoot{[1] - \cite{2011ApJ...730...42L}, [2] - \cite{2011ApJ...726..113I}, [3] - \cite{2011ApJ...743..148B}, [4] - \cite{2008ApJ...673L.185B}, [5] - \cite{2008Sci...322.1348M}, [6] - \cite{2011ApJ...732...61Z}, [7] - \cite{2012ApJ...755...38S}, [8] - \cite{2012ApJ...755L..34C}, [9] - \cite{2010Natur.468.1080M}, [10] - \cite{2008A&A...477L...1D}, [11] - \cite{2002ApJ...575..484G}, [12] - \cite{2004AA...425L..29C}, [13] - \cite{2012arXiv1206.5519F}, [14] - \cite{2010ApJ...715..561A}, [15] - \cite{2005A&A...438L..29C}, [16] - \cite{2010A&A...512A..52B}, [17] - Bonnefoy et al 2013, submitted, [18] - \cite{2008ApJ...689L.153L}, [19] - \cite{2010ApJ...719..497L}, [20] - \cite{2011ApJ...729..139W}., [21] - \cite{2010ApJ...720L..82B}, [22]- \cite{2010A&A...523L...1M}, [23] - \cite{2000ApJ...541..390L}, [24] - \cite{2011MNRAS.416.1430N}, [25] - \cite{1999ApJ...512L..69L}, [26] - \cite{2012arXiv1211.3744C}.}
\tablefoottext{a}{Projected separations with the exeption of HR8799bcd where the separations correspond to the latest orbital fit  \citep{2012ApJ...755L..34C}.\\} 
\tablefoottext{b}{Spectral type assumed here given the comparison with the $\mathrm{L5_{\gamma}}$  2MASSJ035523.51+113337.4 \citep{2012arXiv1206.5519F}.} 

\end{table*}

\end{appendix}
 
\bibliographystyle{aa}

\begin{acknowledgements}
We would like to first to thank our anonymous referee for improving greatly the quality of the manuscrit. We also thank the ESO Paranal staff members for conducting these service mode observations and members involved in the preparation and data analysis of the ESO large program 184.C-0567 for providing the calibration of the true north and instrument platescale corresponding to our January data. We are gratefull to Christiane Helling, Soeren Witte, and Peter Hauschildt for developping and providing the DRIFT-PHOENIX models.  We thank Hubert Klahr, Adam Burrows, and Joshua Schlieder for fruitfull discussions about planet formation/evolutionary models and the identification of young nearby L dwarfs. We also thank Kelly Cruz and Mark Marley for discussions about the classification scheme of young L dwarfs. .The research leading to these results has received funding from the  French ``Agence Nationale de la Recherche'' through project grant ANR10-BLANC0504-01, the ``Programme  National de Physique Stellaire'' (PNPS) of CNRS (INSU), and the European Research Council under the European Community's Seventh  Framework Programme (FP7/2007-2013 Grant Agreement no. 247060). It was also conducted within the Lyon Institute of Origins under  grant ANR-10-LABX-66. 

\end{acknowledgements}


\bibliography{BpicSED}

\end{document}